\documentclass[twocolumn,aps,prl]{revtex4-2}

\usepackage{tabularx}
\usepackage{amssymb}
\usepackage{array}
\usepackage{booktabs}
\usepackage{graphicx}
\usepackage{dcolumn}
\usepackage{bm}
\usepackage{braket}
\usepackage{amsmath}
\usepackage{algorithm}
\usepackage{algpseudocode}
\usepackage{amsfonts}
\usepackage{longtable}
\usepackage{enumitem}
\usepackage{adjustbox}
\usepackage{hyperref}

\begin{document}

\preprint{APS/123-QED}

\title{Quantum AI for Cancer Diagnostic Biomarker Discovery}

\author{Mandeep Kaur Saggi$^{1}$}
\email{drmandeepsaggi@gmail.com}

\author{Amandeep Singh Bhatia$^{1}$}
\email{drasinghbhatia@gmail.com }

\author{Humaira Gowher$^{2}$}
\email{hgowher@purdue.edu}

\author{Sabre Kais$^{1,3}$}
\email{skais@ncsu.edu}

\email{Corresponding author: hgowher@purdue.edu}
\email{Corresponding author: skais@ncsu.edu}

\affiliation{$^{1}$ Department of Electrical and Computer Engineering, North Carolina State University, NC, USA}
\affiliation{$^{2}$ Department of Biochemistry, Purdue University, IN, USA}
\affiliation{$^{3}$Department of Chemistry, North Carolina State University, Raleigh, NC 27695}

\begin{abstract}

Quantum machine learning (QML) offers a promising new paradigm for computational biology by leveraging quantum mechanical principles to enhance cancer classification, biomarker discovery, and bioinformatics diagnostics. In this study, we apply QML to identify subtype specific biomarkers for lung adenocarcinoma (LUAD) and lung squamous cell carcinoma (LUSC), the two predominant forms of non-small cell lung cancer. Our methodology involves a two-phase process: in Phase 1, differential expression analysis and methylation analysis between tumor and normal samples allows us to identify LUAD-specific and LUSC-specific genes, revealing potential prognostic biomarkers for cancer subtypes. Phase 2 focuses on developing a quantum classifier capable of distinguishing between LUAD and LUSC tumors, as well as between tumor and normal samples. This classifier not only enhances diagnostic precision but also demonstrates the quantum advantage in processing large-scale multiomic datasets. We evaluated the performance of a Quantum Neural Network (QNN) model for distinguishing LUAD and LUSC subtypes using three gene sets: Sample$_1$ (Hypermethylated + Downregulated), Sample$_2$ (Hypomethylated + Upregulated), and Sample$_3$ (combination of Sample$_1$ and Sample$_2$). The QNN model was tested across varying parameters, including feature size, number of qubits, layers, epochs, and batch sizes. Our results consistently demonstrated that Sample$_3$, representing the combined gene set, achieved the highest overall predictive performance in all metrics. These results demonstrate that QML provides an effective and scalable approach for biomarker discovery and subtype specific cancer classification. In the classical phase, Kyoto Encyclopedia of Genes and Genomes (KEGG) pathway analysis revealed strong enrichment in neuroactive ligand–receptor interaction, cyclic adenosine monophosphate (cAMP) signaling, glutamatergic and dopaminergic synapses, and axon guidance pathways. Gene Ontology (GO) enrichment analysis highlighted the significant involvement of genes in synaptic signaling, ion channel regulation, and neuronal development. In the quantum phase, KEGG analysis further identified enrichment in cancer-associated pathways, including neurotrophin, MAPK, Ras, and PI3K–Akt signaling, with key genes such as NGFR, NTRK2, and NTF3 suggesting a central role in neurotrophin-mediated oncogenic processes. Our findings highlight the growing potential of quantum computing to advance precision oncology and next-generation biomedical analytics.

	\end{abstract}
	
	\maketitle


\section{INTRODUCTION}
\noindent 

Quantum machine learning (QML) is pioneering a new era in computational biology, unleashing powerful tools that redefine the possibilities for tackling intricate biological challenges. The analyses of large-scale molecular data are beneficial for many aspects of oncology research, including the classification of possible subtypes, stages, and grades of cancer. Several approaches have been proposed to train networks using quantum computing technology more accurately, robustly, and efficiently \cite{1, 35, 36, 37, 38, 39, 40}.
However, hybrid strategies have recently emerged given the current limitations of quantum machines and the constrained number of available qubits. These strategies leverage existing technology to achieve practical and usable solutions \cite{2}.
Quantum-enhanced AI and Machine Learning (ML) methods are gaining attention as promising solutions to medical challenges. Recent advancements in quantum computing and quantum AI have demonstrated their wide-ranging applicability in healthcare. These methods have shown promise in various healthcare and drug-discovery domains and predicting ADME-Tox properties in drug discovery \cite{3} \cite{4} \cite{5}, including rapid genome analysis \cite{6} and sequencing \cite{7}, disease detection \cite{8}, reference crop evapotranspiration classification \cite{9}, for chemistry \cite{10}, Moreover, Quantum Computing (QC) models play a significant role in predicting gene mutations critical for the pathogenesis and diagnosis of specific cancer types, such as the Glioma Tumor Classification \cite{11}.
Cancer sub-type classification is crucial for understanding cancer pathogenesis and developing targeted treatments that can benefit patients the most \cite{11}. 

Lung cancer is the most commonly diagnosed malignant tumor and is a leading cause of cancer-associated mortality \cite{12}. It is the second highest cause of new cancer cases in both genders in the United States and is the second leading cause of cancer deaths in females globally. The most common subtypes of lung cancers are lung squamous cell carcinoma (LUSC) and lung adenocarcinoma (LUAD), classified together as non-small cell lung cancer (NSCLC) \cite{13} \cite{14}. 

The Cancer Genome Atlas (TCGA) LUAD and LUSC cohorts provide diverse omics data that are crucial for cancer research and subtype classification, accessed from the Xena web portal \cite{15}. Specifically, for LUSC and LUAD cancer subtype the dataset includes \textbf{DNA Methylation (DNAme): (485,578 identifiers X 492 samples)} This data helps identify methylation patterns that are specific to LUAD and LUSC subtypes, revealing regulatory changes associated with each cancer type (which means DNA methylation features are present (indicative of significant regulatory changes) or absent (no significant changes) in each subtype to understand their role in tumor biology). Each sample in this dataset represents tumor tissue from an individual patient, with data formatted as rows of gene identifiers and columns of sample beta values.DNA methylation beta values range from 0 to 1 and represent the proportion of methylated signal at a specific locus. Higher beta values indicate higher methylation (hypermethylation), while lower beta values indicate lower methylation (hypomethylation).
\textbf{RNA Sequencing (RNA-seq): 20,531 identifiers X 576 samples}  The gene expression profile was measured experimentally using the Illumina HiSeq 2000 RNA Sequencing platform by the University of North Carolina TCGA genome characterization center, helping to identify genes with differential expression between LUAD and LUSC subtypes. This dataset shows the exon-level transcription estimates, as in RPKM values (Reads Per Kilobase of exon model per Million mapped reads). For LUAD vs. LUSC subtype classification, the dataset includes primary tumor and solid tissue normal samples from which these multiomic features are derived. This approach allows for a thorough examination of molecular differences between the lung subtypes, and enhancing diagnostic accuracy \cite{14}. 

Traditional studies often analyze individual omic features in isolation, focusing on discrete datasets such as DNA Methylation patterns \cite{15}, gene expression profiles \cite{16}, \cite{17},or microRNA levels. In these contexts, a "sample" refers to a biological specimen collected from a patient, such as a piece of tumor tissue or a blood sample. Each sample provides specific molecular data representing that patient's unique biological characteristics. This isolated approach, which examines data from each sample separately, can miss valuable insights from the interactions and correlations between molecular features across different samples.

In recent years, machine learning has been increasingly applied to biomarker discovery \cite{18}. It employs mathematical and computational approaches to train models that can learn patterns from data and perform specific predictive or analytical tasks. Among the most relevant machine learning techniques for biomarker identification are classification and feature selection. Classification is a form of supervised learning in which the algorithm is trained on labeled samples, each represented by a set of features. The goal is to learn a mapping function that can accurately predict the label of new, unseen samples based on their features\cite{18}.

Machine learning and deep learning approaches have been widely applied to TCGA,  LUAD and LUSC cohorts for biomarker discovery, subtype classification, and prognosis prediction \cite{19}. Several studies employed supervised learning algorithms, including random forests, support vector machines, and LASSO, on gene expression and methylation data to identify predictive biomarkers and classify lung cancer subtypes with high accuracy \cite{19, 20, 21}. Deep learning methods, such as autoencoders and hybrid neural networks, have been utilized to reduce high-dimensional omics data and extract key features, resulting in the identification of small sets of robust biomarkers with AUC values often exceeding 0.98 \cite{22, 23}. Integrative multi-omics approaches combining gene expression, miRNA, and methylation data have further improved predictive performance and revealed biologically meaningful interactions \cite{24}.

Despite these successes, several common challenges remain. The high dimensionality of omics data relative to limited sample sizes increases the risk of overfitting, emphasizing the need for careful feature selection. Additionally, deep learning models often lack interpretability, which necessitates the use of explainable AI methods to clarify the biological relevance of identified features. Tumor heterogeneity and microenvironmental effects can further complicate model generalization across different patient cohorts \cite{19, 22, 24}. Multiomic Data: Integrating various data types (e.g., genomics, transcriptomics, proteomics, epigenomics) leads to complex, high-dimensional datasets with different structures and scales. Handling this in classical machine learning is already challenging due to the large feature space and correlations between different data modalities.
In recent years, QML has attracted a great deal of attention due to its great potential in improving algorithms for increasing accuracy, speeding up, and using lesser computational resources for biomarker discovery in cancer research \cite{25} \cite{26}. QML is a rapidly growing research field at the intersection of QC and ML. Executing ML for the analysis of big data needs extensive computational operations. To improve the computing process, quantum methods have found great interest for their significant properties in saving time and computational resources. Quantum computations perform calculating operations more efficiently than classical ones based on quantum mechanical principles including superposition, measurement, and entanglement.By harnessing quantum phenomena, quantum computing devices offer the potential to accelerate the solution of a wide range of complex problems. In this Perspective, we explore the potential impact of quantum computing on computational biology. While acknowledging the current limitations of quantum computing devices, we highlight promising directions for future research in the emerging field of quantum computational biology.

\begin{figure*}	[!ht]
	\centering
	\includegraphics[scale=0.34]{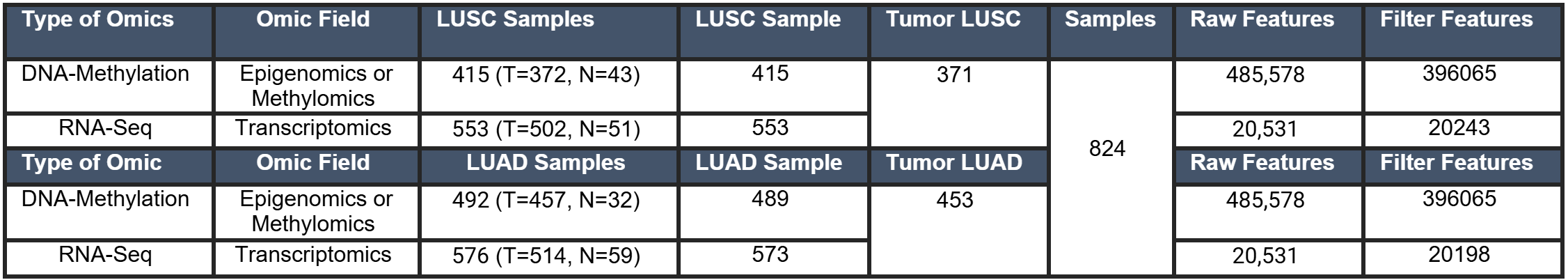}
\caption{\textbf{Summary of TCGA multi-omic datasets used in this study.} The analysis includes two NSCLC subtypes from TCGA: lung adenocarcinoma (LUAD) and lung squamous cell carcinoma (LUSC), comprising a total of 824 patient samples. Multi-omic data include DNA-methylation (epigenomics) and RNA-Seq (transcriptomics). Raw high-dimensional features (485,578 methylation probes and 20,531 genes) were filtered to obtain 396,065 methylation features and ~20k transcriptomic features for downstream analysis.}
\label{Figure1}
\end{figure*}

Lung adenocarcinoma  and lung squamous cell carcinoma are the two major subtypes of non-small cell lung cancer, yet they exhibit overlapping clinical features that complicate accurate subtype classification. Identifying \textbf{LUAD-specific tumor genes} is crucial for precise diagnosis, targeted therapy, and patient stratification. Classical differential gene expression (DGE) analyses using RNA-seq and DNA methylation data allow the discovery of dysregulated genes between tumor and normal tissue, but these approaches are often limited in capturing complex molecular interactions and multi-omic dependencies.

Our research aims to integrate \textbf{multi-omic datasets} including DNA methylation and RNA expression with \textbf{quantum-classical hybrid approaches} to improve the identification of subtype-specific biomarkers. Specifically, we ask:  How can multi-omic integration reveal molecular features and novel biomarkers that effectively distinguish tumor tissue from normal samples?   Can the \textbf{QuBID-Omics framework}, leveraging quantum-inspired machine learning techniques, differentiate LUAD and LUSC subtypes more accurately than classical methods?  Can quantum-classical hybrid models enhance the identification of subtype-specific biomarkers through a more comprehensive analysis of genomic and epigenomic data? By addressing these questions, our study seeks to combine \textbf{classical biological insights with quantum-inspired computational power} to identify robust, clinically relevant LUAD-specific biomarkers, ultimately improving subtype classification and advancing precision oncology.

In our previous study we have developed the MQML-LungSC framework integrates DNA-Methylation, RNA-seq, and miRNA-seq data to explore the interconnectedness of molecular features across a wide range of tumor samples \cite{25, 26}. By examining the relationships between gene expression levels, methylation patterns, and microRNA profiles both within individual samples and across a population of samples, we can identify complex interactions and patterns that differentiate LUAD from LUSC subtypes. This integrated analysis allows for a more comprehensive understanding of the molecular mechanisms underlying these lung subtypes, potentially leading to more accurate classification and the discovery of significant features.

In Fig~.\ref{Figure1}, the development of lung cancer is illustrated by comparing normal cells to tumor cells. However, recent studies have suggested that LUAD and LUSC should be classified and treated as different cancers \cite{27}. Previous studies have utilized traditional feature selection and machine learning methods for cancer diagnosis, detection, and classification, but few have extended them to study potential features and biological pathways to discriminate between LUAD and LUSC \cite{12}. To improve cancer classification accuracy, novel machine learning and feature selection methods have been developed. However, few studies have used overlapping features from different methods for classification, gene expression analysis, and molecular features \cite{13} \cite{14}.

This work proposes a novel lung sub-type classification method that integrates classical DGE/DMR techniques with a quantum classifier. This hybrid approach aims to enhance the accuracy and robustness of sub-type classification, thereby potentially contributing to the development of more effective cancer therapies. We propose a set of hybrid quantum computing and advanced machine learning approaches to apply machine learning on small datasets, such as (primary tumor and normal) sample types in The Cancer Genome Atlas (TCGA). 

\noindent \textbf{Problem Formulation and Motivation:}
Accurate classification of lung cancer subtypes is essential for guiding clinical diagnosis and treatment decisions. The rapid growth of high-dimensional multi-omics data has driven the adoption of machine-learning approaches for cancer analysis, enabling more comprehensive biomarker discovery. However, integrating and analyzing heterogeneous multi-omics data at scale remains challenging due to data complexity, dimensionality, and computational limitations. This study aims to develop a hybrid classical–quantum learning framework for non-small cell lung cancer (NSCLC) classification. Specifically, we (i) classify LUAD and LUSC cohorts under tumor–normal and tumor–tumor subtype settings, (ii) identify significant multi-omics features from RNA-seq and DNA methylation data and analyze their regulatory consistency (hypomethylation–upregulation and hypermethylation–downregulation), and (iii) integrate top-ranked biomarkers into a quantum neural network to evaluate diagnostic performance, scalability, and model stability by progressively increasing feature dimensionality on a quantum simulator.

\begin{figure*}	[!ht]
	\centering
	\includegraphics[scale=0.3]{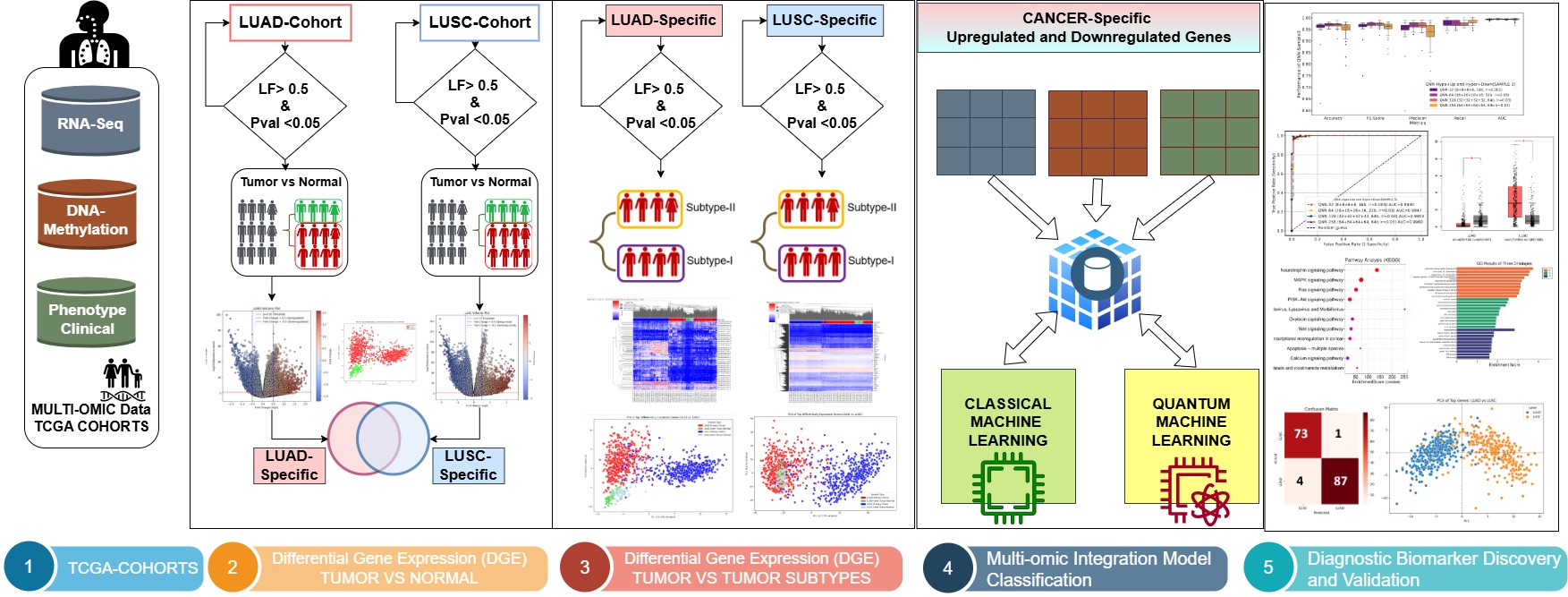}
	\caption{ Overview of the proposed QUBID framework for LUAD vs. LUSC classification. The pipeline consists of five phases: (1) multi-omics data collection (DNA methylation, RNA-seq, and clinical data) with preprocessing and filtering; (2) differential analysis (DGE/DMR) between tumor and normal samples; (3) subtype-specific analysis (LUAD vs. LUSC) to identify cancer-specific genes; (4) feature ranking, multi-omics integration, and classification using a Quantum Neural Network (QNN) with comparison to classical machine learning models; and (5) validation of identified biomarkers through functional enrichment (GO/KEGG), performance metrics, and visualization analyses.}
    \label{Figure2}
\end{figure*}
The remainder of the article is organized as follows. Section II presents the Materials and Methods, including the dataset, processes, and methods. Section III outlines the proposed methodology of our framework, detailing the implementation process and describing the multi-omic quantum machine learning approach for Phase1, Phase2, Phase3, Phase4 and Phase5. Section IV provides the results and performance analysis, Section V provides model compilation parameters, and final section concludes the paper and explores future research directions.
\section{MATERIAL AND METHODS}
This section provides a comprehensive overview of the methods utilized in our study. We describe the feature engineering process, feature selection methods, and proposed quantum neural network model for diagnosing lung dataset subtypes LUAD and LUSC. Each subsection below elaborates on these methods in detail. 
\subsection{\textbf{Study Population}}
Multi-omics datasets for Lung Adenocarcinoma (LUAD) and Lung Squamous Cell Carcinoma (LUSC) were obtained from the publicly available TCGA Multi-Omics Cancer Benchmark repository (http://cancergenome.nih.gov/). This study integrated DNA methylation, RNA-seq transcriptomic data, and associated clinical information to investigate lung cancer subtype classification and molecular feature identification within a quantum–classical framework. The DNA methylation dataset (Illumina HumanMethylation450 platform) included 492 LUAD and 415 LUSC patients, each with 485,578 CpG probes represented as beta values (0–1). The RNA-seq dataset comprised 576 LUAD and 553 LUSC samples, each containing 20,531 gene expression features. Gene expression profiles were generated using the Illumina HiSeq 2000 platform and processed as Level 3 RSEM normalized counts (log2(x+1) transformed). Gene identifiers were mapped to HUGO symbols using the UCSC Xena probeMap (hg19, Gencode V24lift37).

\subsection{\textbf{Data Loading and Data Pre-processing}}
In the First phase, MQML takes any number of omic measures such as genomic, epigenomic, and transcriptomic datasets as input. Due to the inherent complexity and size of the dataset, pre-processing steps were necessary to ensure data quality and reduce computational burden. The LUAD and LUSC, Lung datasets are acquired from the Multi-Omics Cancer Benchmark TCGA Pre-processed Data, It consists of four types of omics datasets for a case study of lung squamous cell carcinoma.: RNA-seq, DNAme, and Clinical patient information.    In the second phase, each omic has several gene features column-wise and patient samples row-wise. The raw data of each Omic1$_{Subtype-I}$ and Omic1$_{Subtype-II}$ is combined column-wise, with patient samples row-wise. Then, features that contain a sum of 0 are removed. To extract and select the survival/clinical attributes of both subtypes, combine the survival and clinical attributes based on sample type, i.e., diagnostic subtype-I and subtype-II.

\subsection{\textbf{Feature Engineering and Feature Selection}}
In the second phase of our study, we performed comprehensive feature engineering and statistical feature selection to refine multi-omic datasets for downstream analysis including RNA-seq and DNA methylation data. Missing or zero-valued entries were removed, and expression values were standardized using z-score normalization to ensure comparability across samples. 

Specifically, clinical samples corresponding to LUAD subtypes were first stratified into Subtype-I and Subtype-II, resulting in separate matrices for each omic layer, denoted as Omic1.1${Subtype-I}$ and Omic1.2${Subtype-II}$. To identify molecular features that significantly discriminate between tumor and normal samples, we applied classical statistical tests, including unpaired t-tests, to compare the mean expression or methylation levels of each gene between the two groups. This step allowed the detection of differentially expressed genes (DEGs) or differentially methylated regions (DMRs) that are indicative of tumor-specific alterations. Multiple testing correction was applied using the Benjamini-Hochberg procedure to control the false discovery rate, ensuring that only statistically robust features were retained. Genes meeting the adjusted p-value threshold ($ \le 0.05 $) and exhibiting biologically meaningful log2 fold changes were flagged as significant, providing a preliminary but stringent set of candidate biomarkers. Significant genes were further categorized as upregulated or downregulated based on fold-change thresholds, and their variance across samples was computed to prioritize highly variable, biologically informative features. For RNA-seq data, top upregulated and downregulated genes were extracted for downstream analyses, including PCA visualization, hierarchical clustering, and volcano plot representation. This approach enabled the identification of subtype-specific molecular signatures while reducing noise from non-informative features.

\section{METHODOLOGY}
We present a pioneering quantum-based biomarker identification and diagnostic framework for multi-omics lung cancer (QUBID) to differentiate between subtypes.

\subsection{Overview Steps for the QUBID Framework} 
\subsection{Phase 1: Tumor Cancer-specific genes between Tumor vs Normal}
In this step, Identifying differential DNA methylation regions (DMRs) and RNA expression genes (DEGs) in the LUAD  and LUSC cohort.

\begin{itemize}[noitemsep, topsep=1pt, left=10pt]
\item \textbf{Step 1.1:} LUAD tumor vs LUAD normal (Cohort I) — \textbf{(F1)} 
    \begin{itemize}[noitemsep, topsep=1pt, left=15pt]
        \item \textbf{Step 1.1.1:} LUAD Tumor vs LUAD Normal (denoted as \text{LUAD}$_{TN}$) for RNA-Seq from TCGA.
        \item \textbf{Step 1.1.2:} LUAD Tumor vs LUAD Normal (denoted as \text{LUAD}$_{TN}$) for DNA-Methylation from TCGA.
    \end{itemize}
\end{itemize}
\begin{itemize}[noitemsep, topsep=1pt, left=10pt]
\item \textbf{Step 1.2:} LUSC Tumor vs LUSC Normal (Cohort II) — \textbf{(F2)} 
    \begin{itemize}
        \item \textbf{Step 1.2.1:} LUSC Tumor vs LUSC Normal (denoted as \text{LUSC}$_{TN}$) for RNA-seq from TCGA.
        \item \textbf{Step 1.2.2:} LUSC Tumor vs LUSC Normal (denoted as \text{LUSC}$_{TN}$) for DNA-Methylation from TCGA.       
    \end{itemize}
\end{itemize}
\subsection{Phase 2: Tumor Cancer-specific genes between Tumor (Type-I) vs Tumor (Type-II}
\begin{itemize}[noitemsep, topsep=1pt, left=10pt]
\item \textbf{Step 2.1:} Luad Tumor vs Lusc Tumor (I-II subtypes)------(Cohort1 and Cohort2) — \textbf{(F3)} 
    \begin{itemize}
        \item \textbf{Step 2.1.1:} LUAD Tumor vs LUSC Tumor (denoted as \text{LUAD}$_{TT}$) for RNA-Seq from TCGA.
        \item \textbf{Step 2.1.2:} LUAD Tumor vs LUSC Tumor (denoted as \text{LUAD}$_{TT}$) for DNA-Methylation from TCGA.
    \end{itemize}
\end{itemize}
\subsection{Phase 3: Integrating with (Up+hypo) and (Down+Hyper) for Luad cohort}
\begin{figure}[!ht]
	\centering
	\includegraphics[scale=0.65]{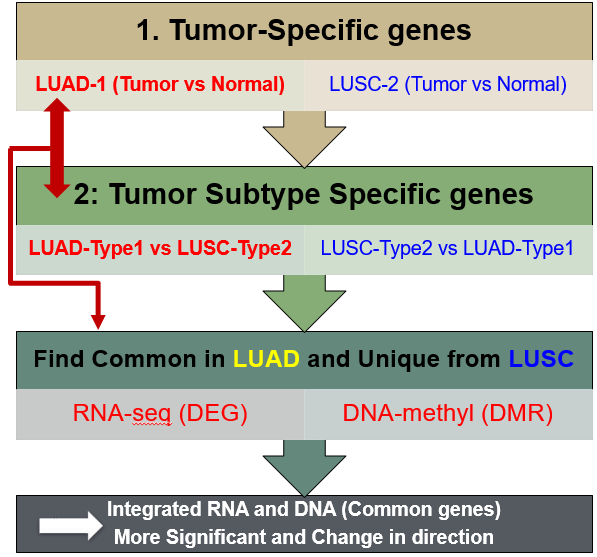}
	\caption{Phase1-Phase2 Workflow Process with Integration of RNA and DNA}
     \label{Figure3}
\end{figure}

\begin{itemize}[noitemsep, topsep=1pt, left=10pt]
\item \textbf{Step 3.1:} Find Common and Uncommon genes using overlapping between Cohort-I, Cohort-II and Cohort-III as shown in Fig~.\ref{Figure3},.
\item \textbf{Step 3.2:} Find F1, F2, F3 Comparison Identification algorithm using cross-type luad-biased gene sets using direction and significance level from statistical parameters. Divide the genes into categories) based on Pvalue and log Fold change:
\item \textbf{Step 3.3:} Identification of top-ranked integrated common genes from RNA and DNA for common \text{LUAD}$_{Tumor}$ and \text{LUSC}$_{Tumor}$ (lung subtype) clinical patients, including the attributes: \text{Gene}, \text{CpG}, \text{Ensembl\_ID}, $\text{DNA}_{\Delta\beta}$, \text{DNA\_AdjPvalue}, \text{DNA\_Significance}, \text{DNA\_Direction}, \text{DNA\_is\_promoter}, \text{RNA\_Log2FC}, \text{RNA\_AdjPvalue}, and \text{RNA\_Direction} for feature sets F1, F2, and F3. Based on these parameters, the \text{CombinedEffect} is computed and represented by the \text{Final\_Category} along with its \text{Interpretation} as $\text{Sample}_{1}\{\text{Hypermethylated} + \text{Downregulated}\}$, $\text{Sample}_{2}\{\text{Hypomethylated} + \text{Upregulated}\}$, $\text{Sample}_{3}\{\text{Hypermethylated} + \text{Downregulated and Hypomethylated} + \text{Upregulated}\}$.
 save into .CSV file.
\end{itemize}
\subsection{Phase 4: Integration with Quantum (QNN) Variants with Sample\_1, Sample\_2, and Sample\_3 using different parameters}
\begin{itemize}[noitemsep, topsep=1pt, left=10pt]
    \item \textbf{Step 4.1:} Evaluation of $\text{Sample}_{1}$, $\text{Sample}_{2}$, and $\text{Sample}_{3}$ using different QNN variants $\text{QNN}_{32}$, $\text{QNN}_{64}$, $\text{QNN}_{128}$, and $\text{QNN}_{256}$ corresponding to DNA-RNA feature combinations (32--32, 64--64, 128--128; total dimensions 64, 128, 256) as shown in (Table~\ref{tab:Table1}), (Table~\ref{tab:Table2}), and (Table~\ref{tab:Table3}). 
    \item \textbf{Step 4.2:} Model parameters were optimized using gradient-based methods under varying learning rates, batch sizes, and circuit depths.
    \item \textbf{Step 4.3:}$\text{Sample}_{3}$ represents a combination of $\text{Sample}_{1} + \text{Sample}_{2}$.
    \item \textbf{Step 4.4:} Identify the top biomarkers, evaluate performance metrics for all $\text{QNN}$ variants, and generate visualization plots.
    \item \textbf{Step 4.5:} Evaluating with the final and Comparison with classical methods.
\end{itemize}

\subsection{Phase 5: Functional Enrichment Analysis Using Pathways and GO Tools for Steps 3.3 and 4}
In this phase, the top-ranked genes identified in Steps 3.3 and 4 were subjected to functional enrichment analysis using Gene Ontology (GO) and pathway tools. This analysis was performed to understand the biological significance of the identified biomarkers by examining their involvement in key biological processes, molecular functions, cellular components, and KEGG pathways. These annotations help reveal the underlying mechanisms associated with LUAD and LUSC subtypes and provide insights into potential diagnostic and therapeutic targets.

\subsection{Overview of Phase1-Phase2 developement}
\noindent \textbf{RNA-Seq Differential Gene Expression (DGE) and DNA-methylation Differentially Methylated Regions (DMRs) computational biology Methods} 
In this step, RNA-Seq data were analyzed using Differential Gene Expression (DGE) to identify the most significant genes between tumor and normal samples in Cohort-I and Cohort-II, as well as between tumor subtypes in Cohort-II. Statistical significance was determined using t-tests with p-value thresholds ($p < 0.05$) and log fold change criteria ($\text{LogFC} > 1$ or $0.5$ for upregulated, $\text{LogFC} < -1$ or $-0.5$ for downregulated) as Case1 and Case2. Differentially Methylated Regions (DMRs) analysis was applied to DNA methylation data, with regions showing adjusted beta values (DB) $> 0.5$ considered hypermethylated and DB $< -0.5$ considered hypomethylated as Case 1. This workflow enabled identification of top-ranked features for integration across RNA expression and DNA methylation, allowing evaluation of correlations between methylation changes and gene expression. In total, 60,488 RNA features and 485,577 DNA features were analyzed. Top-ranked features were filtered into three sets (F1, F2, F3) to enable stepwise evaluation and integration of RNA expression and DNA methylation changes, allowing assessment of whether methylation changes consistently correlate with gene expression across patients.
\begin{figure}[!ht]
	\centering
	\includegraphics[scale=0.4]{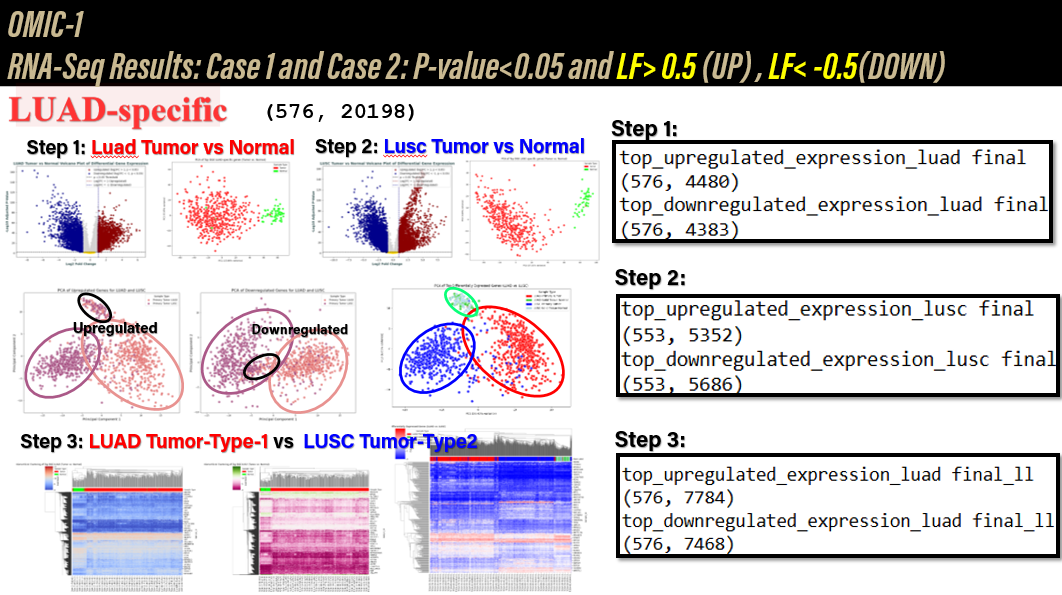}
    \includegraphics[scale=0.37]{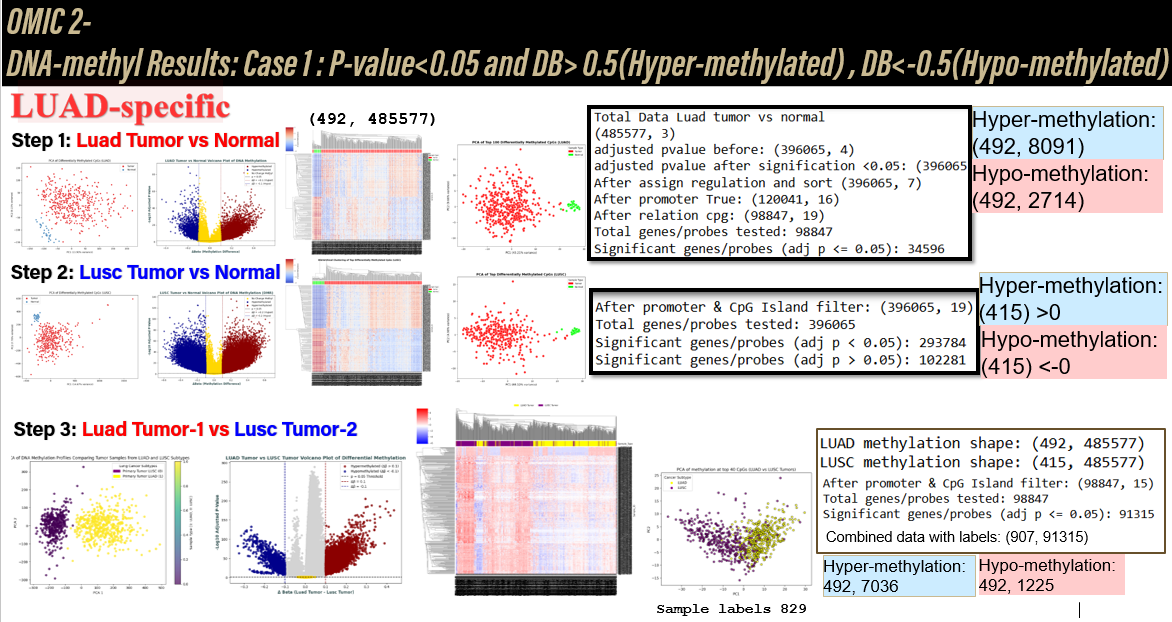}
    \includegraphics[scale=0.37]{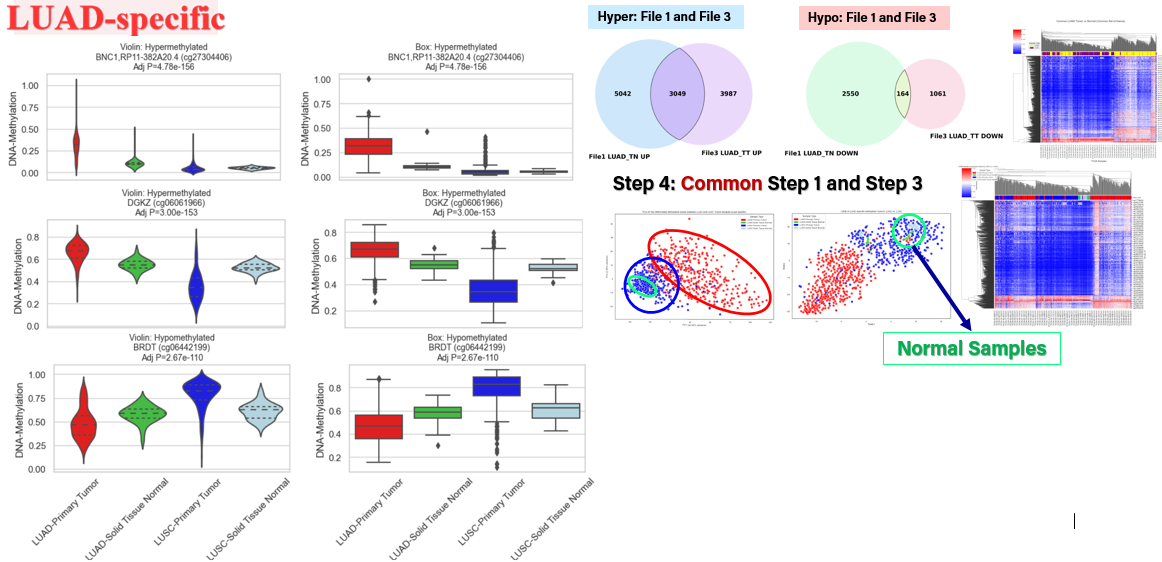}
	\caption{Phase1-Phase2 Workflow Process results of identifying genes}
    \label{Figure4}
\end{figure}

\subsection{Overview of Phase-1, Phase-2 and Phase-3 theoretical explanation for development and Integration}
In this step as shown in Fig~.\ref{Figure4},, we provide the detail theoretical overview of phase-1 and phase-2 DGE and DMR method to identify LUAD-specific using three key comparisons:
\begin{itemize}[noitemsep, topsep=2pt, left=10pt]
  \item \textbf{Set 1:} LUAD tumor vs LUAD normal (denoted as LUAD\_TN)
  \item \textbf{Set 2:} LUSC tumor vs LUSC normal (denoted as LUSC\_TN)
  \item \textbf{Set 3:} LUAD tumor vs LUSC tumor (denoted as LUAD vs LUSC \_TT)
\end{itemize}
Each gene is evaluated using:
\begin{itemize}[noitemsep, topsep=2pt, left=10pt]
  \item Adjusted p-value from statistical testing (after FDR correction)
  \item Delta-beta value ($\Delta\beta$) representing methylation difference
  \item Sign consistency of $\Delta\beta$ between LUAD\_TN and LUADvsLUSC
\end{itemize}

We define the LUAD-specific methylation gene set as:

\[
\mathcal{G}_{LUAD}^{spec} = 
\left\{
g \ \middle| \
\begin{aligned}
&\text{AdjP}_{g}^{LUAD\_TN} < \alpha,\quad |\Delta\beta_{g}^{LUAD\_TN}| > \delta \\
&\land \text{AdjP}_{g}^{LUSC\_TN} \geq \alpha \\
&\land \text{sign}(\Delta\beta_{g}^{LUAD\_TN}) = \text{sign}(\Delta\beta_{g}^{LUADvsLUSC}), \\
&\quad \text{AdjP}_{g}^{LUADvsLUSC} < \alpha
\end{aligned}
\right\}
\]

\noindent Where:
\begin{itemize}[noitemsep, topsep=2pt, left=10pt]
  \item $\alpha = 0.05$ is the FDR-adjusted p-value threshold
  \item $\delta = 0.2$ is the biological threshold for methylation difference
  \item $\text{sign}(x) \in \{-1, 0, +1\}$ indicates the direction of methylation change
\end{itemize}

We divide the full set into upregulated (hypermethylated) and downregulated (hypomethylated) subsets.\\

\noindent \text{Objective 1: LUAD-Specific Hypermethylated Genes}
Define:
\[
\begin{aligned}
U_1 &= \left\{ g \mid \Delta\beta_{g}^{LUAD\_TN} > \delta, \quad \text{AdjP}_{g}^{LUAD\_TN} < \alpha \right\} \\
U_2 &= \left\{ g \mid \Delta\beta_{g}^{LUSC\_TN} > \delta, \quad \text{AdjP}_{g}^{LUSC\_TN} < \alpha \right\} \\
U_3 &= \left\{ g \mid \Delta\beta_{g}^{LUADvsLUSC} > 0, \quad \text{AdjP}_{g}^{LUADvsLUSC} < \alpha \right\}
\end{aligned}
\]

Then the LUAD-specific hypermethylated gene set is:
\[
\boxed{
\mathcal{G}_{LUAD}^{UP} = (U_1 \setminus U_2) \cap U_3
}
\]

\noindent \text{Objective 2: LUAD-Specific Hypomethylated Genes}
\[
\begin{aligned}
D_1 &= \left\{ g \mid \Delta\beta_{g}^{LUAD\_TN} < -\delta, \quad \text{AdjP}_{g}^{LUAD\_TN} < \alpha \right\} \\
D_2 &= \left\{ g \mid \Delta\beta_{g}^{LUSC\_TN} < -\delta, \quad \text{AdjP}_{g}^{LUSC\_TN} < \alpha \right\} \\
D_3 &= \left\{ g \mid \Delta\beta_{g}^{LUADvsLUSC} < 0, \quad \text{AdjP}_{g}^{LUADvsLUSC} < \alpha \right\}
\end{aligned}
\]

Then the LUAD-specific hypomethylated gene set is:
\[
\boxed{
\mathcal{G}_{LUAD}^{DOWN} = (D_1 \setminus D_2) \cap D_3
}
\]

\noindent  The Final LUAD-specific DGE/DMR gene set is the union:
\[
\boxed{
\mathcal{G}_{LUAD}^{spec} = \mathcal{G}_{LUAD}^{UP} \cup \mathcal{G}_{LUAD}^{DOWN}
}
\]

The final results in CSV file contains the biological interpretation
\begin{itemize}
  \item Genes in $\mathcal{G}_{LUAD}^{UPregulated}$ are hypermethylated in LUAD only.
  \item Genes in $\mathcal{G}_{LUAD}^{DOWNregulated}$ are hypomethylated in LUAD only.
  \item These genes are not differentially methylated in LUSC and show subtype-specific differences.
\end{itemize}

\subsection{Luad-Specific based Phase-3 Biological/Computational for RNA/DNA Integration Common Genes different categories}

The goal of this analysis is to mathematically classify genes according to their differential expression patterns between LUAD and LUSC, and to identify LUAD-biased molecular signatures that may drive subtype-specific tumor biology. 
In this step, we analyze three differential expression comparisons:
\begin{itemize}
    \item $F1$: LUAD (Tumor vs. Normal)
    \item $F2$: LUSC (Tumor vs. Normal)
    \item $F3$: LUAD (Tumor vs. Tumor subtype)
\end{itemize}

Each dataset provides:
\begin{itemize}
    \item $\log_2(\text{Fold Change})$: logarithmic expression ratio between conditions
    \item $p_{\text{adj}}$: adjusted $p$-value (false discovery rate control)
\end{itemize}

\noindent \text{Expression Direction Encoding: } Each gene’s expression direction is encoded as:
\[
D_i =
\begin{cases}
+1, & \text{if } \log_2(FC_i) > 0.1 (Upregulated)\\
-1, & \text{if } \log_2(FC_i) < -0.1 Downregulated)\\
0,  & \text{if } |\log_2(FC_i)| \le 0.1 (No change)
\end{cases}
\]

where
\[
D_i \in \{-1,0,+1\}
\]
represents downregulated, neutral, and upregulated states, respectively.  
This discretization removes small, biologically irrelevant fold changes.\\

\noindent \text{Statistical Significance:} A gene $g$ is considered significant in dataset $k$ ($k \in \{F1,F2,F3\}$) if:
\[
S_{g,k} =
\begin{cases}
1, & \text{if } p_{\text{adj},g,k} < 0.05 \\
0, & \text{otherwise}
\end{cases}
\]

LUAD significance is defined as:
\[
S_{\text{LUAD},g} = \max(S_{g,F1}, S_{g,F3})
\]

and LUSC significance as:
\[
S_{\text{LUSC},g} = S_{g,F2}
\]

\noindent \textbf{Cross-type Gene Classification Logic:} This produces categorical gene-level groups. We compare the direction and significance of LUAD (\(D_{\mathrm{LUAD}}\)) and LUSC (\(D_{\mathrm{LUSC}}\)) as follows:
Let \(D_{\mathrm{LUAD}}(g)\) = direction from \(F_{1}\) (or \(F_{3}\) if missing), and  \(D_{\mathrm{LUSC}}(g)\) = direction from \(F_{2}\).

\textbf{(a) Opposite direction} \(\big(D_{\mathrm{LUAD}} \times D_{\mathrm{LUSC}} = -1\big)\)

\[
\begin{cases}
\text{LUAD-specific}, & \text{if } S_{\mathrm{LUAD}} = 1 \text{ and } S_{\mathrm{LUSC}} = 0, \\
\text{Opposite both significant}, & \text{if } S_{\mathrm{LUAD}} = 1 \text{ and } S_{\mathrm{LUSC}} = 1, \\
\text{Opposite weak}, & \text{if } S_{\mathrm{LUAD}} = 0 \text{ and/or } S_{\mathrm{LUSC}} = 0.
\end{cases}
\]

\textbf{(b) Same direction} \(\big(D_{\mathrm{LUAD}} \times D_{\mathrm{LUSC}} = +1\big)\)

\[
\begin{cases}
\text{Shared both significant}, & S_{\mathrm{LUAD}} = 1, \; S_{\mathrm{LUSC}} = 1, \\
\text{LUAD-dominant shared}, & S_{\mathrm{LUAD}} = 1, \; S_{\mathrm{LUSC}} = 0, \\
\text{LUSC-dominant shared}, & S_{\mathrm{LUAD}} = 0, \; S_{\mathrm{LUSC}} = 1, \\
\text{Shared non-sig}, & S_{\mathrm{LUAD}} = 0, \; S_{\mathrm{LUSC}} = 0.
\end{cases}
\]

\textbf{(c) Undefined or mixed:}
Neutral / No strong direction data.

\noindent \text{Significance Strength Ratio:} To quantify bias strength between LUAD and LUSC, the code computes:\[
\text{SigStrength}_{k,g} = -\log_{10}\!\left(p_{\text{adj},g,k}\right)
\] Then, the LUAD vs. LUSC bias ratio is:

\[
R_g =
\frac{
\max\!\left(-\log_{10}\!\left(p_{\text{adj},F1,g}\right),\,
-\log_{10}\!\left(p_{\text{adj},F3,g}\right)\right)
}{
-\log_{10}\!\left(p_{\text{adj},F2,g}\right) + \varepsilon
}
\]

where \(\varepsilon = 10^{-10}\) avoids division by zero. If \( R_g > 10 \), gene \( g \) is considered LUAD-biased. This ratio measures how much stronger LUAD’s significance is compared to LUSC’s. To measure relative significance strength between LUAD and LUSC:

\begin{itemize}
\item If \( R > 10 \), the gene is considered LUAD-biased (LUAD effect \( \gg \) LUSC effect).
\item This bias metric is applied within the shared, opposite, and neutral groups to derive refined LUAD-biased categories.
\end{itemize}

However, sometimes both LUAD and LUSC are significant, or one is only slightly weaker. Therefore, instead of using a binary decision of significant or not, we quantify how strong the LUAD signal is relative to the LUSC signal. Below summarizes the final LUAD-biased gene sets categories/classification identified from the analysis.

\begin{center}
\small
\begin{tabular}{l c}
\hline
Category & Count \\
\hline
Shared direction (both significant) & 12,264 \\
Neutral / No strong direction & 4,898 \\
Opposite but both significant (complex) & 1,198 \\
LUAD-dominant shared (only LUAD sig) & 1,086 \\
LUAD-specific (sig in LUAD, not in LUSC, opp dir) & 617 \\
Shared direction (non-sig) & 89 \\
LUSC-dominant shared (only LUSC sig) & 62 \\
Opposite but weak significance & 29 \\
\hline
\end{tabular}
\end{center}

\begin{figure*}[!ht]
	\centering
	\includegraphics[scale=0.35]{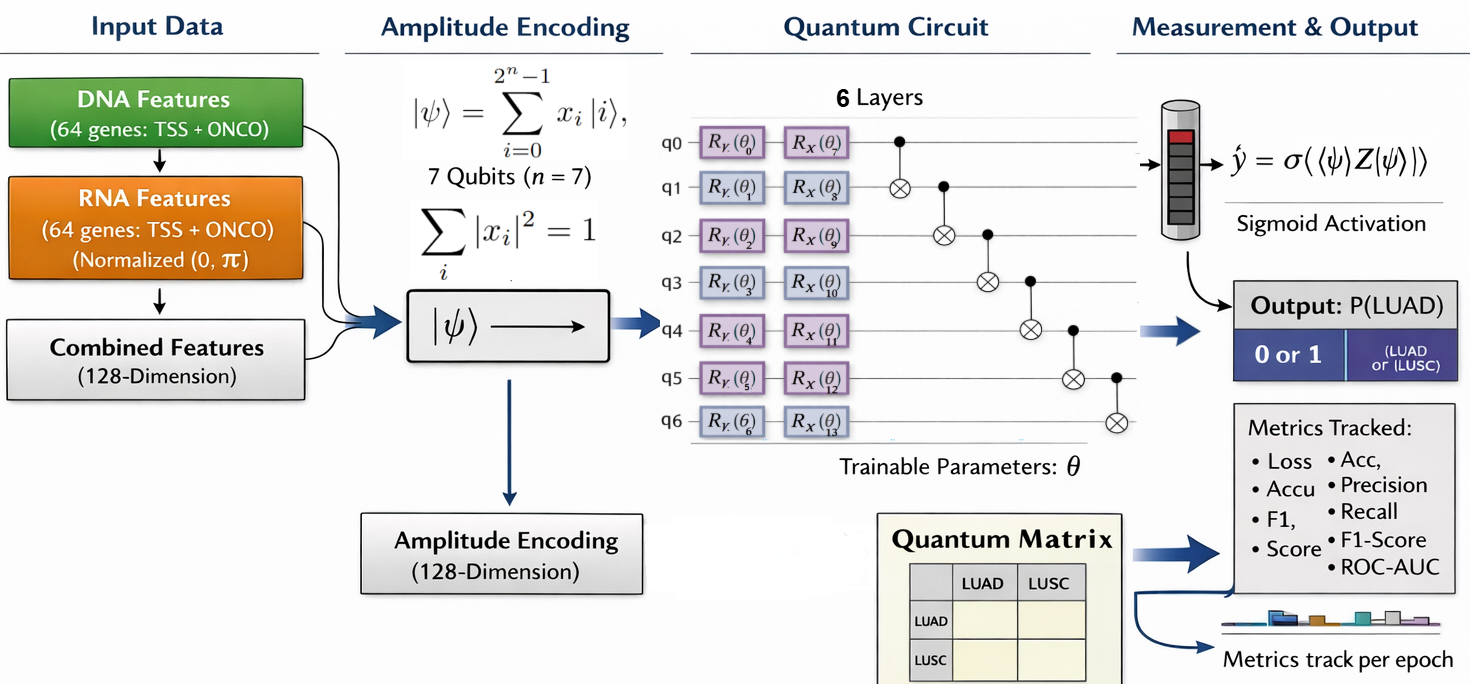}
	\caption{ \textbf{Phase 4: Quantum Model Workflow Process}. Schematic representation of the Phase 4 quantum modeling framework, illustrating the sequential workflow from input data preparation and feature encoding to quantum circuit construction, parameter optimization, model training, and final prediction output generation.}
     \label{Figure5}
\end{figure*}

\begin{table}[ht!]
\centering
\caption{QNN Performance Metrics for Sample 1 (Hypermethylated + Downregulated Genes)
}
\begin{tabular}{lcccccc}
\hline
Metric & QNN64& QNN64 & QNN64 & QNN128 & QNN256 & QNN256 \\
\hline
Genes & 32+32 & 32+32 & 32+32 & 64+64 & 128+128 & 128+128 \\
Total & 64 & 64 & 64 & 128 & 256 & 256 \\
Qubits & 6 & 6 & 6 & 7 & 7 & 8 \\
Layers & 6 & 6 & 6 & 6 & 6 & 6 \\
Epochs & 32 & 32 & 50 & 50 & 50 & 50 \\
LeRate & 0.03 & 0.003 & 0.03 & 0.03 & 0.03 & 0.03 \\
Tr-Acc & 0.9545 & 0.9363 & 0.9575 & 0.9469 & 0.9575 & 0.9499 \\
Te-Acc & 0.9455 & 0.9394 & 0.9576 & 0.9394 & 0.9515 & 0.9333 \\
Tr-F1 & 0.9593 & 0.9426 & 0.9620 & 0.9529 & 0.9614 & 0.9549 \\
Te-F1 & 0.9524 & 0.9444 & 0.9626 & 0.9462 & 0.9565 & 0.9418 \\
Tr-Prec & 0.9415 & 0.9324 & 0.9465 & 0.9291 & 0.9588 & 0.9458 \\
Te-Prec & 0.9184 & 0.9551 & 0.9375 & 0.9263 & 0.9462 & 0.9082 \\
Tr-Rec & 0.9779 & 0.9530 & 0.9779 & 0.9779 & 0.9641 & 0.9641 \\
Te-Rec & 0.9890 & 0.9341 & 0.9890 & 0.9670 & 0.9670 & 0.9780 \\
Tr-AUC & 0.9873 & 0.9639 & 0.9907 & 0.9790 & 0.9896 & 0.9757 \\
Te-AUC & 0.9964 & 0.9846 & 0.9881 & 0.9874 & 0.9936 & 0.9954 \\
\hline
\end{tabular}
\label{tab:Table1}
\end{table}

\begin{table}[ht!]
\centering
\caption{QNN Model Metrics for Sample 2 (Hypomethylated + Downregulated Genes)}
\begin{tabular}{lccccccc}
\hline
Metric & QNN64 & QNN64 & QNN64 & QNN64  & QNN128 & QNN256 & QNN256 \\
\hline
Genes& 32+32 & 32+32 & 32+32 & 32+32 & 64+64 & 128+128 & 128+128 \\
Total& 64 & 64 & 64 & 64 & 128 & 256 & 256 \\
Qubits& 6 & 6 & 6 & 6 & 7 & 8 & 8 \\
Layers& 6 & 6 & 5 & 6 & 6 & 6 & 6 \\
Epochs& 32 & 50 & 50 & 50 & 50 & 50 & 50 \\
LeRate& 0.003 & 0.003 & 0.03 & 0.03 & 0.03 & 0.03 & 0.03 \\
Tr-Acc & 0.9272 & 0.9196 & 0.9272 & 0.9575 & 0.9408 & 0.9484 & 0.9590 \\
Te-Acc & 0.9333 & 0.9091 & 0.9394 & 0.9455 & 0.9455 & 0.9455 & 0.9576 \\
Tr-F1 & 0.9351 & 0.9292 & 0.9353 & 0.9621 & 0.9478 & 0.9518 & 0.9636 \\
Te-F1 & 0.9405 & 0.9215 & 0.9462 & 0.9514 & 0.9529 & 0.9492 & 0.9622 \\
Tr-Prec & 0.9153 & 0.8992 & 0.9132 & 0.9441 & 0.9195 & 0.9767 & 0.9420 \\
Te-Prec & 0.9255 & 0.8800 & 0.9263 & 0.9362 & 0.9100 & 0.9767 & 0.9468 \\
Tr-Rec & 0.9558 & 0.9613 & 0.9586 & 0.9807 & 0.9779 & 0.9282 & 0.9862 \\
Te-Rec & 0.9560 & 0.9670 & 0.9670 & 0.9670 & 1.0000 & 0.9231 & 0.9780 \\
Tr-AUC & 0.9693 & 0.9734 & 0.9732 & 0.9934 & 0.9859 & 0.9890 & 0.9928 \\
Te-AUC & 0.9688 & 0.9690 & 0.9715 & 0.9835 & 0.9710 & 0.9905 & 0.9884 \\
\hline
\end{tabular}
\label{tab:Table2}
\end{table}

\subsection{Phase 4: Overview of Quantum (QNN) model  }

In this study, we developed a \textbf{Quantum Neural Network (QNN)} framework for the classification of \textbf{lung adenocarcinoma (LUAD)} and \textbf{lung squamous cell carcinoma (LUSC)} using integrated \textbf{DNA methylation} and \textbf{RNA expression} features. Our methodology combines classical preprocessing, feature selection, quantum-inspired amplitude encoding, and parametrized quantum circuits for supervised learning, Fig~.\ref{Figure5}.

\noindent \textbf{Feature Selection:} Gene annotations were loaded from a curated dataset containing \textbf{gene names} and \textbf{biological interpretations}. Features were selected based on:
\begin{enumerate}
    \item \textbf{Sample 1.}: DNA hypermethylation and RNA downregulation.
    \item \textbf{Sample 2.}: DNA hypomethylation and RNA upregulation.
\end{enumerate}

A non-overlapping top-$N$ selection was applied:
\begin{align}
\text{TopGenes} &= \text{pick\_top\_unique\_no\_overlap} \big( G, I, N, \notag \\
&\quad \text{exclude\_genes} \big)
\end{align}

where $G$ is the gene annotation dataframe, $I$ is the interpretation category, $N=64$, and \texttt{exclude\_genes} prevents duplication. The final sets included DNA genes: \texttt{dna\_tss + dna\_onco}, and RNA genes: \texttt{rna\_tss + rna\_onco}, yielding $128$ features per omics type.

\noindent \textbf{Integration and Cleaning: } Integrated multi-omics data
\[
X_{\text{DNA}} \in \mathbb{R}^{m \times 128}, \quad X_{\text{RNA}} \in \mathbb{R}^{m \times 128}
\]
with $m$ samples. Infinite values were replaced with NaN, and missing values were imputed with the median:
\[
X_{ij} = 
\begin{cases} 
\text{median}(X_j) & \text{if } X_{ij} \in \{\text{NaN}, \pm \infty\} \\
X_{ij} & \text{otherwise}
\end{cases}
\]

\noindent \textbf{Train-Test Split:} Data were split into $80\%$ training and $20\%$ testing:
\[
\begin{split}
(X_{\text{train}}, X_{\text{test}}, &\, y_{\text{train}}, y_{\text{test}}) = \\
& \text{train\_test\_split}(X_{\text{DNA}}, X_{\text{RNA}}, y, \text{test\_size}=0.2)
\end{split}
\]

\noindent \textbf{RNA Normalization:}
RNA expression was normalized to $[0, \pi]$:
\[
X_{\text{RNA}}^{\text{norm}} = \pi \cdot \frac{X_{\text{RNA}} - \min(X_{\text{RNA}})}{\max(X_{\text{RNA}}) - \min(X_{\text{RNA}}) + \epsilon}
\]

\noindent \textbf{Feature Concatenation:} Normalized RNA features were concatenated with DNA features:
\[
X_{\text{train}} = [X_{\text{DNA, train}} \, \Vert \, X_{\text{RNA, train}}^{\text{norm}}] \in \mathbb{R}^{m_{\text{train}} \times 256}
\]
\[
X_{\text{test}} = [X_{\text{DNA, test}} \, \Vert \, X_{\text{RNA, test}}^{\text{norm}}] \in \mathbb{R}^{m_{\text{test}} \times 256}
\]

\noindent \textbf{Quantum Input Encoding}
Amplitude encoding maps a classical vector $x \in \mathbb{R}^d$ into a quantum state:
\[
|\psi\rangle = \sum_{i=0}^{2^n-1} x_i \, |i\rangle, \quad \sum_i |x_i|^2 = 1
\]
with $n = \lceil \log_2(d) \rceil$ qubits. Batch encoding was performed using TensorCircuit’s \texttt{vmap}.

\noindent \textbf{Quantum Circuit Architecture}

The QNN uses $n = 6$ qubits and $L = 6$ layers. Each layer of the parameterized quantum circuit (PQC) consists of:

\begin{enumerate}
    \item \textbf{Single-qubit rotations:} Each qubit undergoes parameterized rotations around the $y$ and $x$ axes:
 \[
\begin{split}
U_\text{rot}^{(l)}(i) &= R_y(\theta^{(l)}_{y,i}) \, R_x(\theta^{(l)}_{x,i}), \\
&\quad i = 0, \dots, n-1
\end{split}
\]

\[
\begin{split}
R_y(\theta) &= 
\begin{bmatrix} 
\cos(\theta/2) & -\sin(\theta/2) \\ 
\sin(\theta/2) & \cos(\theta/2) 
\end{bmatrix}, \\
R_x(\theta) &= 
\begin{bmatrix} 
\cos(\theta/2) & -i \sin(\theta/2) \\ 
-i \sin(\theta/2) & \cos(\theta/2) 
\end{bmatrix}.
\end{split}
\]

    \item \textbf{Entanglement:} Multi-qubit correlations are introduced via all-to-all CNOT gates:
    \[
    U_\text{ent} = \prod_{i=0}^{n-2} \prod_{j=i+1}^{n-1} \text{CNOT}_{i \to j},
    \]
    where each CNOT gate acts as
    \[
    \text{CNOT}_{i,j} |q_i q_j\rangle = |q_i\rangle \otimes X^{q_i} |q_j\rangle.
    \]
    
    \item \textbf{Measurement:} The final output of the quantum circuit is obtained by measuring the expectation value of the Pauli-Z operator on the last qubit:
    \[
    f_\text{QC}(x; \theta) = \langle \psi_\text{final} | Z_{n-1} | \psi_\text{final} \rangle,
    \]
    where $|\psi_\text{final}\rangle = U_\text{ent} U_\text{rot} \dots U_\text{rot} |\psi_x\rangle$ is the state after applying all layers to the amplitude-encoded input $|\psi_x\rangle$.
\end{enumerate}

\noindent \textbf{Hybrid Quantum-Classical Neural Network}
The quantum circuit was embedded into a PyTorch model:

\[
\hat{y}_{\text{final}} = \sigma(W \hat{y} + b)
\]

where $\sigma$ is the sigmoid function. Trainable parameters $\theta \in \mathbb{R}^{2L \times n}$ were optimized during training.

\noindent \textbf{Training Procedure}

\noindent \textbf{Loss Function}
Binary cross-entropy:
\[
\mathcal{L} = -\frac{1}{m} \sum_{i=1}^m \left[ y_i \log(\hat{y}_i) + (1 - y_i) \log(1 - \hat{y}_i) \right]
\]

\noindent \textbf{Optimization}
Parameters were updated using Adam optimizer ($\alpha = 0.03$):
\[
\theta \gets \theta - \alpha \frac{\partial \mathcal{L}}{\partial \theta}
\]

Training was performed for $50$ epochs with mini-batches of $64$ samples.

\begin{figure*}	[!ht]
	\centering
	\includegraphics[scale=0.20]{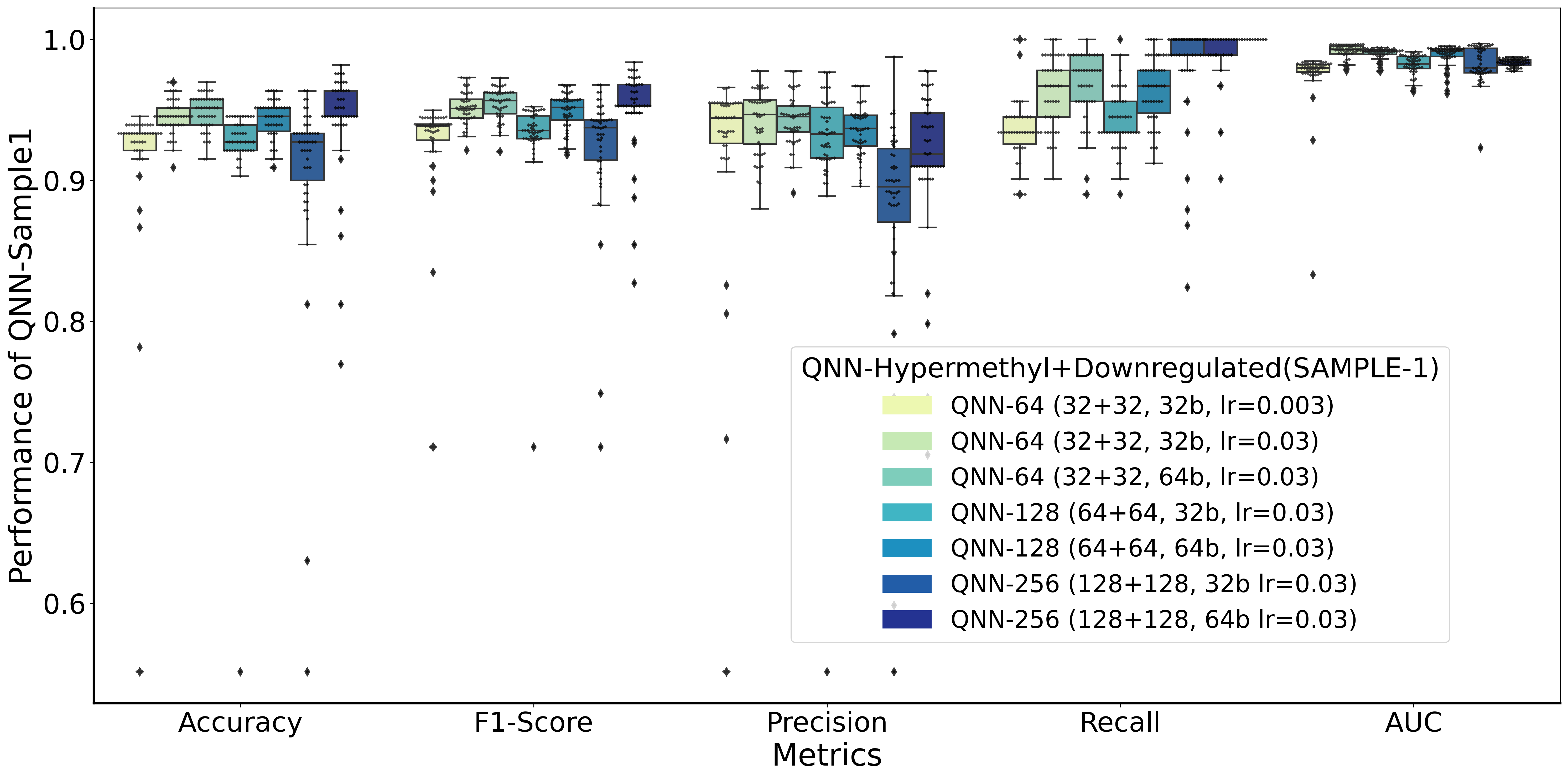}
    \includegraphics[scale=0.34]{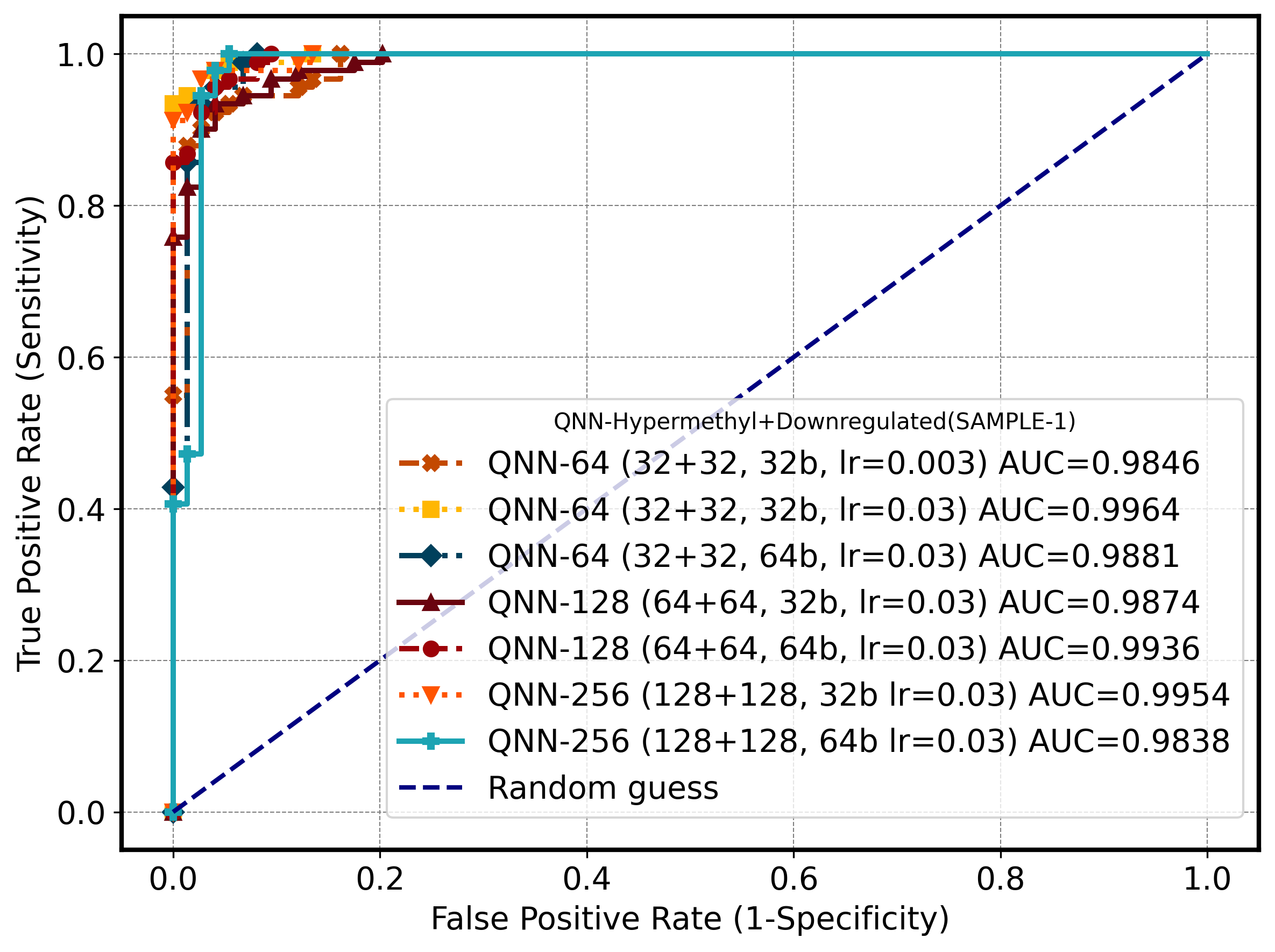}\\
    \includegraphics[scale=0.20]{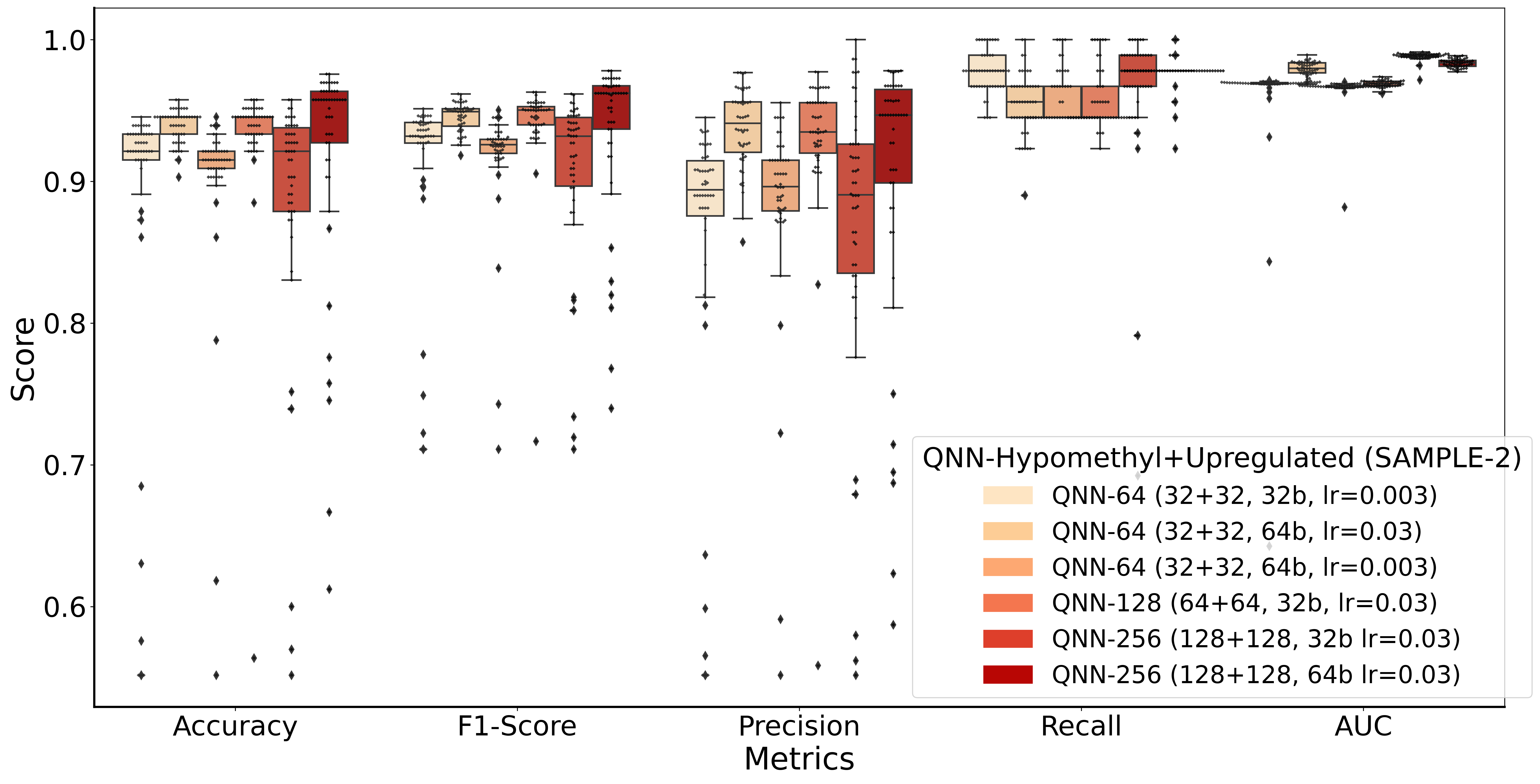}
     \includegraphics[scale=0.34]{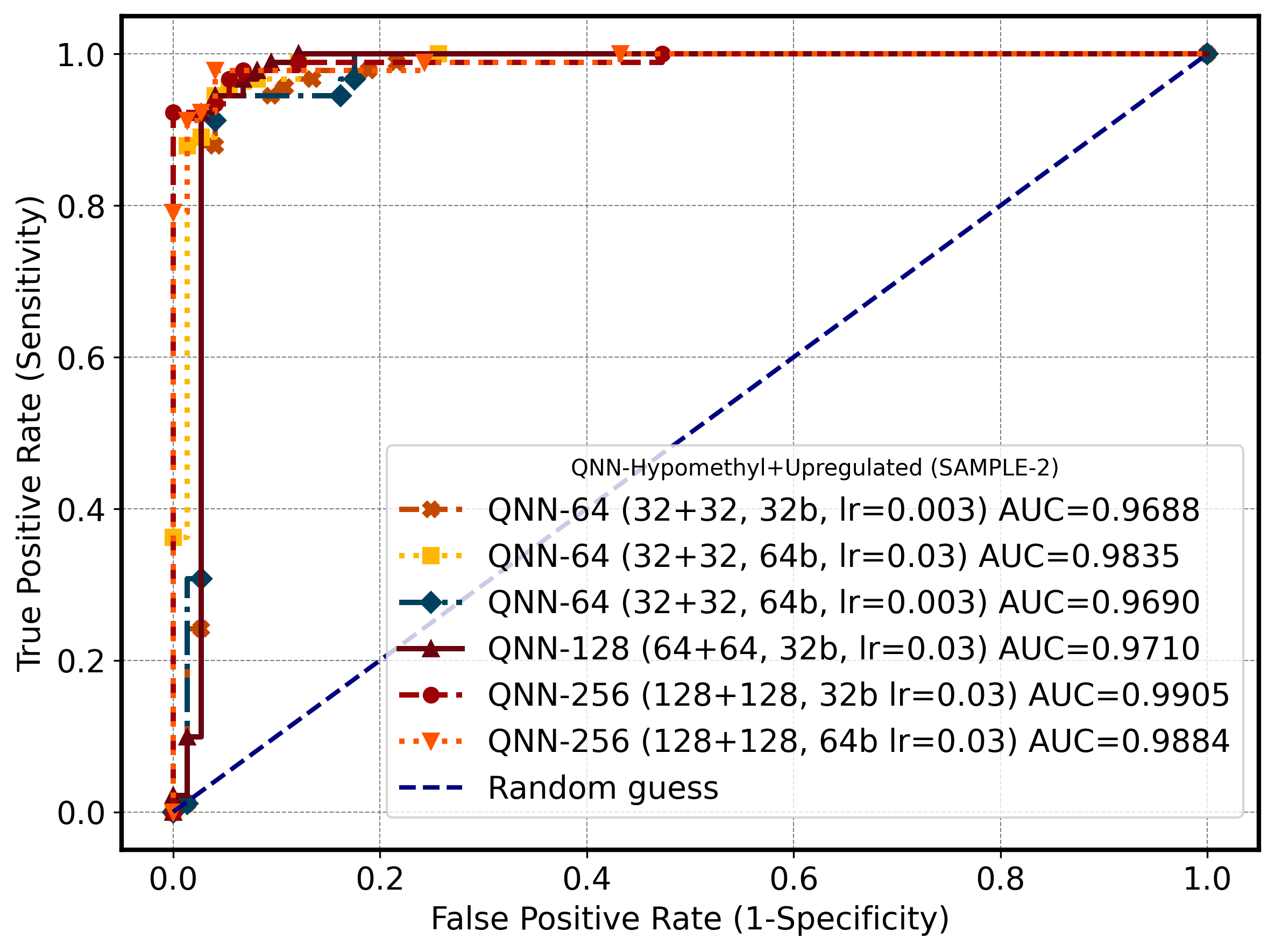}\\
    \includegraphics[scale=0.20]{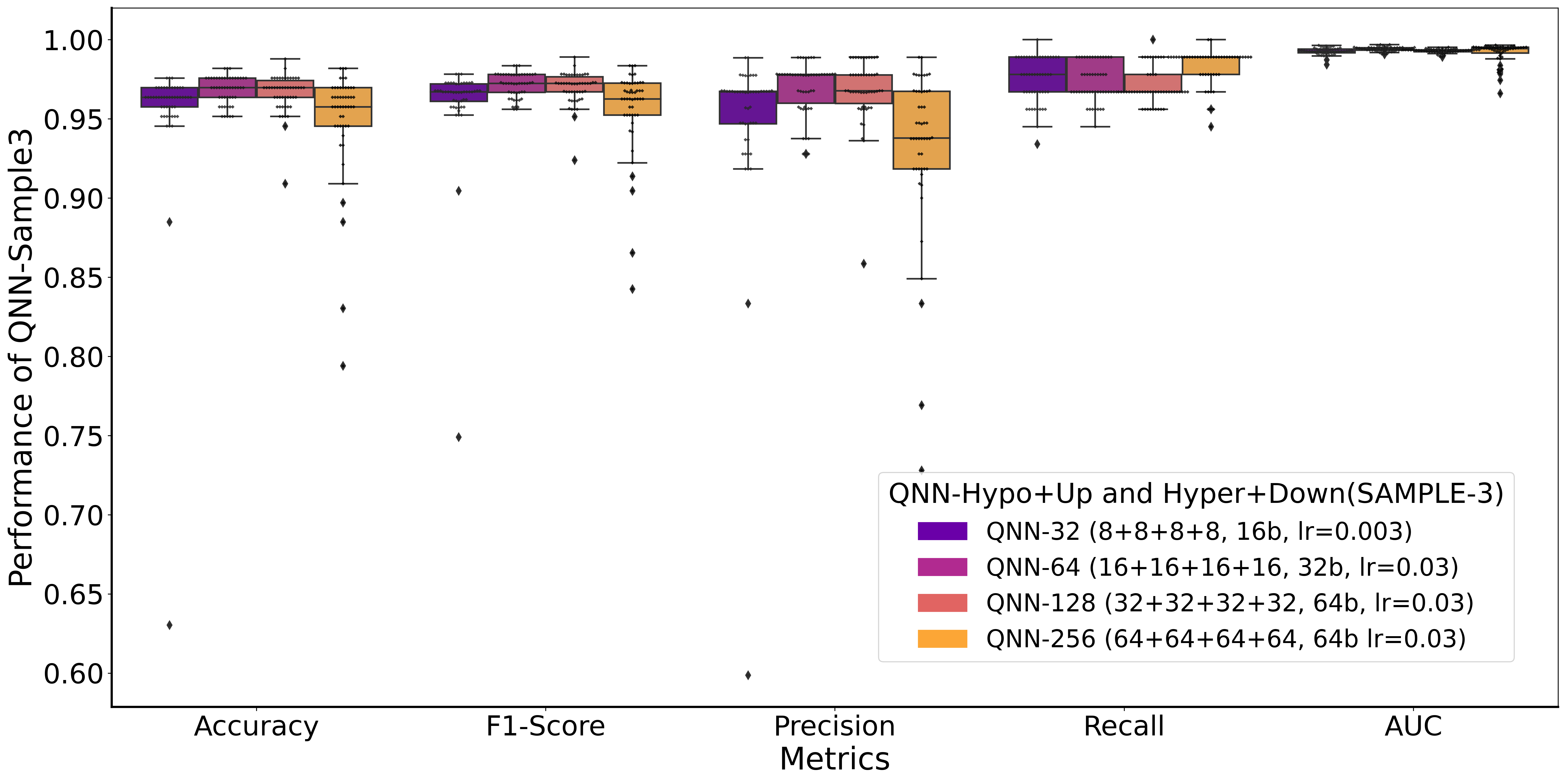}
    \includegraphics[scale=0.34]{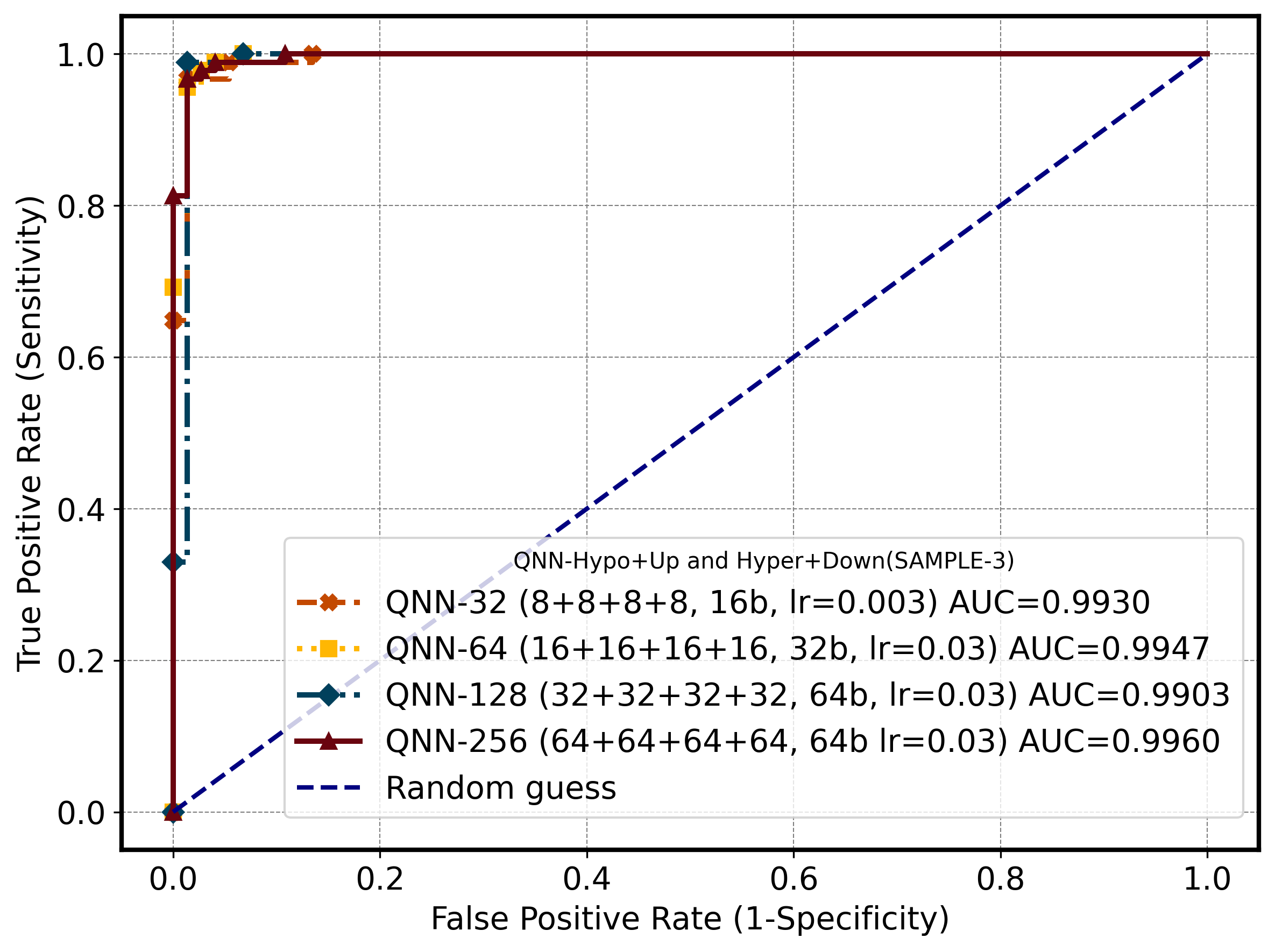}\\
	\caption{Performance of QNN-model with Sample1, Sample2 and Sample 3 using metrics for luad and lung subtypes diagnostic classification}
 \label{Figure6}
\end{figure*}

\section{RESULTS and DISCUSSION}
In this section, we will present the performance of our proposed QUBID framework, in detail as shown in Fig~.\ref{Figure2}. The overall workflow of this study was designed to systematically evaluate the contribution of multi-omics features derived from DNA methylation and RNA-seq data for LUAD versus LUSC lung cancer subtype classification using a QNN model. First, LUAD-specific genes were identified and ranked from both omics layers using a prior filtering strategy. Based on their epigenetic and transcriptional behavior, the genes were organized into three experimental groups. In Sample 1, genes exhibiting hypermethylation together with downregulated expression were selected to represent epigenetically silenced signatures. In Sample 2, genes showing hypomethylation together with upregulated expression were selected to capture epigenetically activated signatures. In Sample 3, genes from Sample 1 and Sample 2 were combined to create an integrated dataset containing both silenced and activated regulatory patterns. This approach enabled the exploration of how mixed epigenetic states influence model learning and prediction performance. This ranking ensured that only biologically relevant and subtype-informative genes were considered for downstream modeling. From these ranked lists, matched subsets of DNA methylation features (Feature-1) and RNA-seq features (Feature-2) were selected in equal numbers to maintain balanced multi-omics representation. Following gene selection, the two feature sets were concatenated to form a unified multi-omics input vector. Different feature sizes were constructed by progressively increasing the number of selected genes from both modalities (e.g., 32–32, 64–64, 128–128, and 256–256), resulting in total feature dimensions of 64, 128, 256, and 512, respectively. This step allowed systematic investigation of how feature dimensionality influences classification performance. Each combined feature vector was normalized and prepared for quantum encoding prior to model training.

In the quantum modeling phase, the integrated multi-omics feature vectors were mapped onto quantum states using a qubit-based encoding scheme, where the number of qubits was determined by the total feature dimension through a power-of-two mapping ($2^n$). A parameterized quantum circuit with multiple variational layers was then constructed, enabling the model to learn nonlinear relationships between DNA methylation and RNA expression patterns. The QNN was trained using a supervised learning strategy with LUAD and LUSC labels, and its parameters were optimized using gradient-based methods under different learning rates, batch sizes, and circuit depths.
We evaluated the performance of our QNN for classifying multi-omic data using combined DNA and RNA features. The experiments were performed by progressively increasing the number of features from 64 to 256, while monitoring standard classification metrics, including accuracy (ACC), loss, F1-score, precision, recall, and area under the ROC curve (AUC) as shown in Fig~.\ref{Figure6}. All models were trained with varying learning rates, epoch, and batch sizes, and the corresponding results are summarized below.

\subsection{Performance of Phase-3 Identify Luad-specific tumor biomarkers}
\begin{figure*}
	\centering        
	\includegraphics[scale=0.13]{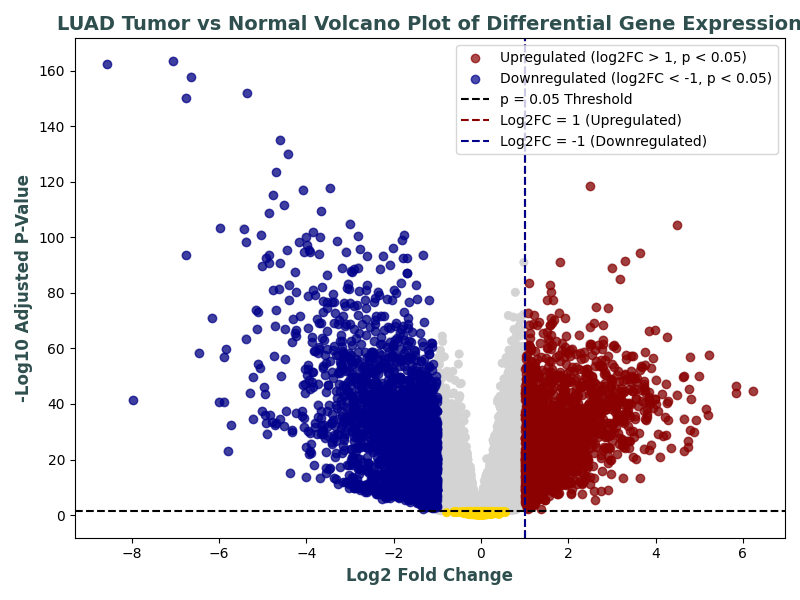}
    \includegraphics[scale=0.11]{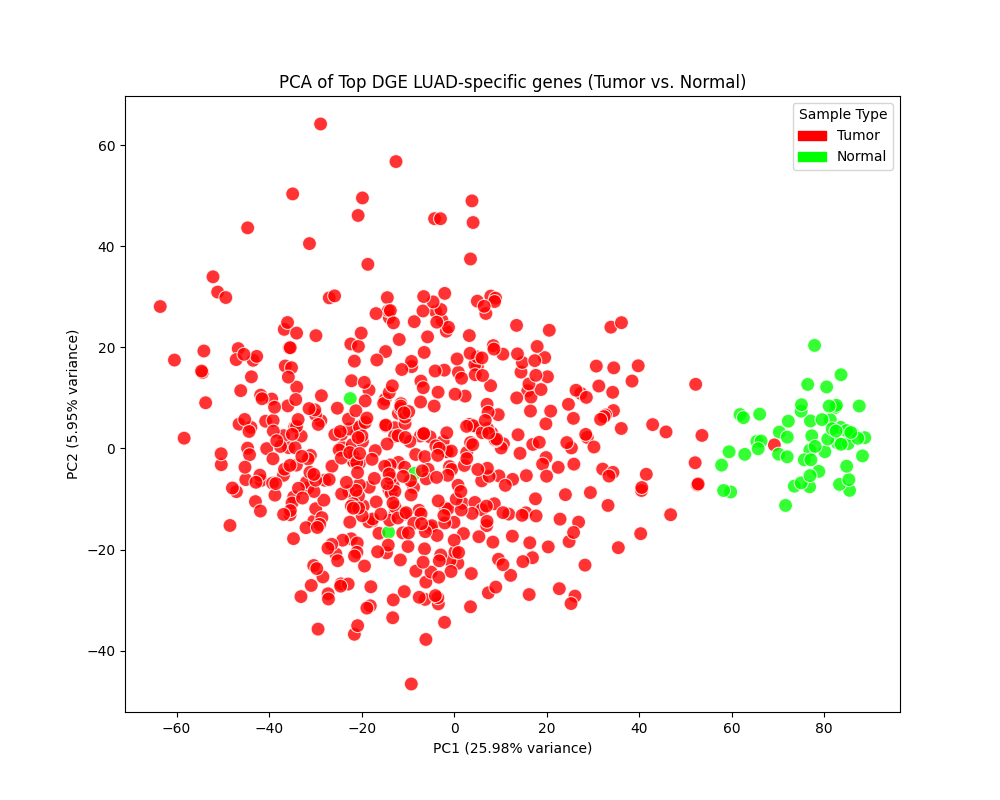}
	\includegraphics[scale=0.14]{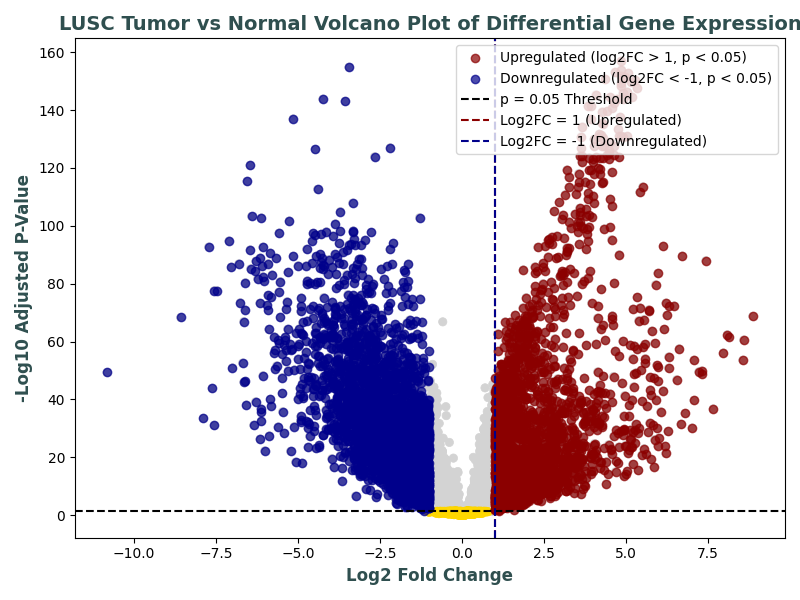}
    \includegraphics[scale=0.11]{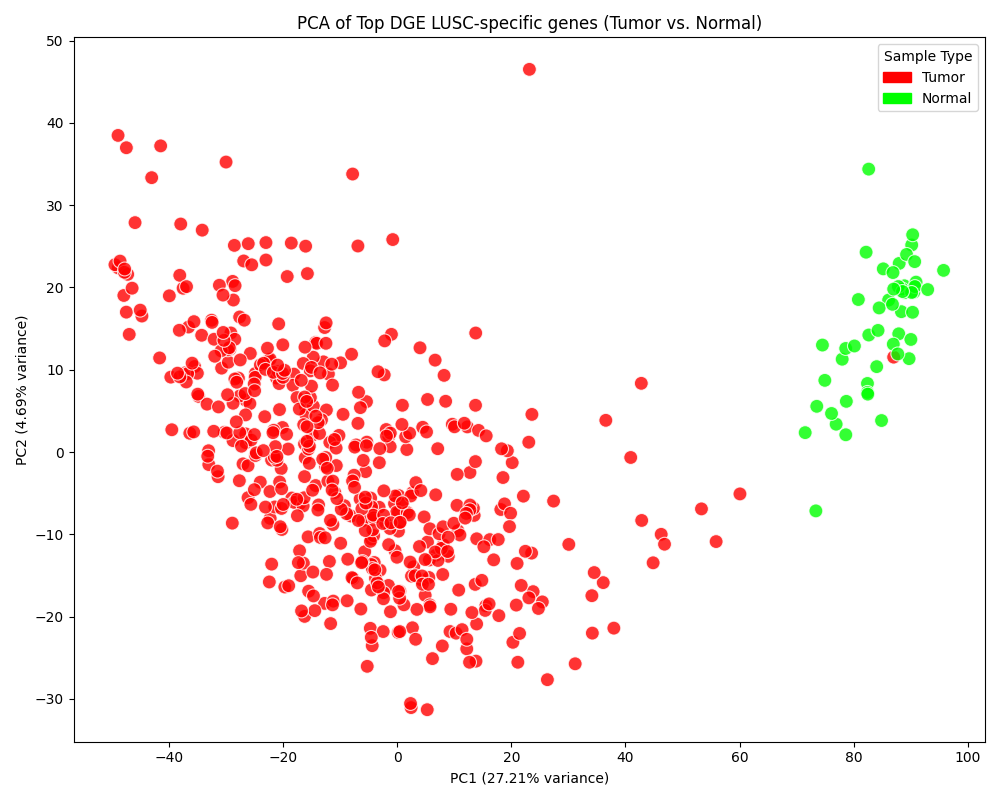}\\    
    \includegraphics[scale=0.13]{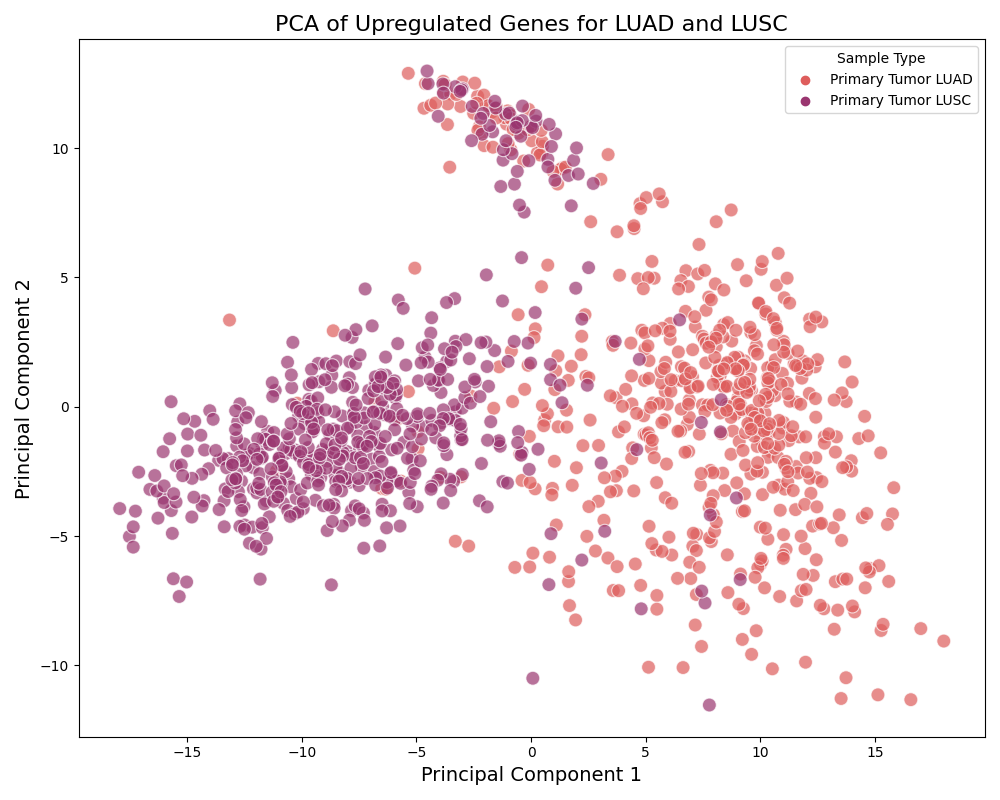}
    \includegraphics[scale=0.13]{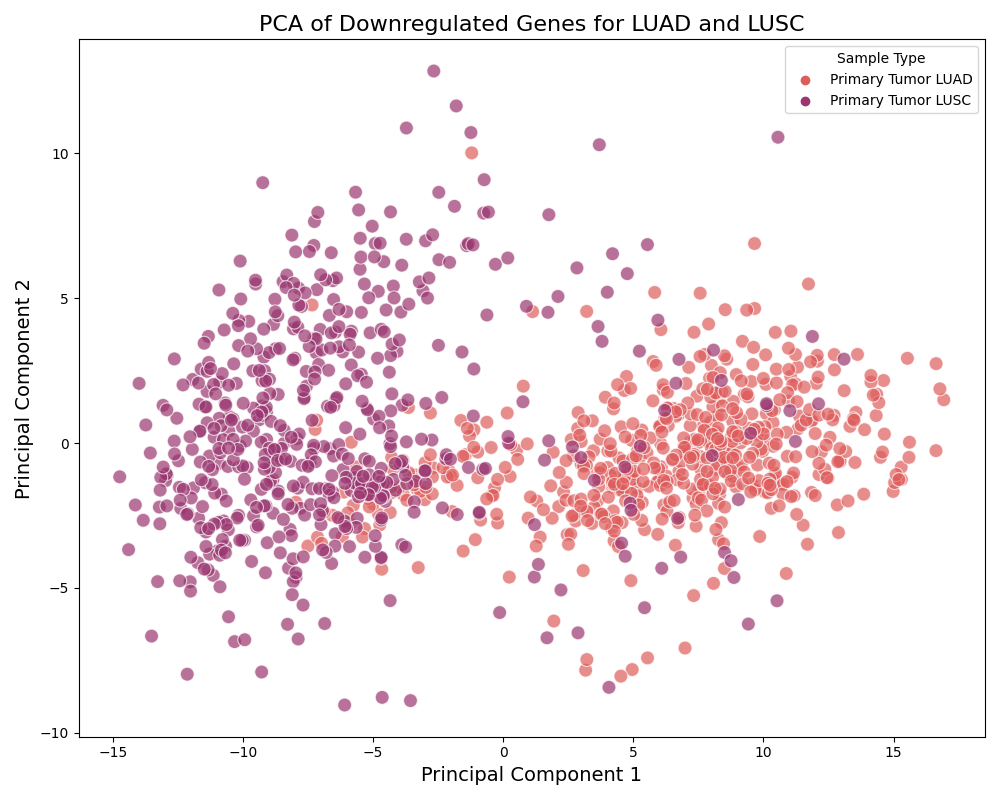}
    \includegraphics[scale=0.14]{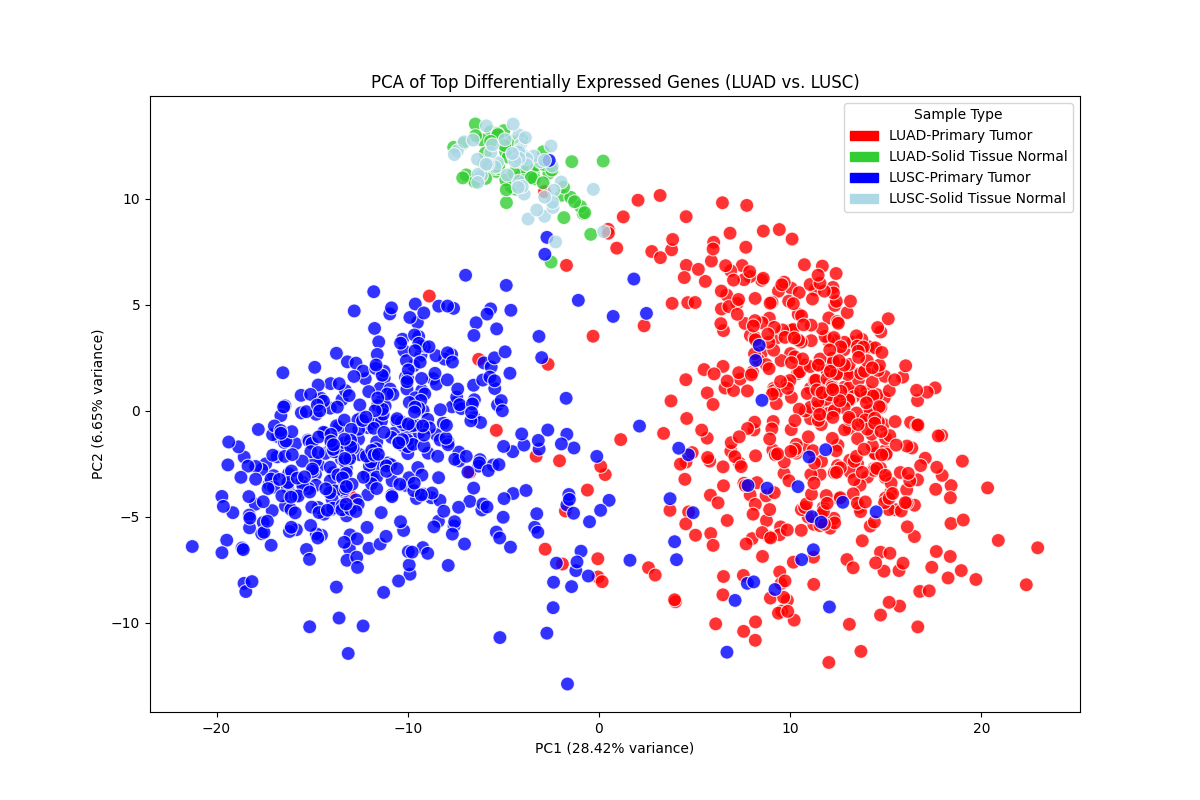}\\
    \includegraphics[scale=0.10]{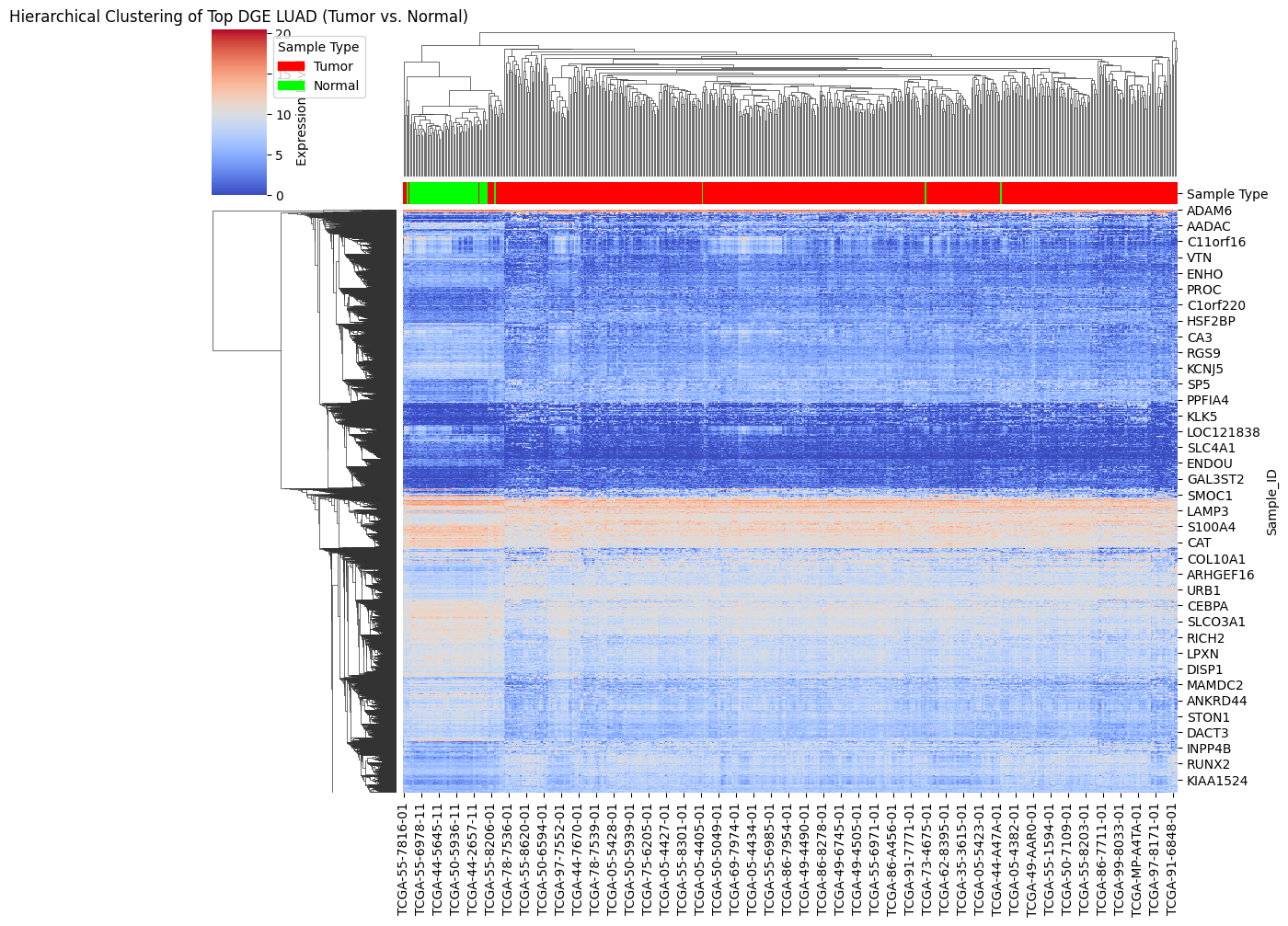}
    \includegraphics[scale=0.10]{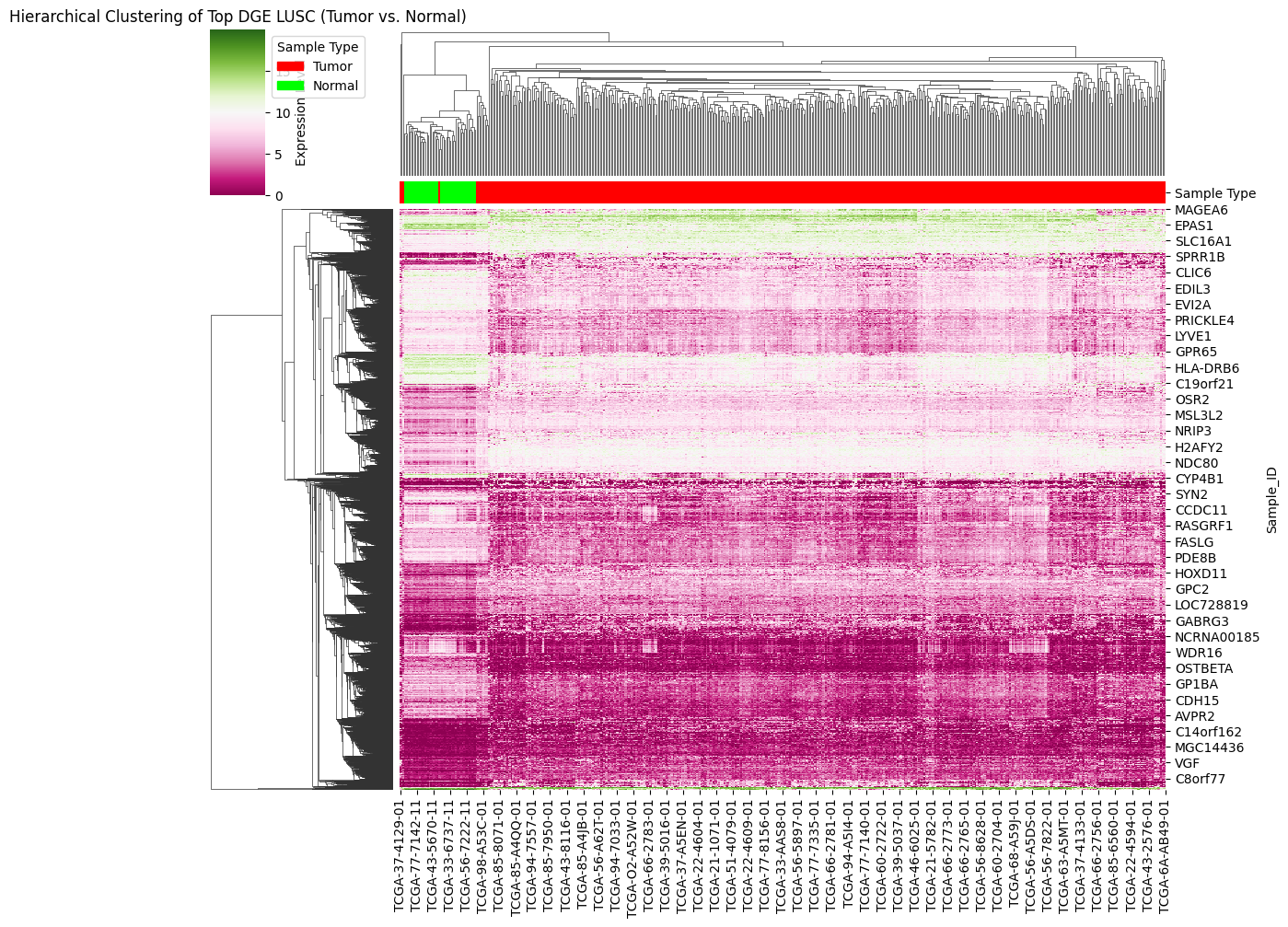}
    \includegraphics[scale=0.09]{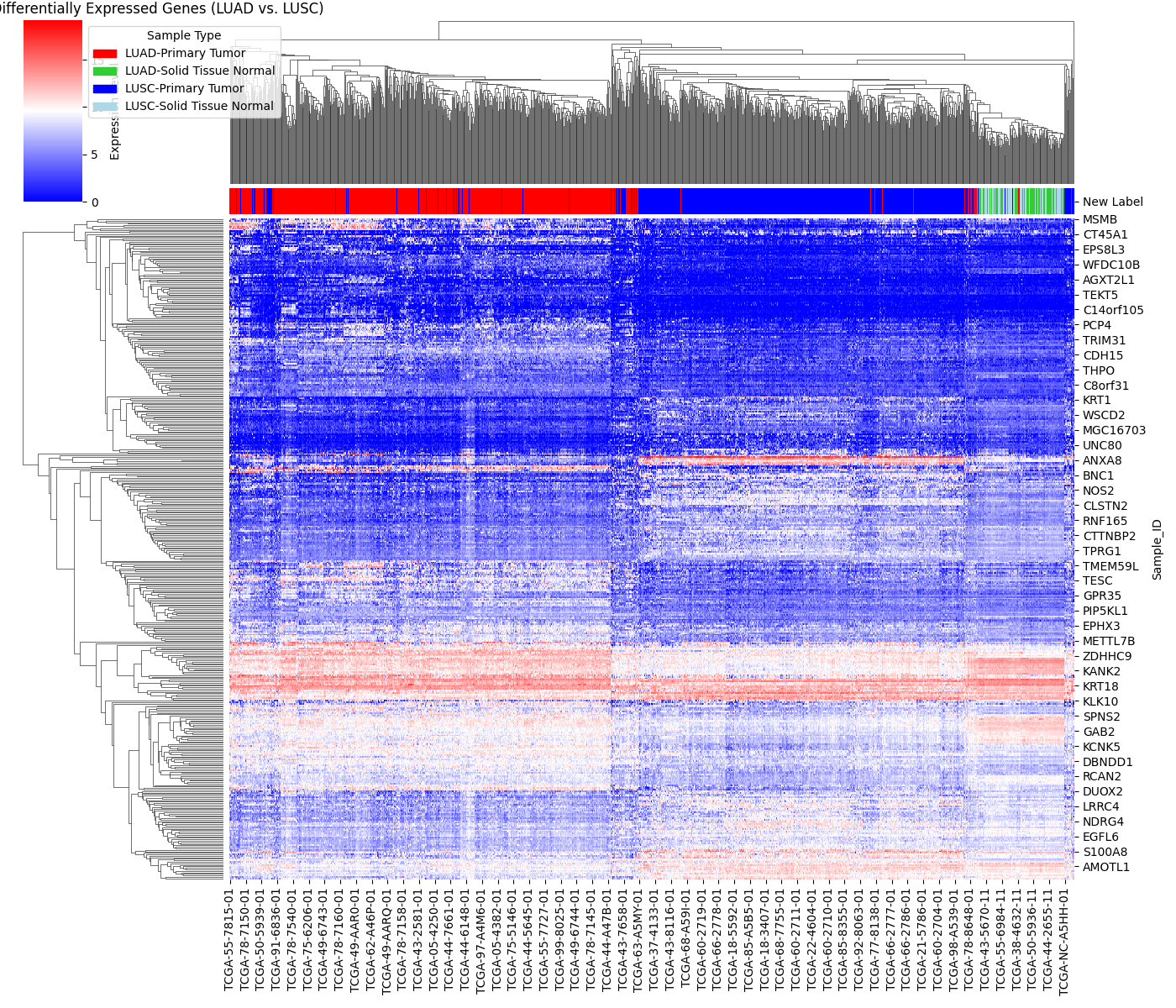}
    \caption{\textbf{Performance of differential gene expression (DGE) analysis for: CASE-1 }
    (a) LUAD tumor vs. normal, (b) LUSC tumor vs. normal, and (c) LUAD tumor vs. LUSC tumor. \textbf{LUAD-specific} Case 1 criteria were applied: p-value $<$ 0.05 and Log Fold Change (LF) $>$ 1 for upregulated genes and $<$ -1 for downregulated genes. The results are visualized using a volcano plot, PCA plot, and hierarchical clustering heatmap.}
	\label{Figure7}
\end{figure*}

\begin{figure*}
	\centering        
	\includegraphics[scale=0.13]{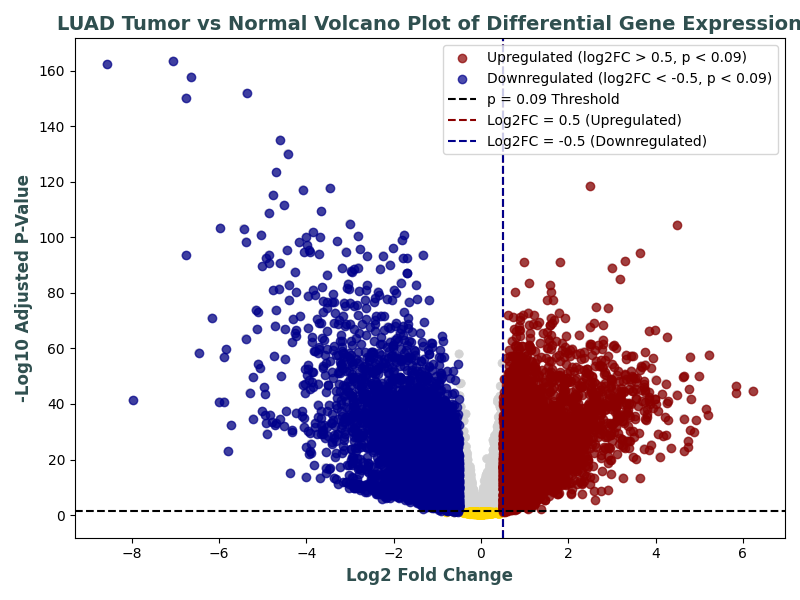}
    \includegraphics[scale=0.11]{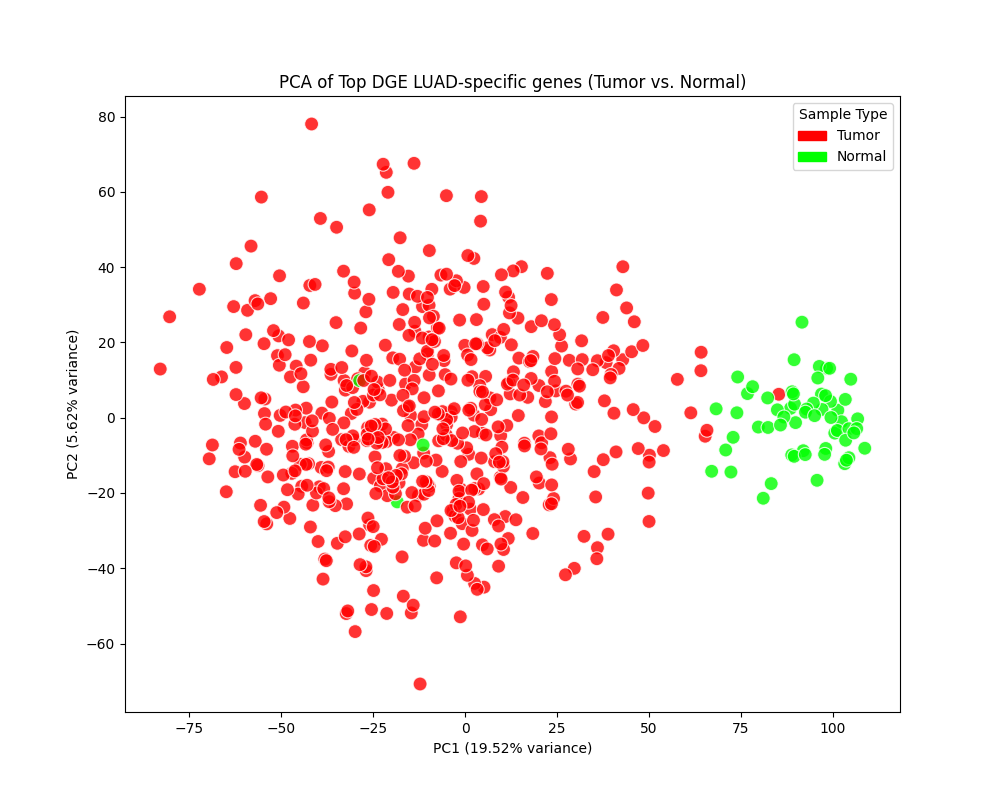}
	\includegraphics[scale=0.14]{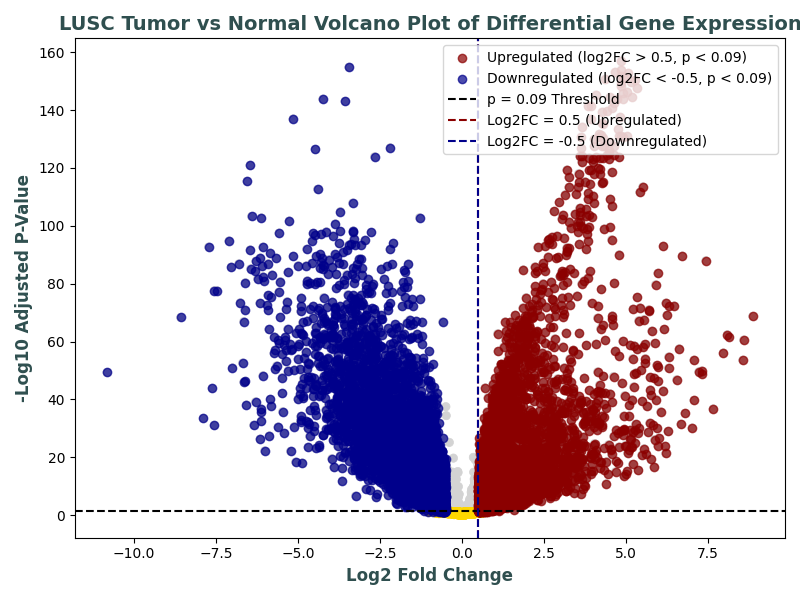}
    \includegraphics[scale=0.11]{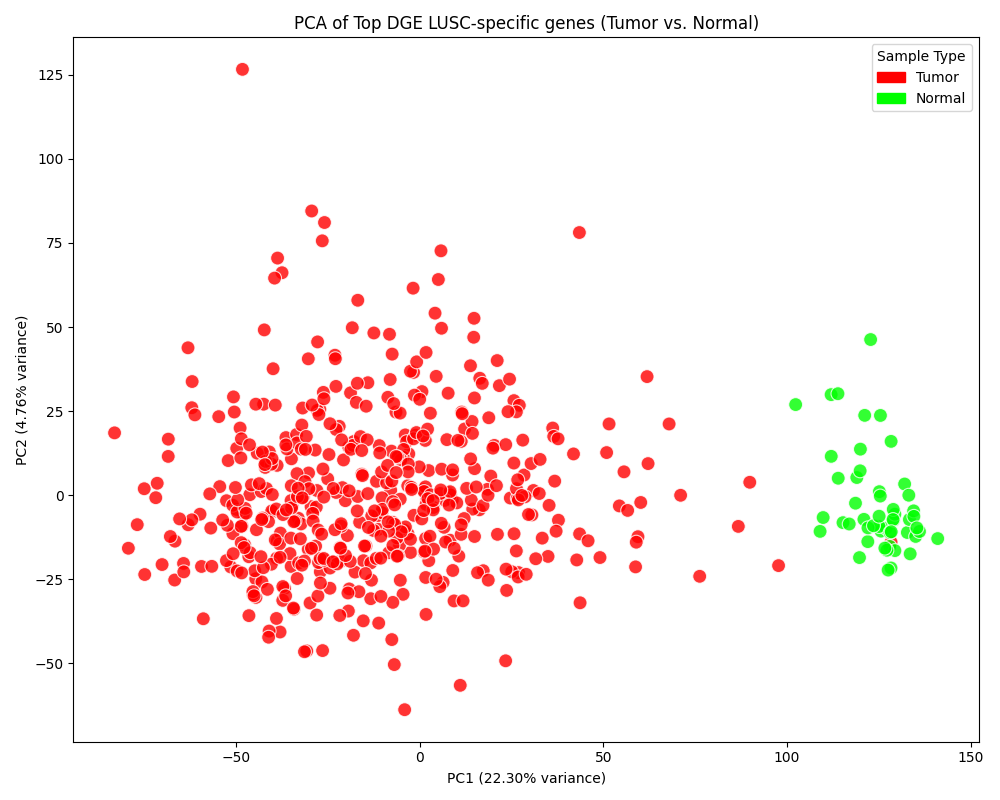}\\
    
    \includegraphics[scale=0.13]{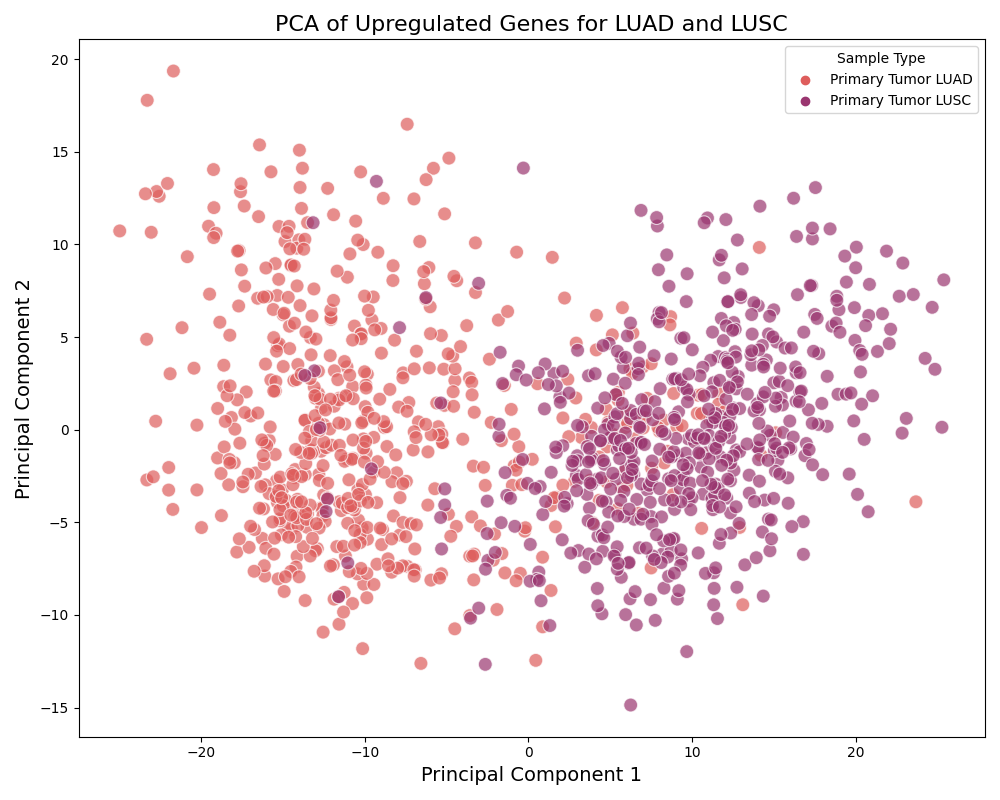}
    \includegraphics[scale=0.13]{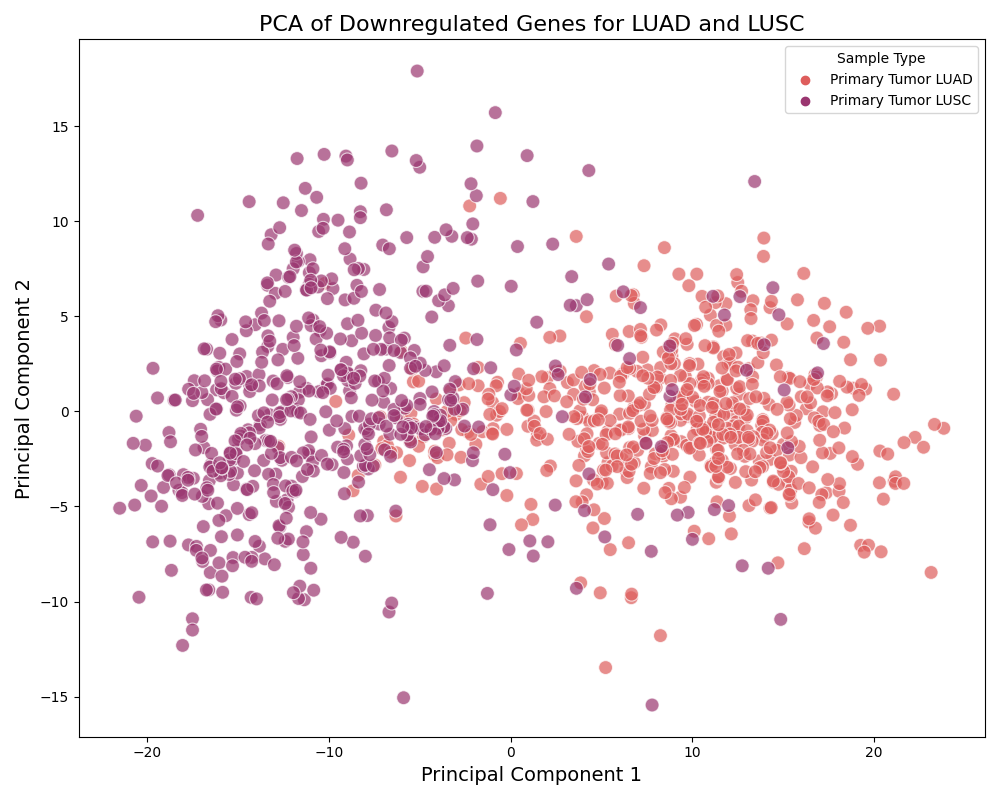}
    \includegraphics[scale=0.14]{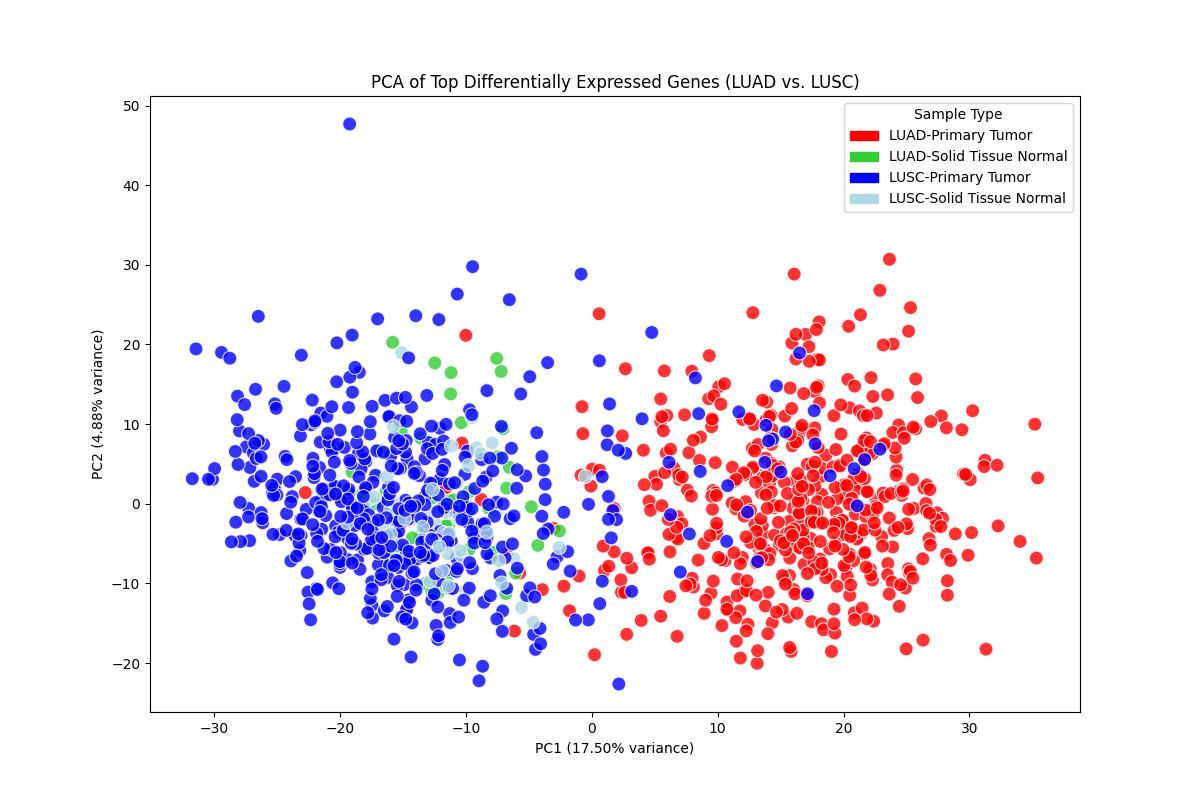}\\

    \includegraphics[scale=0.10]{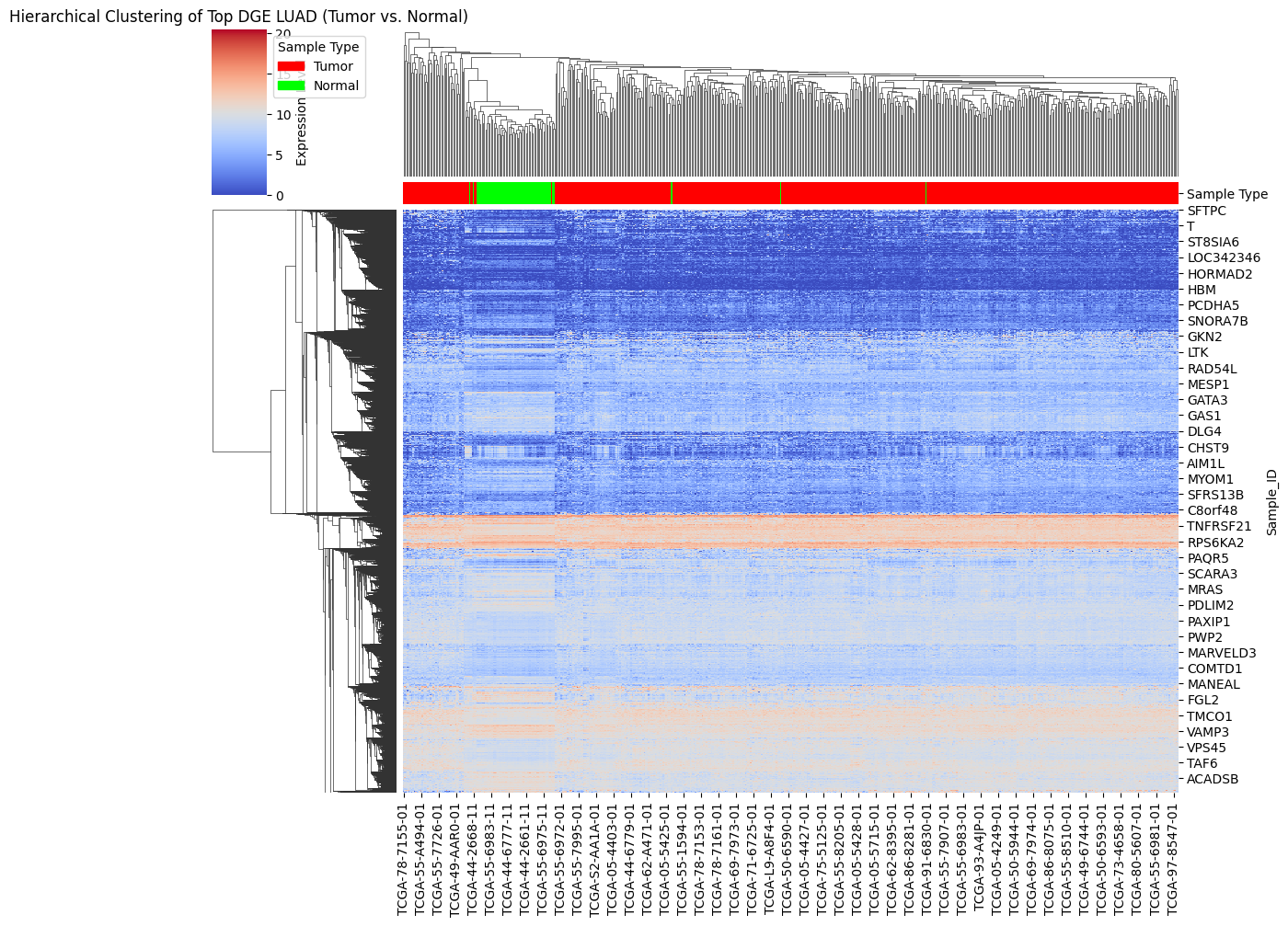}
    \includegraphics[scale=0.10]{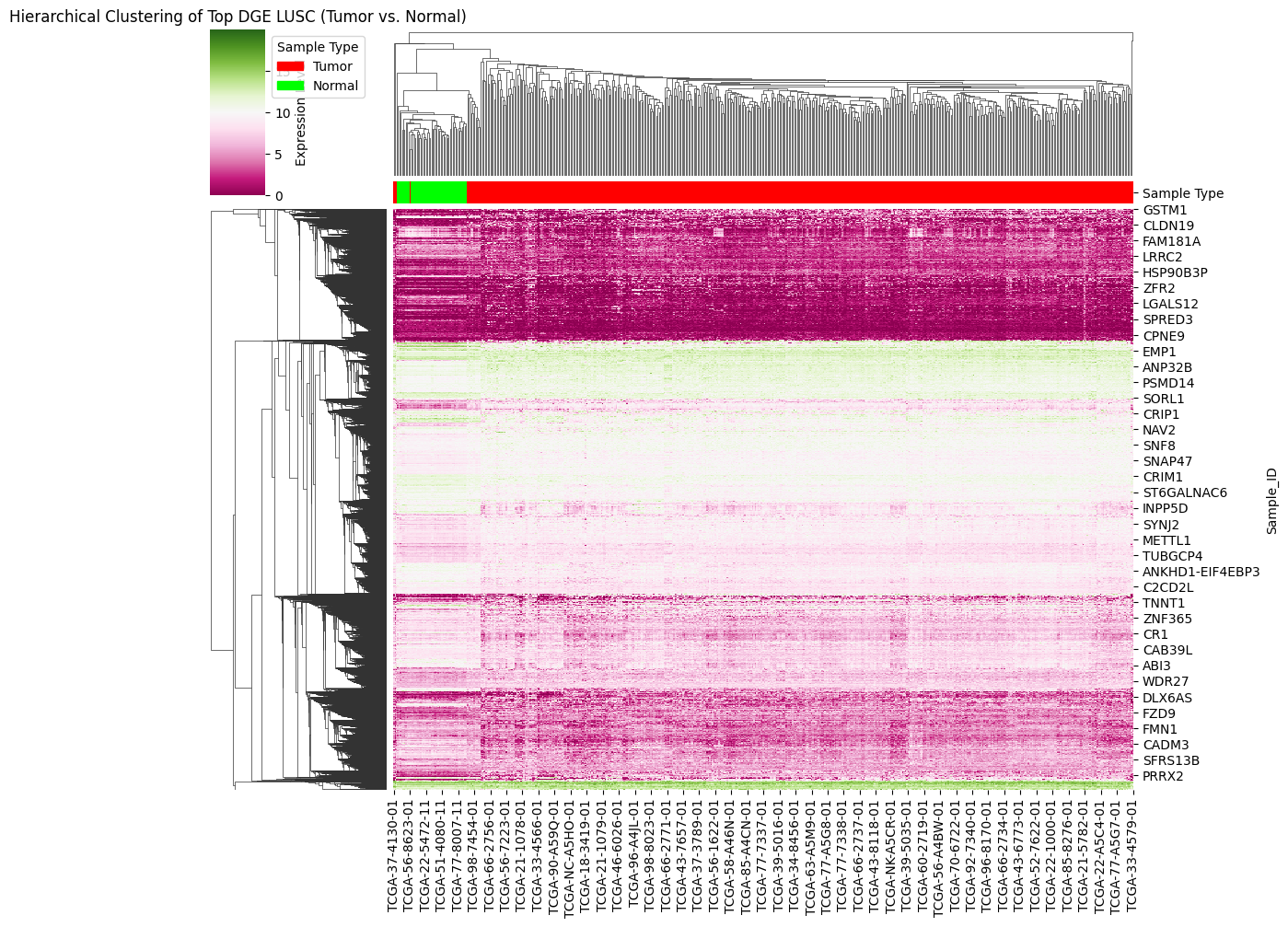}
    \includegraphics[scale=0.09]{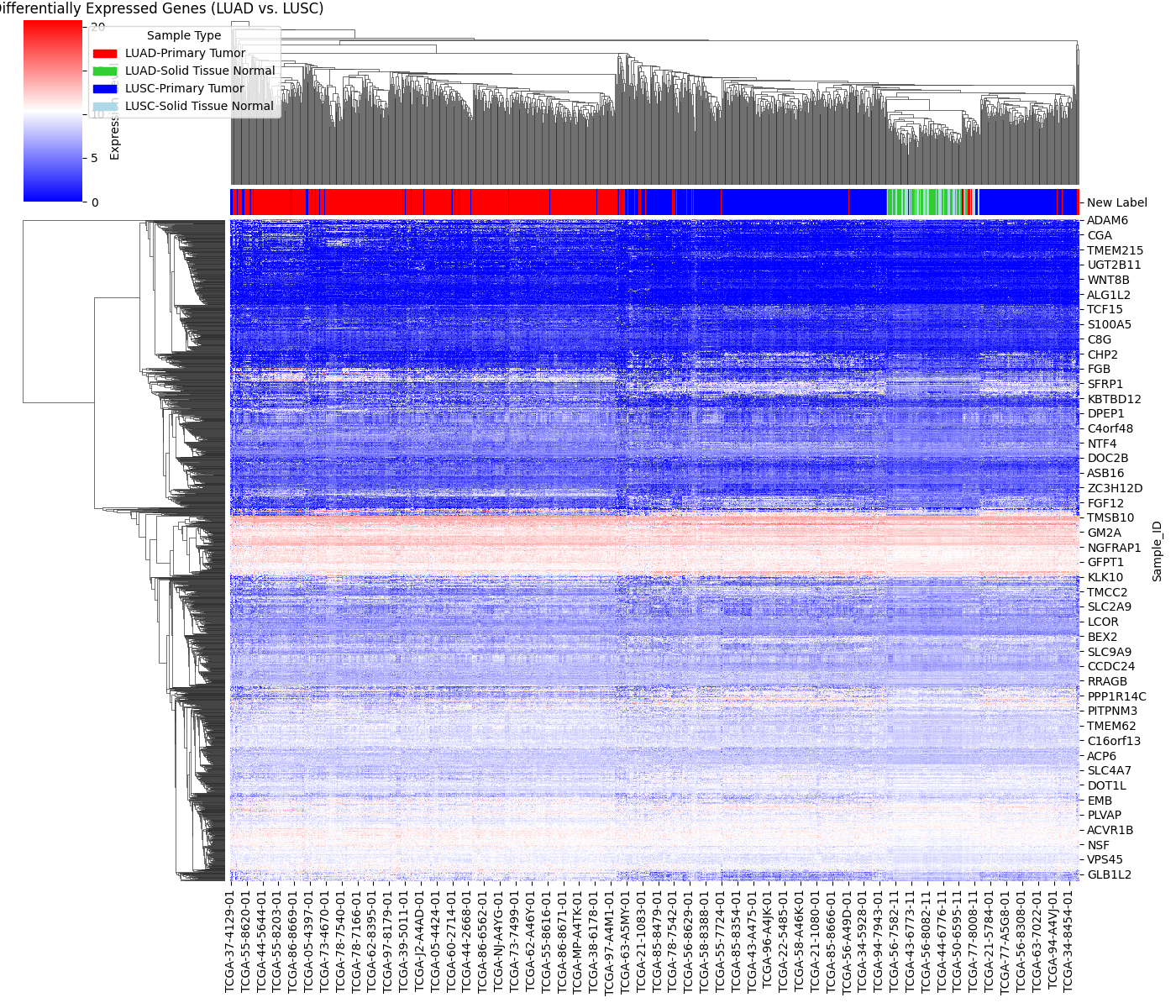}

    \caption{\textbf{Performance of differential gene expression (DGE) analysis for: CASE-2} (a) LUAD tumor vs. normal, (b) LUSC tumor vs. normal, and (c) LUAD tumor vs. LUSC tumor. \textbf{LUAD-specific} }Case 2 criteria were applied: p-value $<$ 0.09 and Log Fold Change (LF) $>$ 0.5 for upregulated genes and $<$ -0.5 for downregulated genes. The results are visualized using a volcano plot, PCA plot, and hierarchical clustering heatmap.
    
	\label{Figure8}
\end{figure*}

\begin{figure*}
	\centering        
	\includegraphics[scale=0.13]{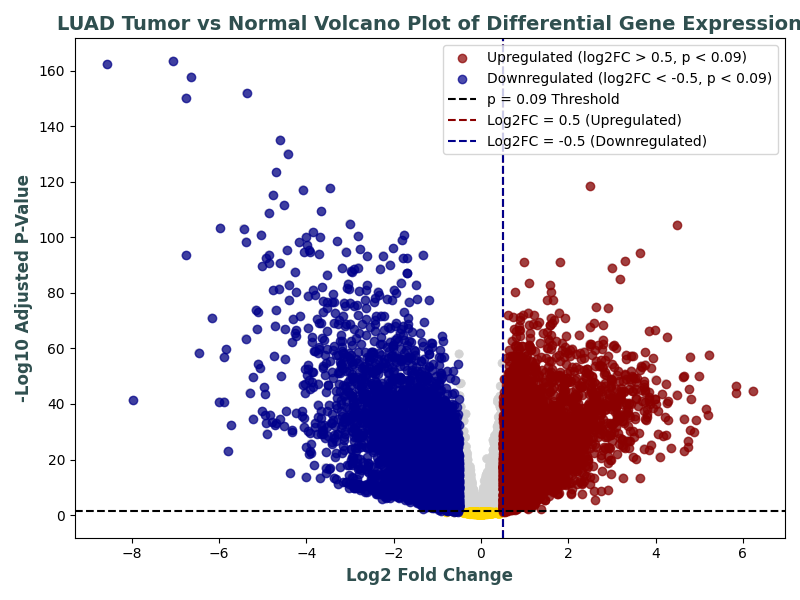}
    \includegraphics[scale=0.11]{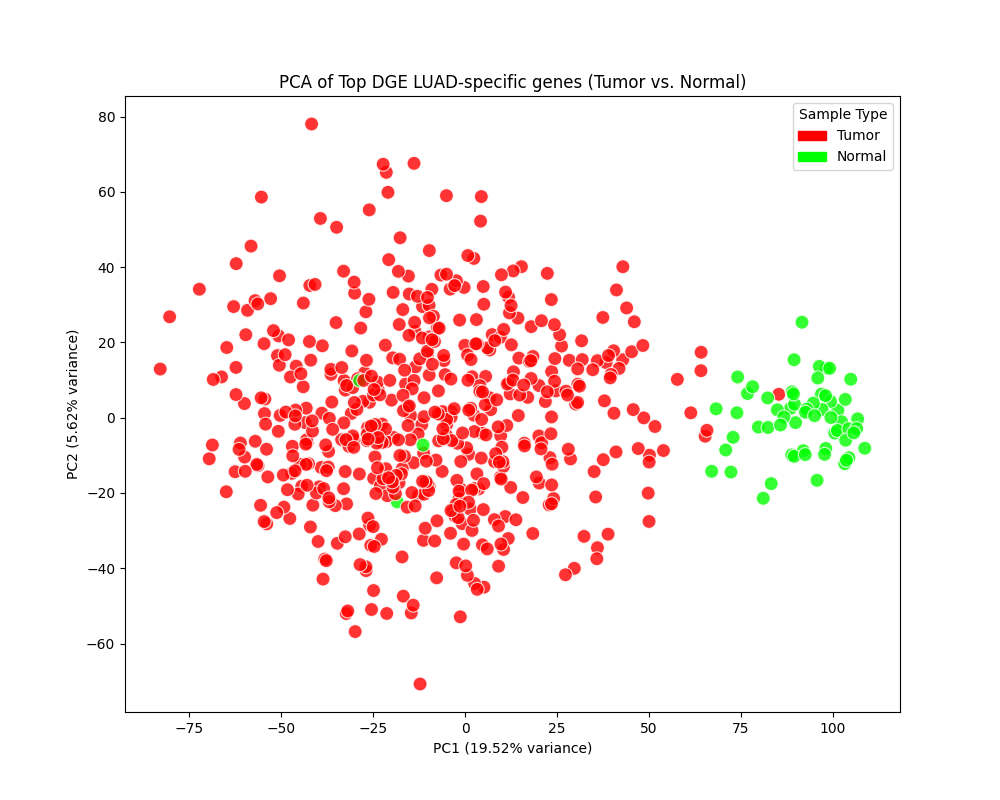}
	\includegraphics[scale=0.14]{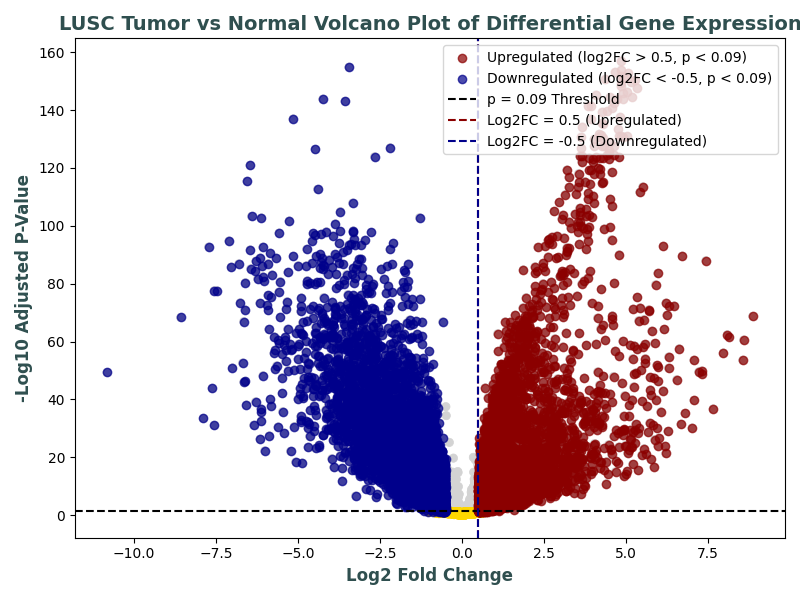}
         \includegraphics[scale=0.11]{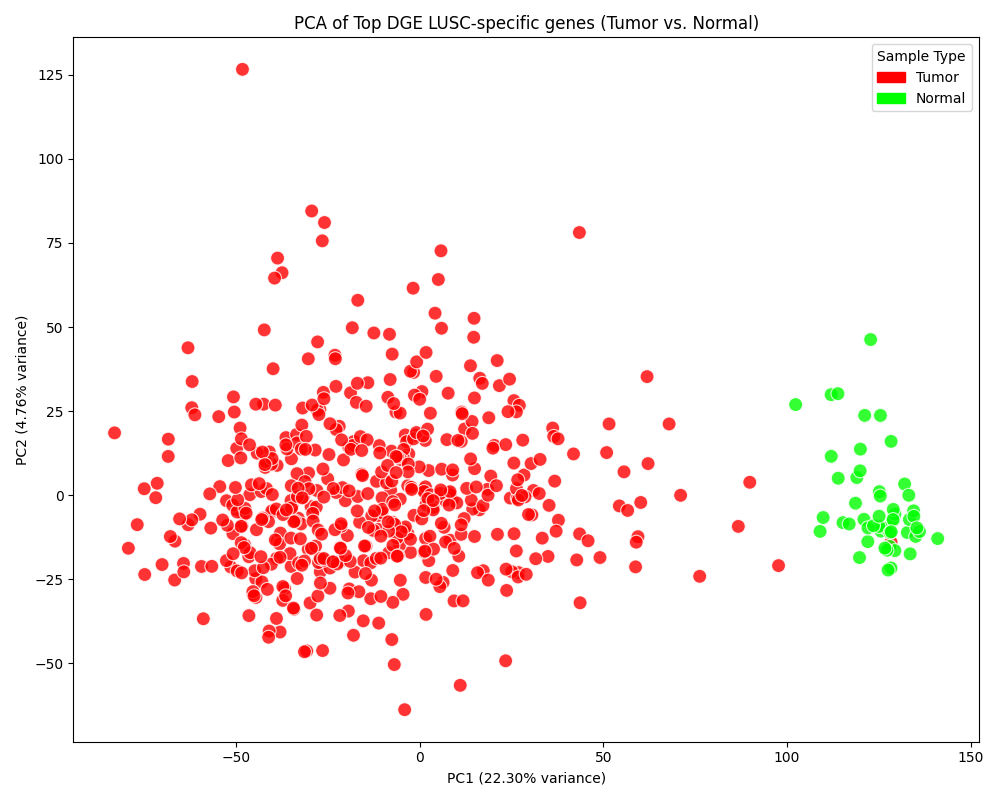}\\
    \includegraphics[scale=0.13]{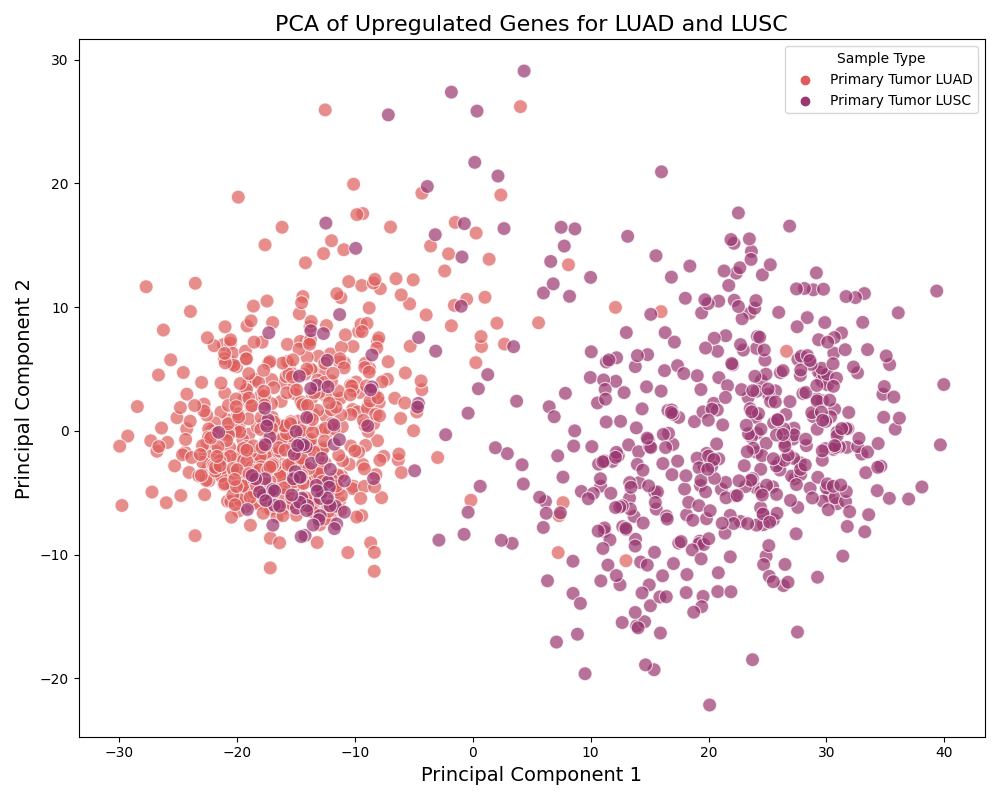}
    \includegraphics[scale=0.13]{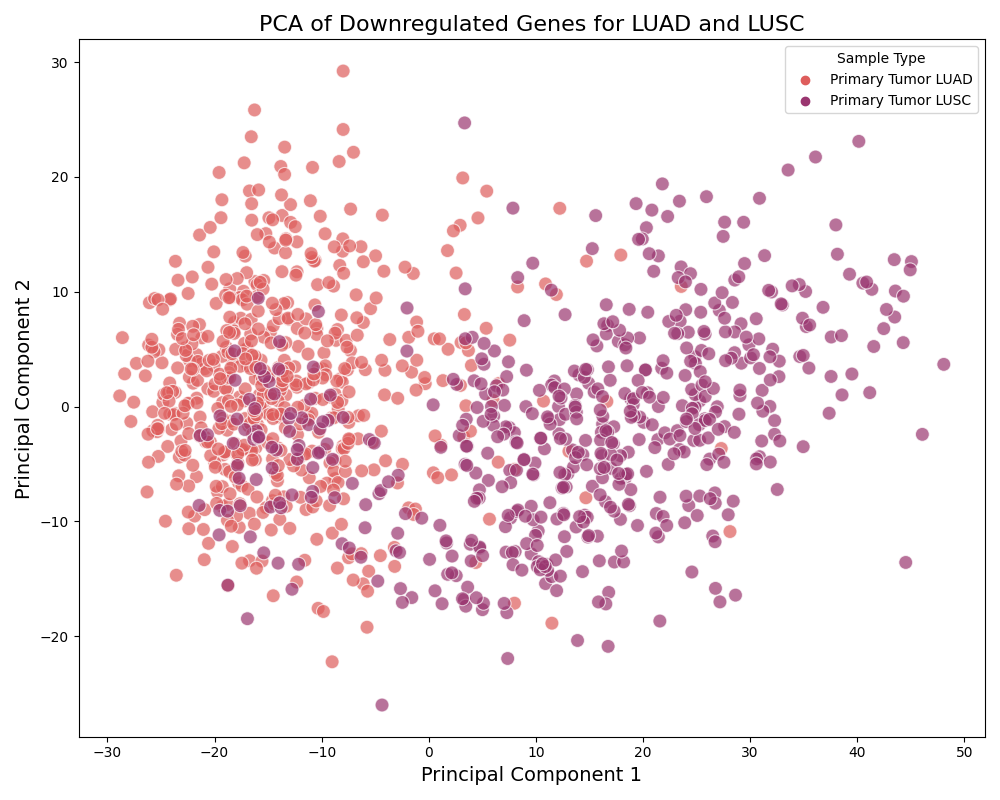}
    \includegraphics[scale=0.14]{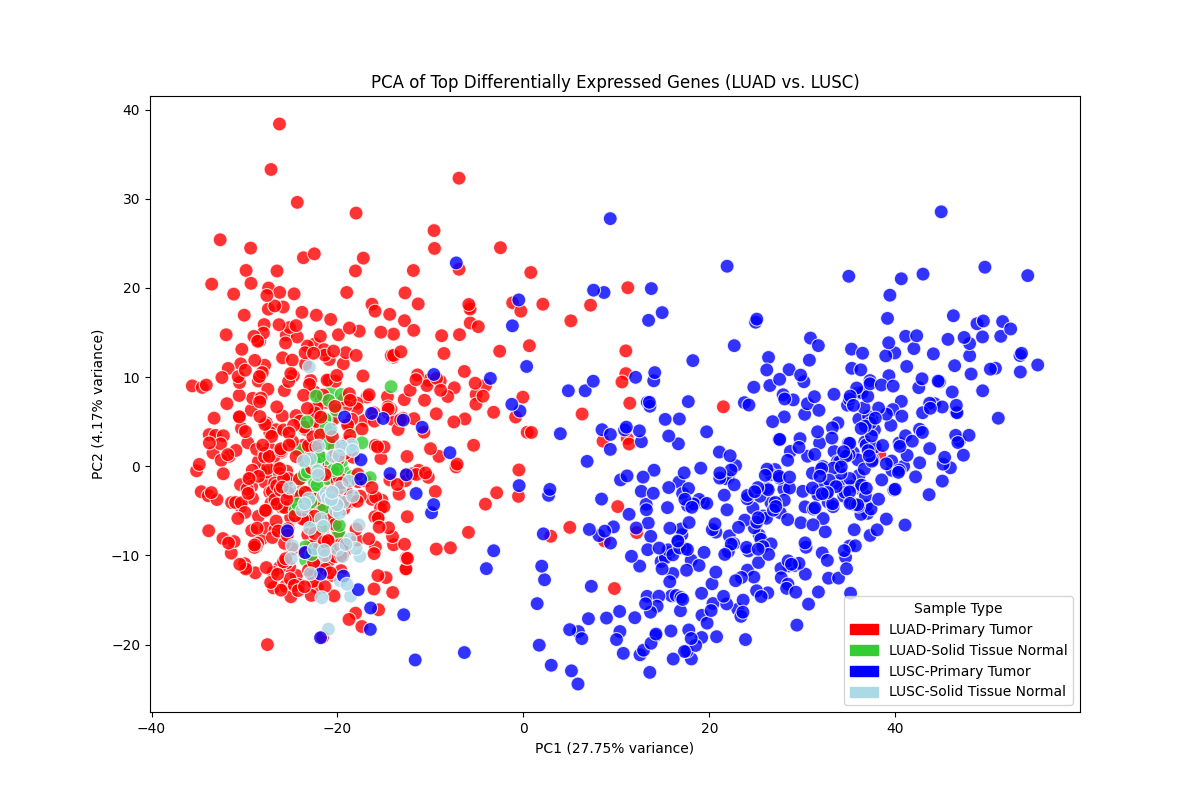}\\

    \includegraphics[scale=0.10]{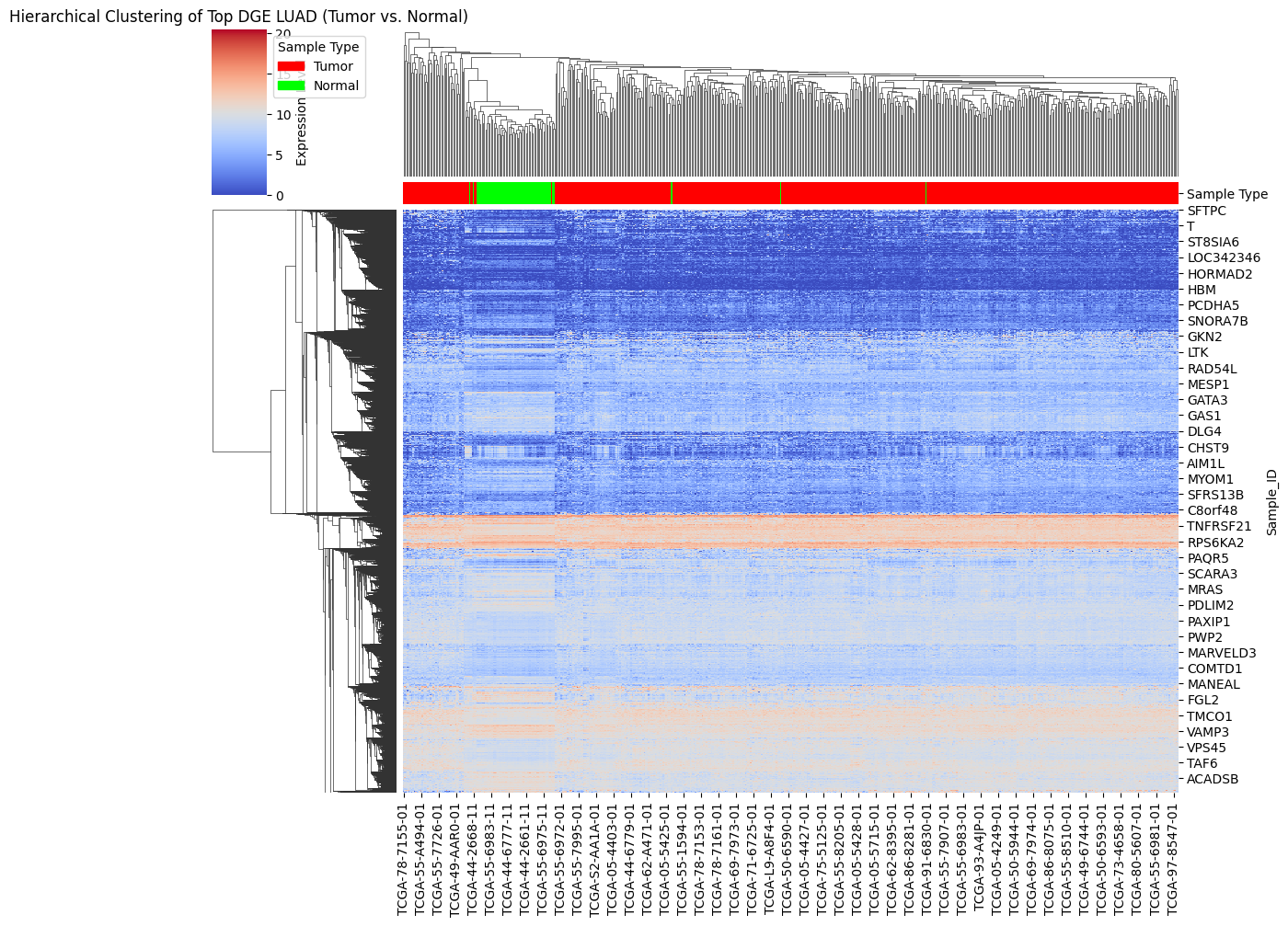}
    \includegraphics[scale=0.10]{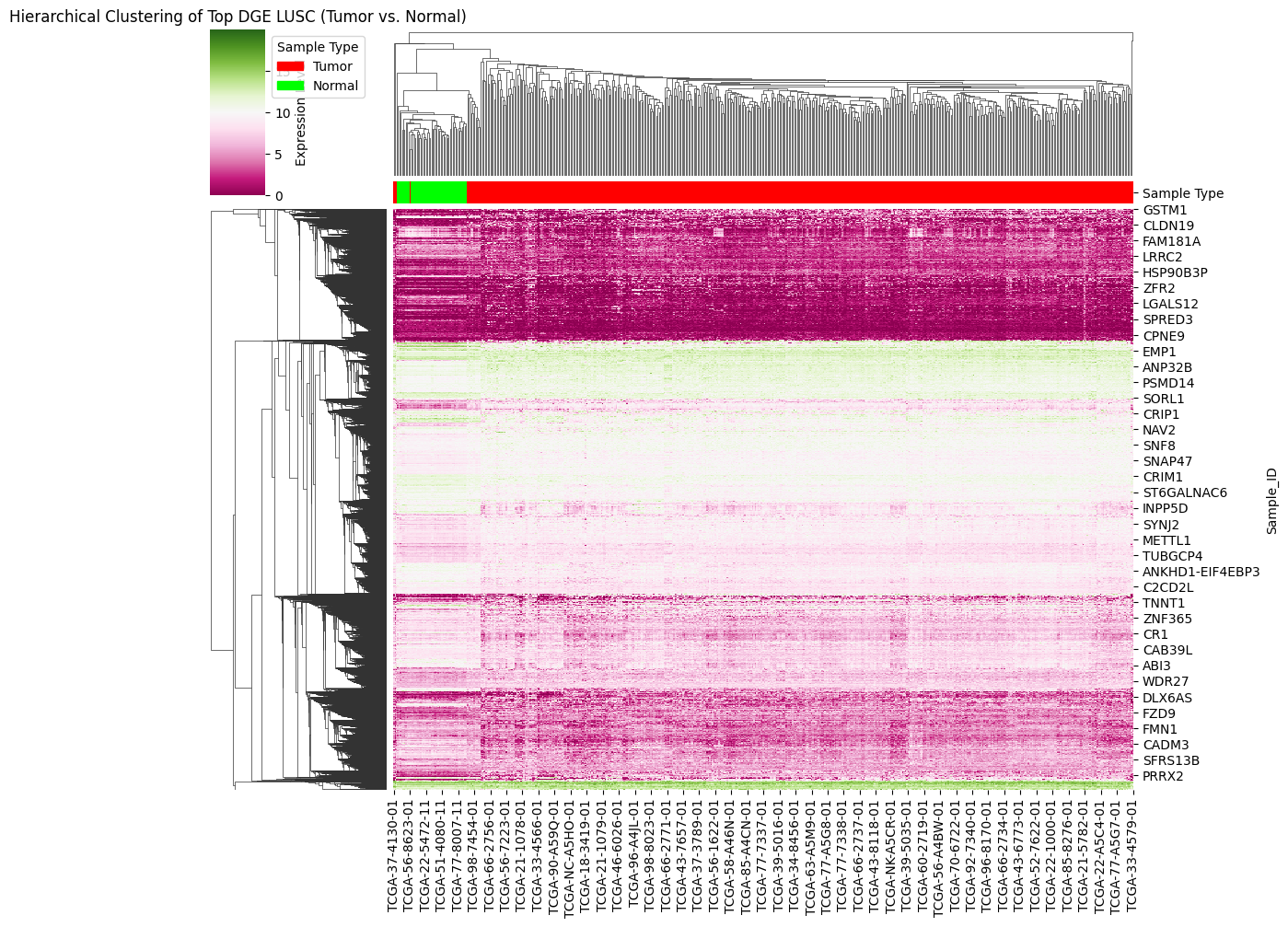}
    \includegraphics[scale=0.09]{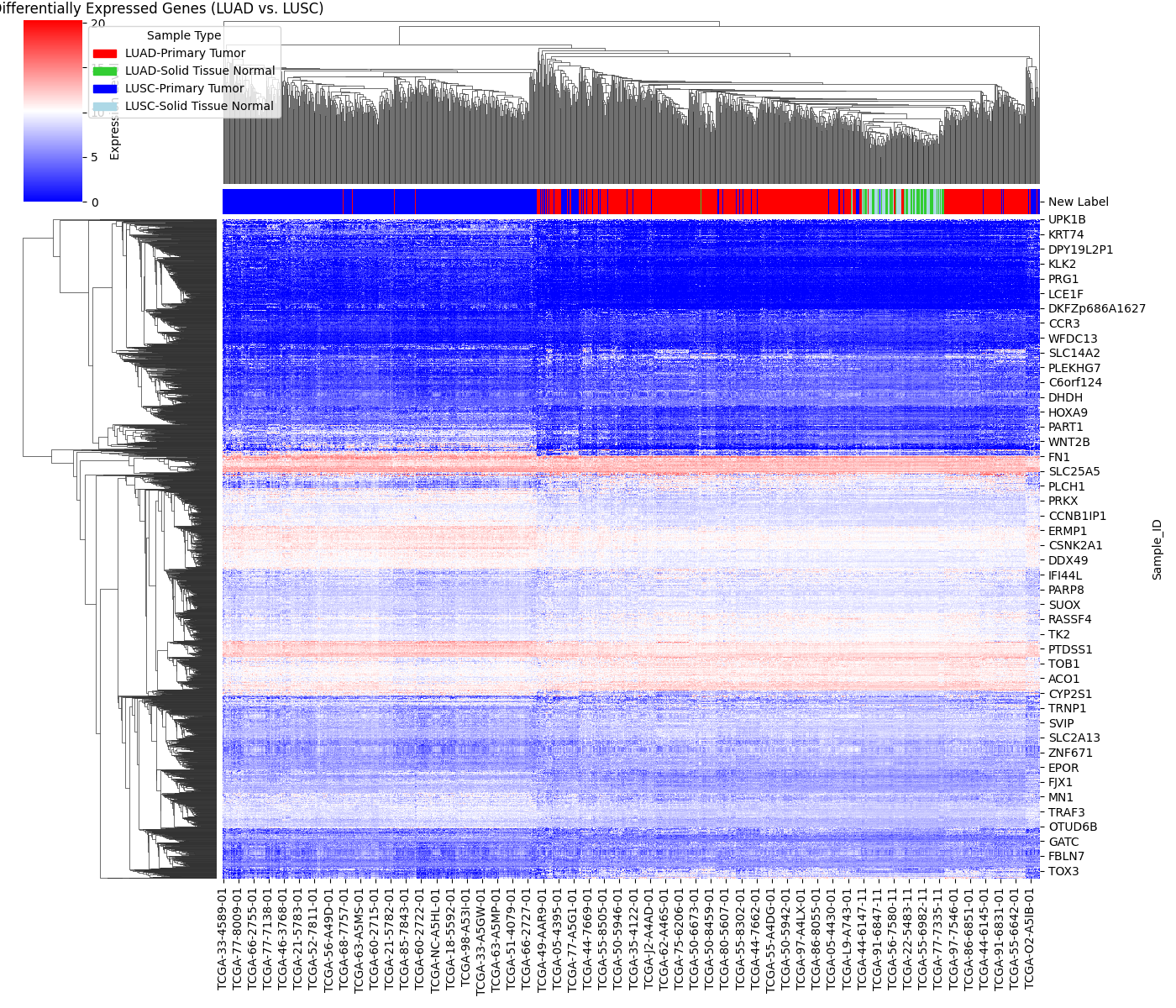}

    \caption{\textbf{Performance of differential gene expression (DGE) analysis for: CASE-1} (a) LUSC tumor vs. normal, (b) LUAD tumor vs. normal, and (c) LUSC tumor vs. LUAD tumor. \textbf{LUSC-specific} Case 1 criteria were applied: p-value $<$ 0.05 and Log Fold Change (LF) $>$ 1 for upregulated genes and $<$ -1 for downregulated genes. The results are visualized using a volcano plot, PCA plot, and hierarchical clustering heatmap.}

	\label{Figure9}
\end{figure*}

\begin{figure*}
	\centering        
	\includegraphics[scale=0.13]{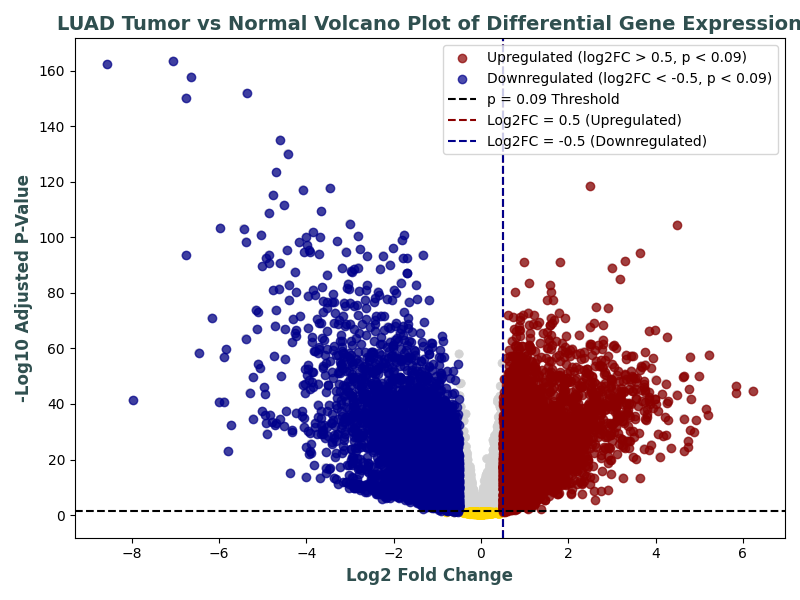}
        \includegraphics[scale=0.11]{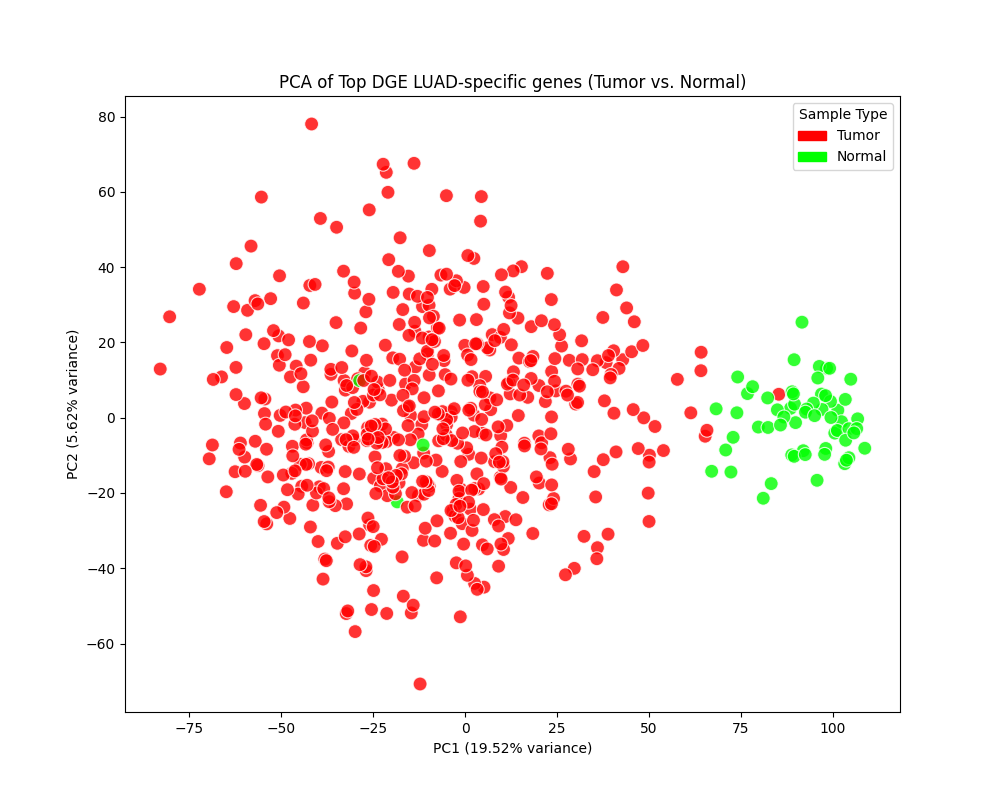}
	\includegraphics[scale=0.14]{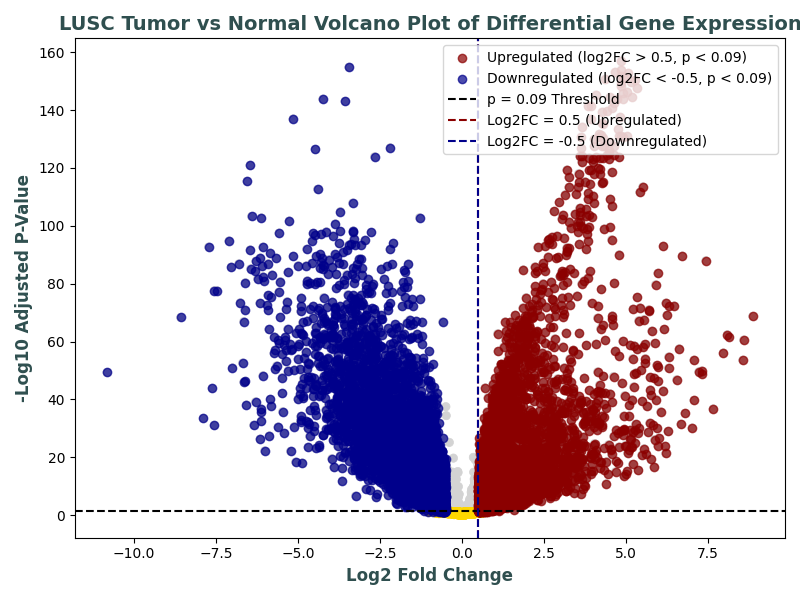}
         \includegraphics[scale=0.11]{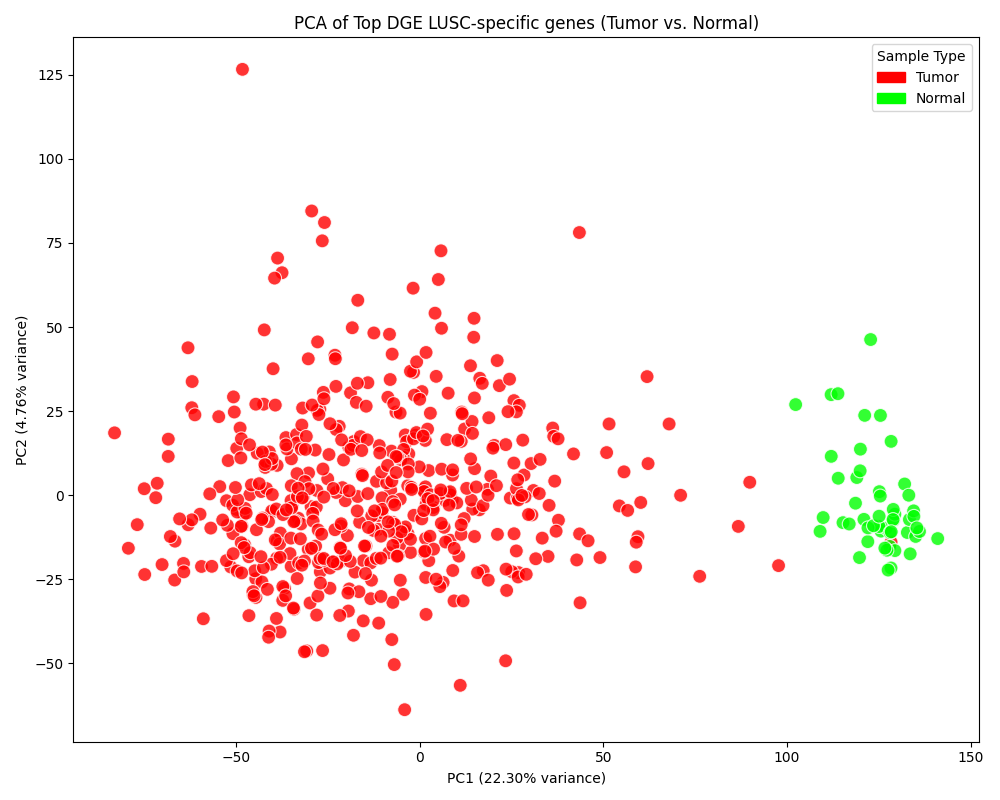}\\
    
    \includegraphics[scale=0.13]{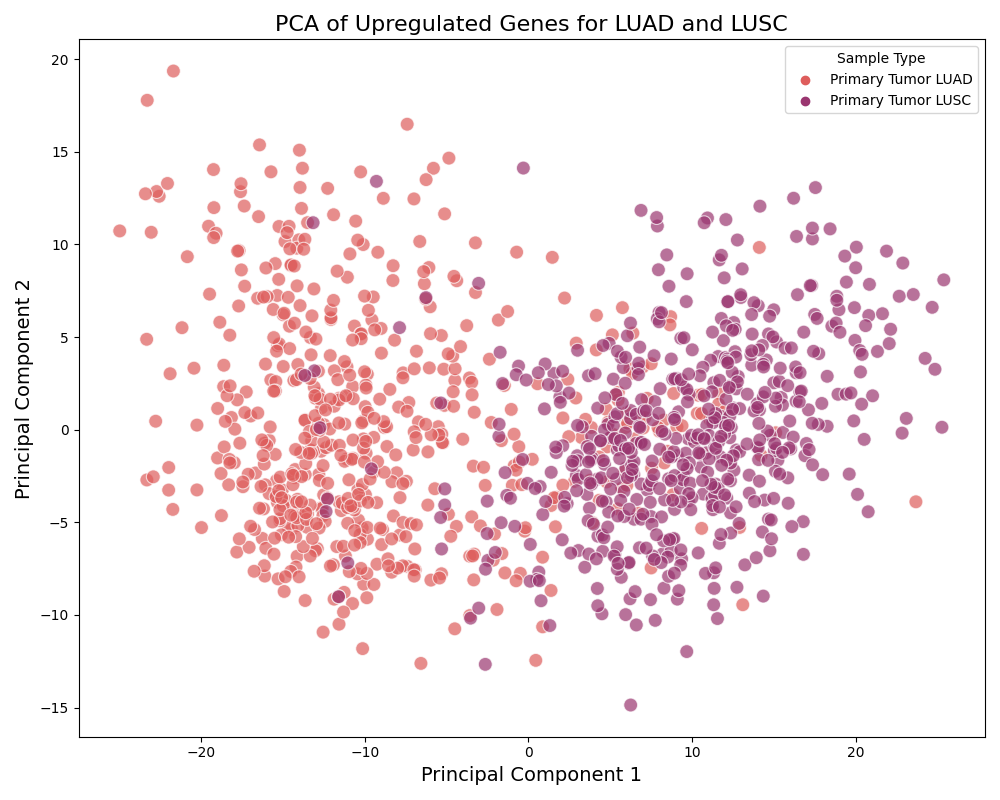}
    \includegraphics[scale=0.13]{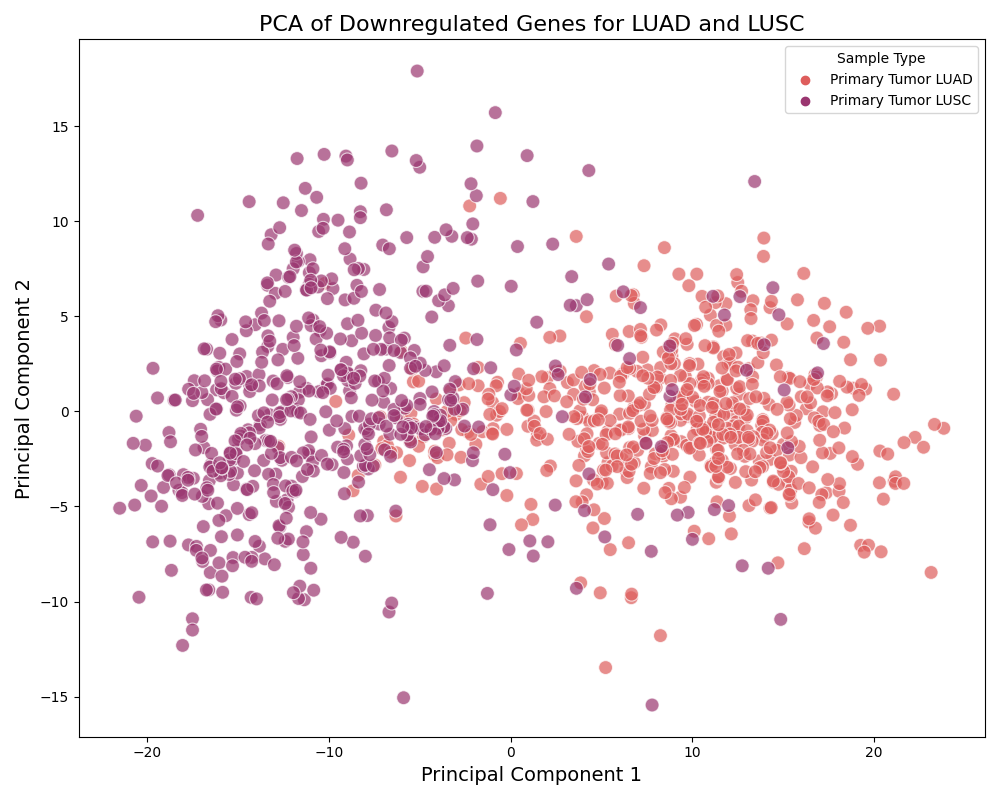}
    \includegraphics[scale=0.14]{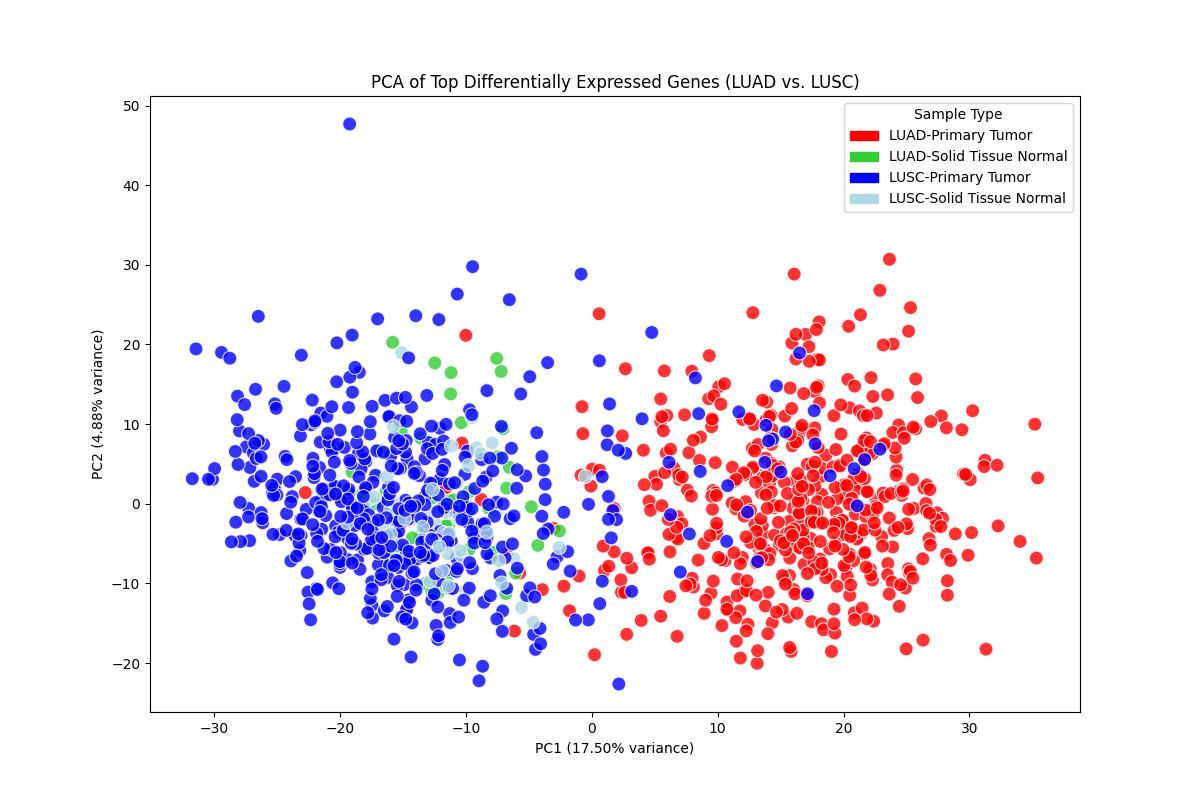}\\

    \includegraphics[scale=0.10]{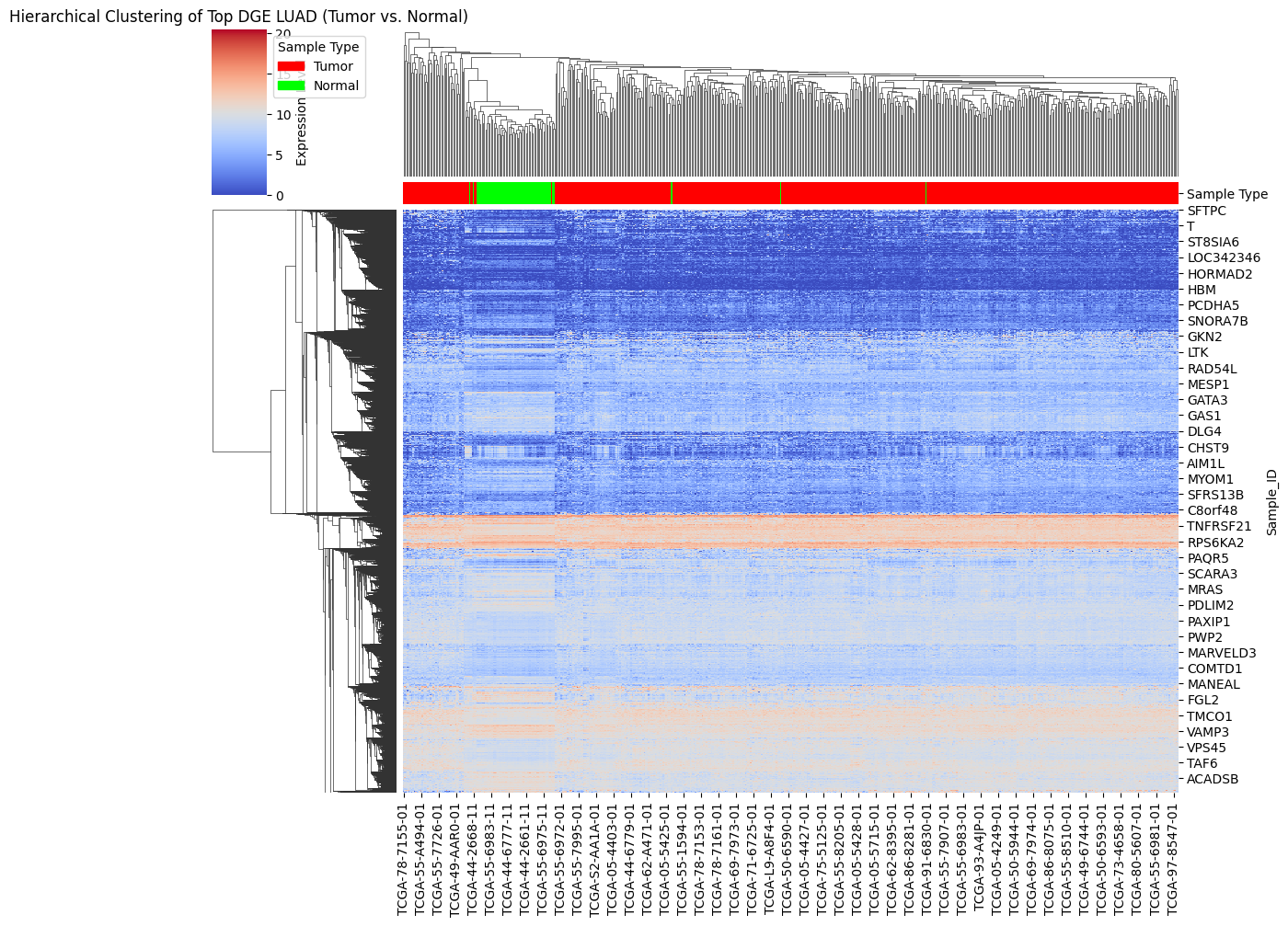}
    \includegraphics[scale=0.10]{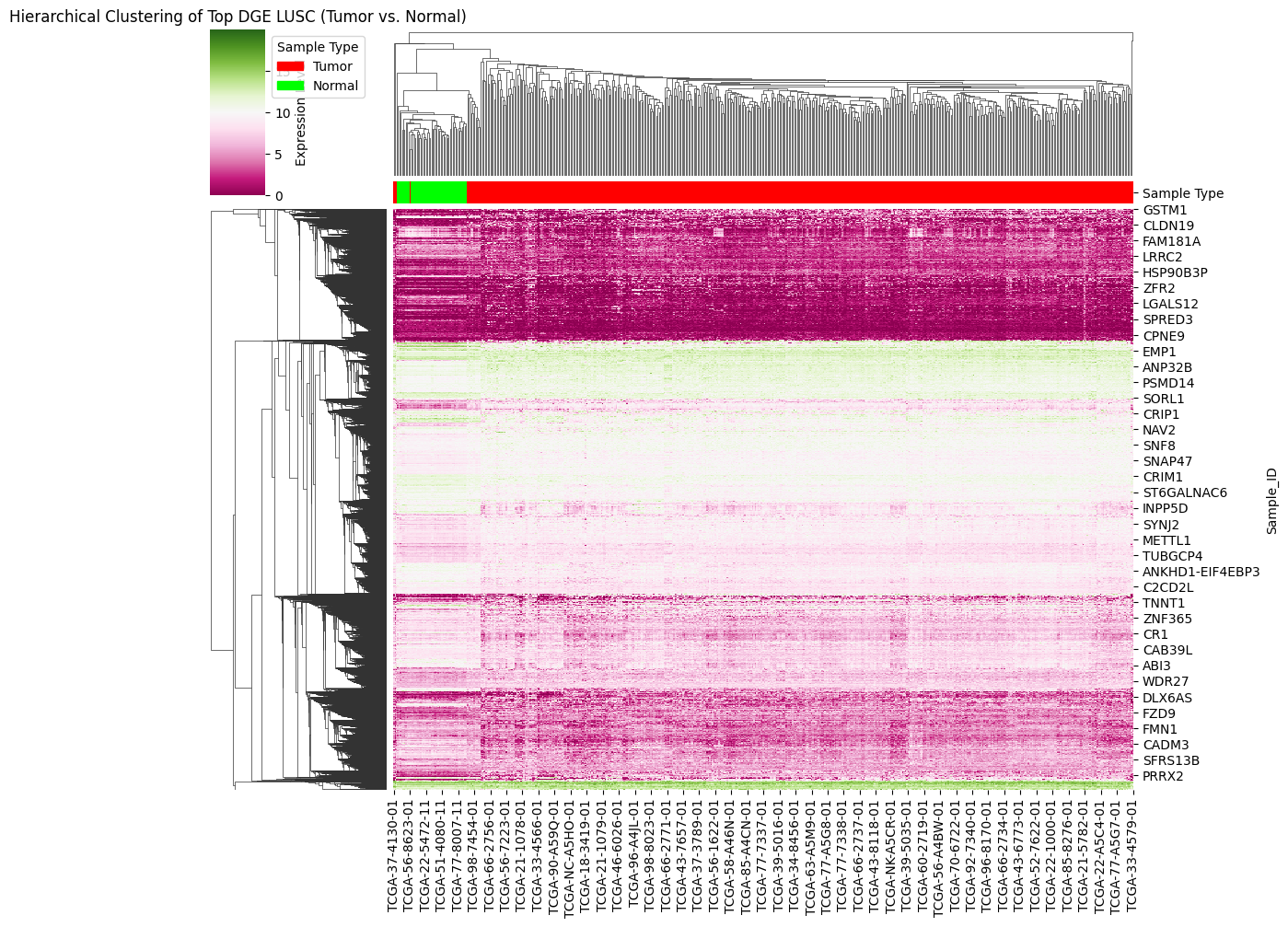}
    \includegraphics[scale=0.09]{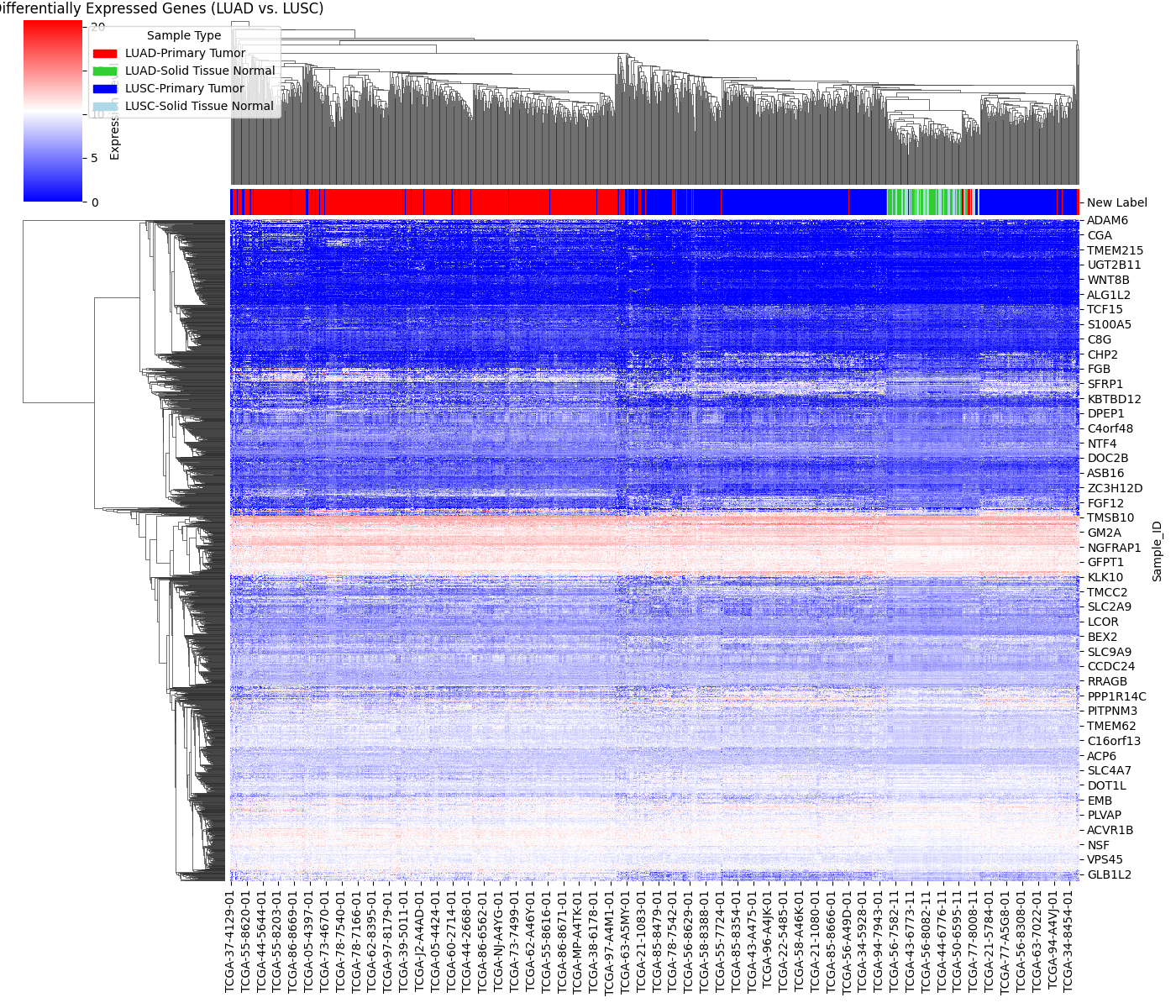}

    \caption{\textbf{Performance of differential gene expression (DGE) analysis for: CASE-2} (a) LUSC tumor vs. normal, (b) LUAD tumor vs. normal, and (c) LUSC tumor vs. LUAD tumor. \textbf{LUSC-specific} Case 2 criteria were applied: p-value $<$ 0.09 and Log Fold Change (LF) $>$ 0.5 for upregulated genes and $<$ -0.5 for downregulated genes. The results are visualized using a volcano plot, PCA plot, and hierarchical clustering heatmap.}

	\label{Figure10}
\end{figure*}

\noindent \text{Case 1:} LUAD-Specific Genes (p-value $<$ 0.05, LF $>$ 1 for Upregulated, LF $<$ -1 for Downregulated) shown as Fig~.\ref{Figure7}. \noindent \text{Case 2:} LUAD-Specific Genes (p-value $<$ 0.09, LF $>$ 0.5 for Upregulated, LF $<$ -0.5 for Downregulated) shown as Fig~.\ref{Figure8}. \noindent \text{Case 1:} LUSC-Specific Genes (p-value $<$ 0.05, LF $>$ 1 for Upregulated, LF $<$ -1 for Downregulated) shown as Fig~.\ref{Figure9}. \noindent \text{Case 2:} LUSC-Specific Genes (p-value $<$ 0.09, LF $>$ 0.5 for Upregulated, LF $<$ -0.5 for Downregulated) shown as Fig~.\ref{Figure10}.

\begin{figure*}
	\centering        
	\includegraphics[scale=0.15]{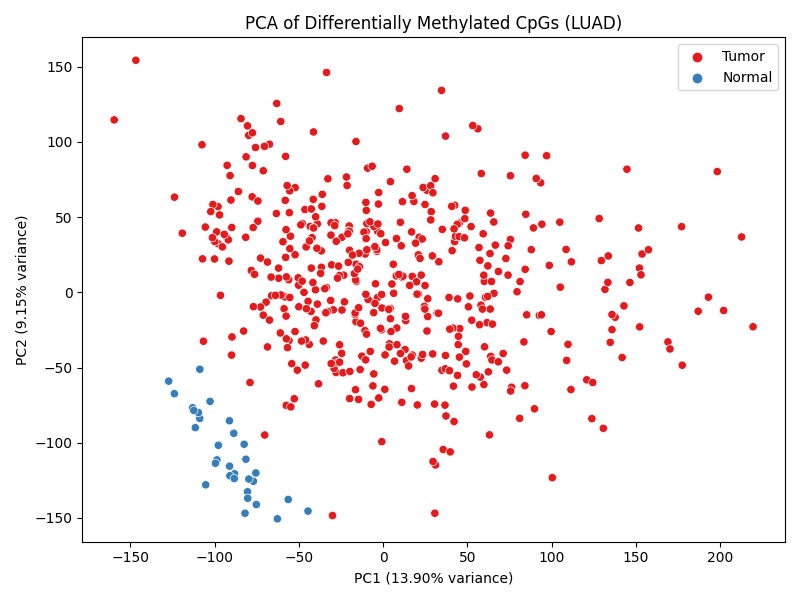}
        \includegraphics[scale=0.15]{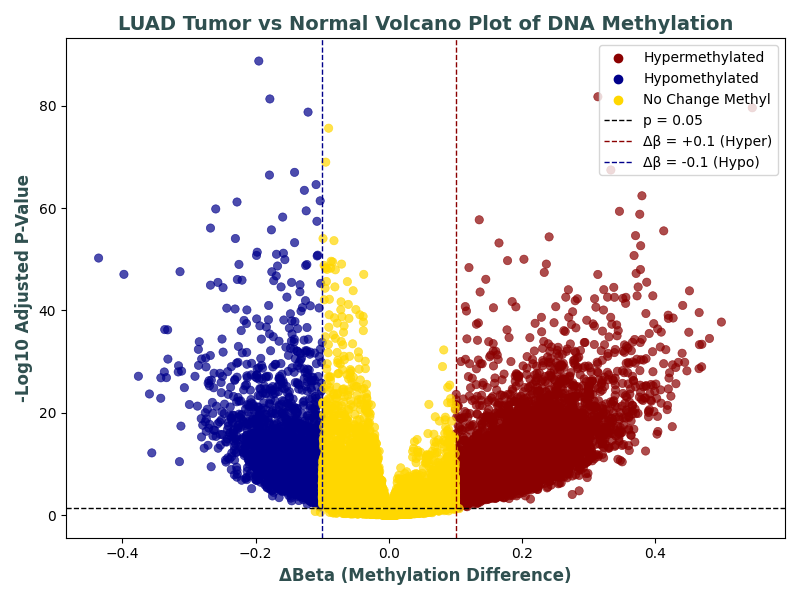}
	\includegraphics[scale=0.10]{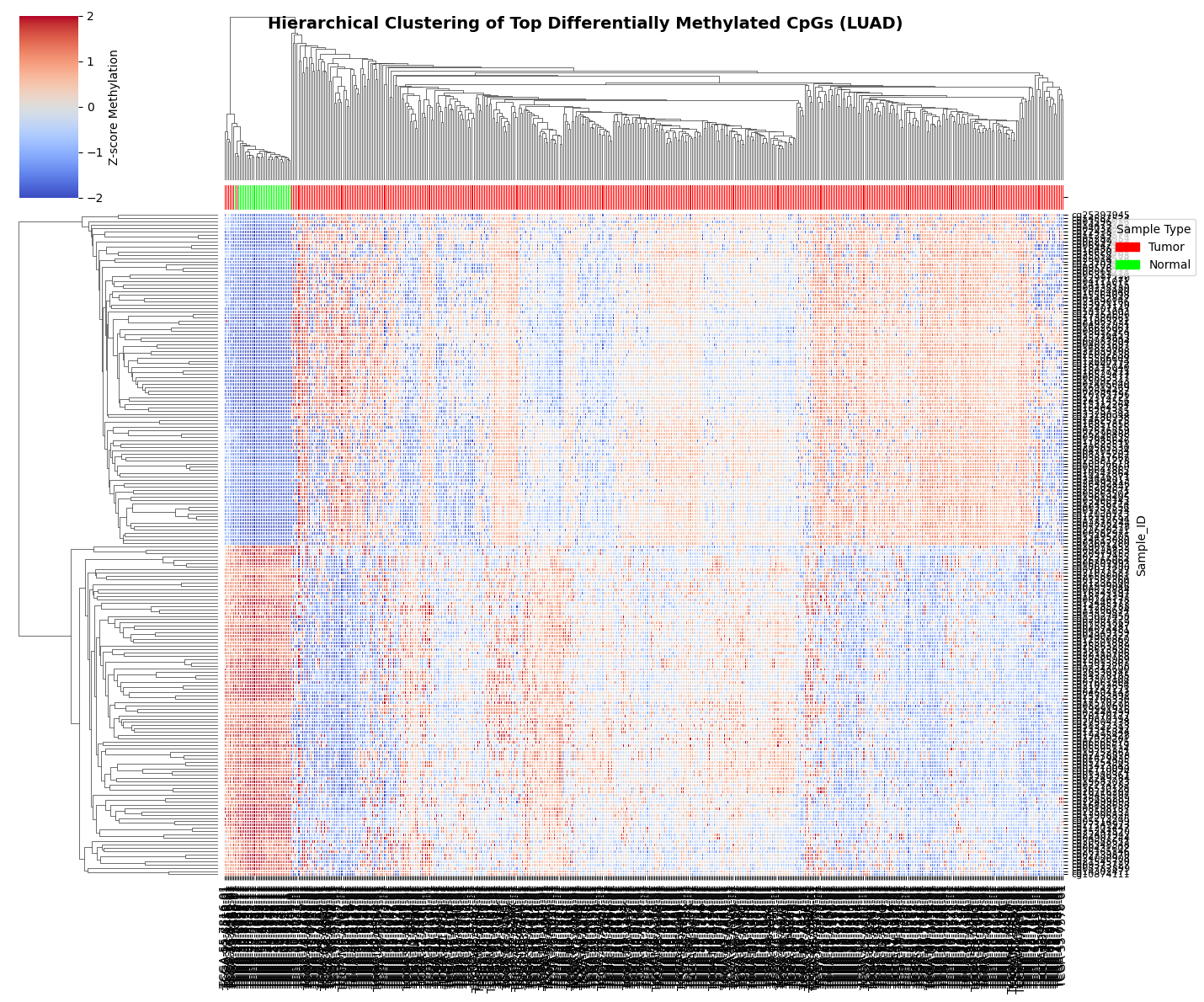}
         \includegraphics[scale=0.15]{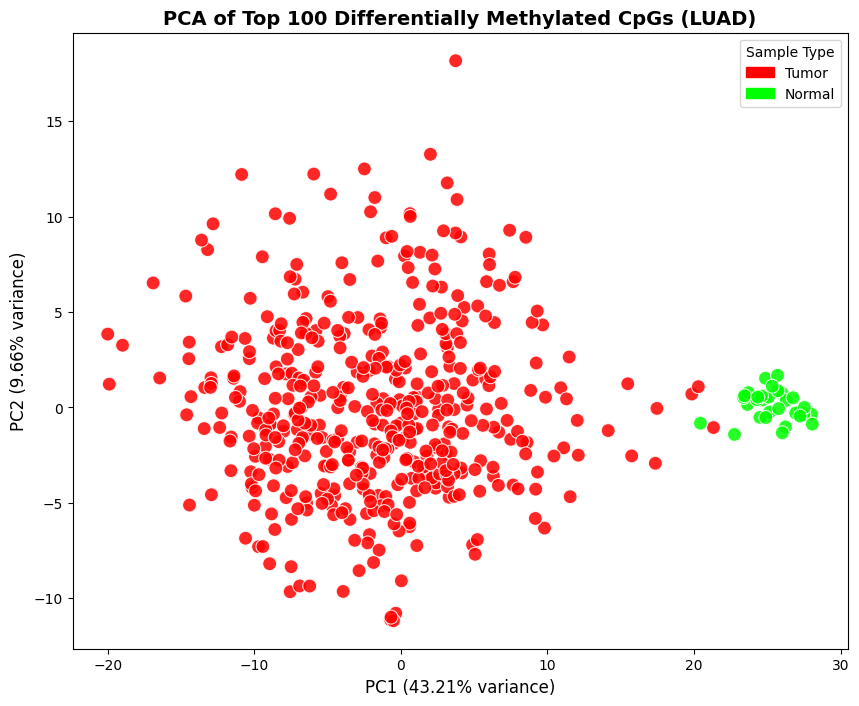}

         	\includegraphics[scale=0.15]{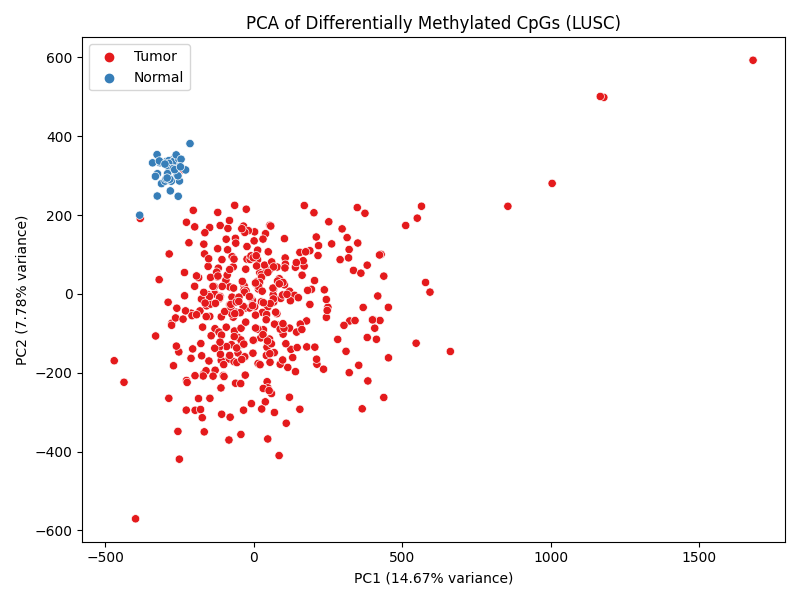}
        \includegraphics[scale=0.15]{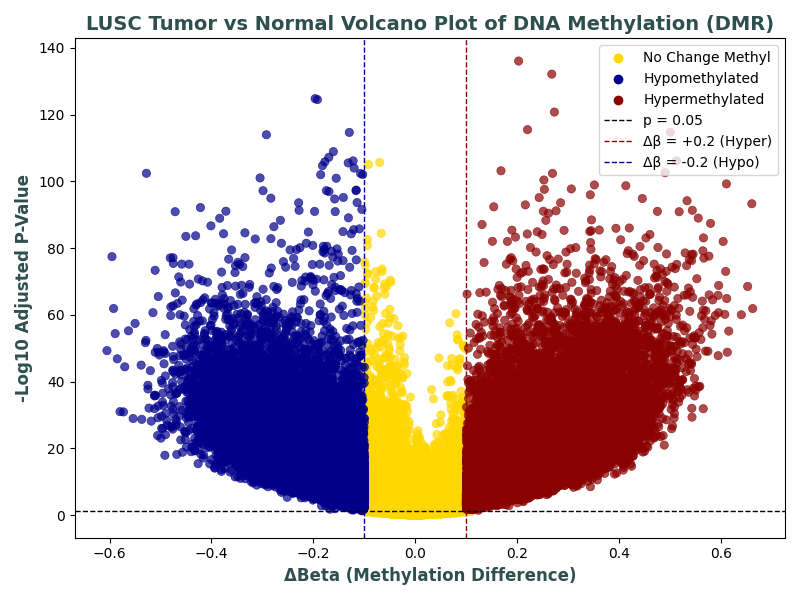}
	\includegraphics[scale=0.10]{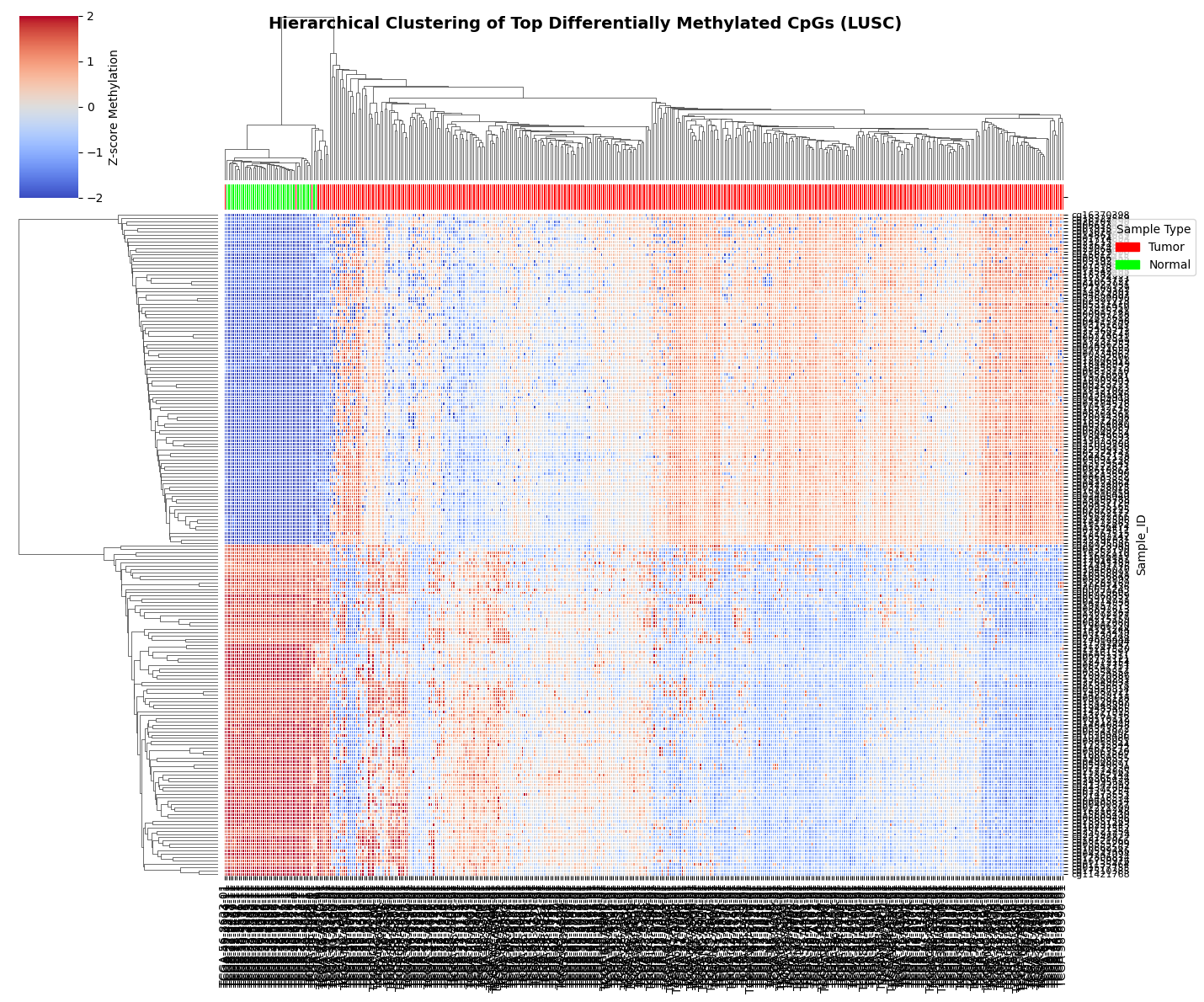}
         \includegraphics[scale=0.15]{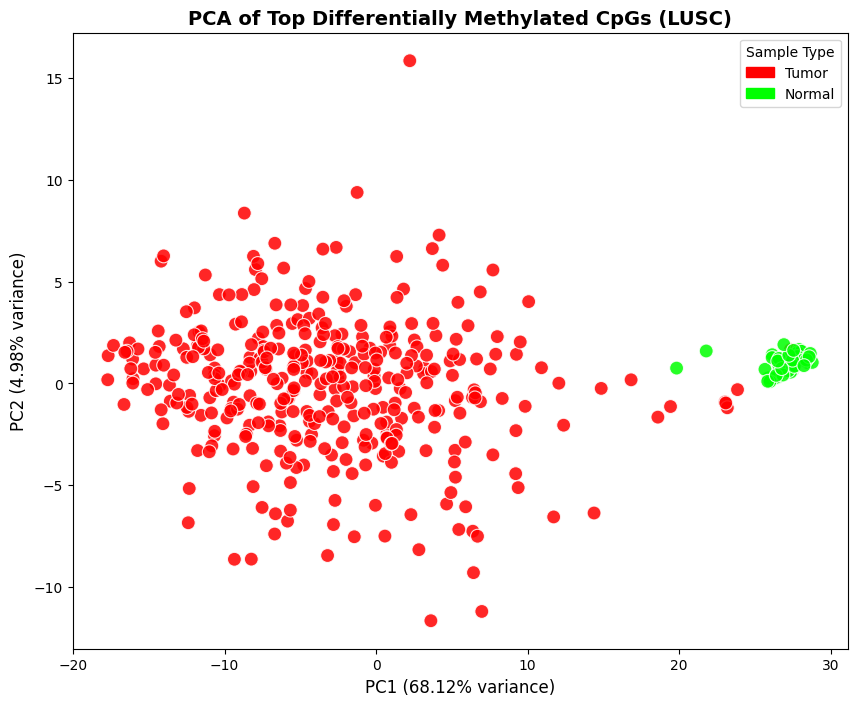}

        \includegraphics[scale=0.15]{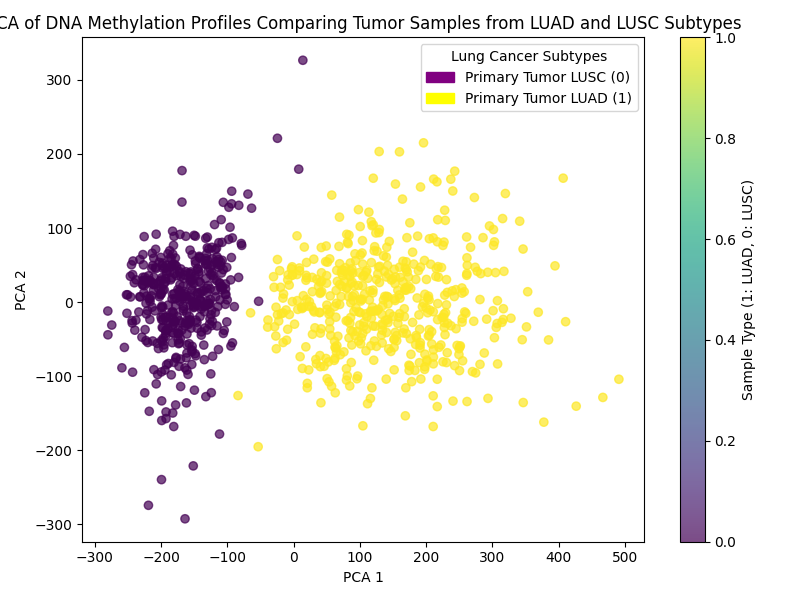}
         \includegraphics[scale=0.15]{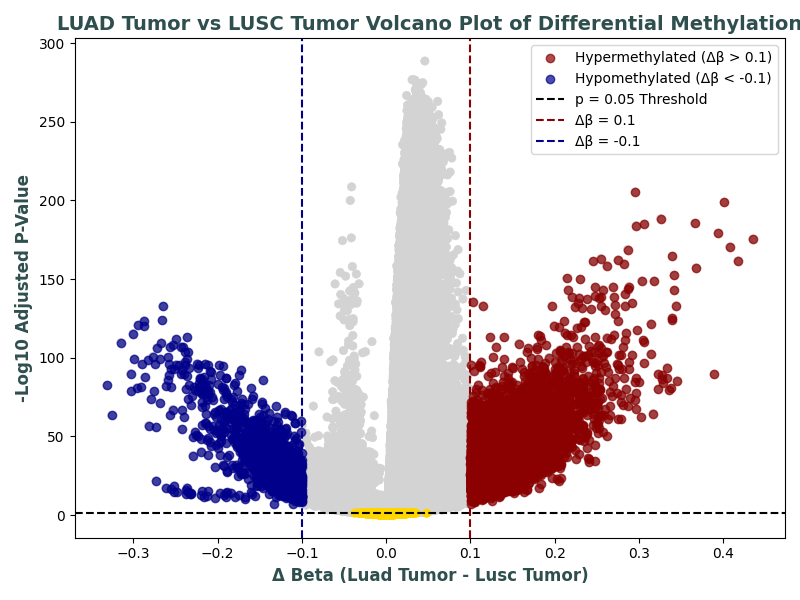}
        \includegraphics[scale=0.10]{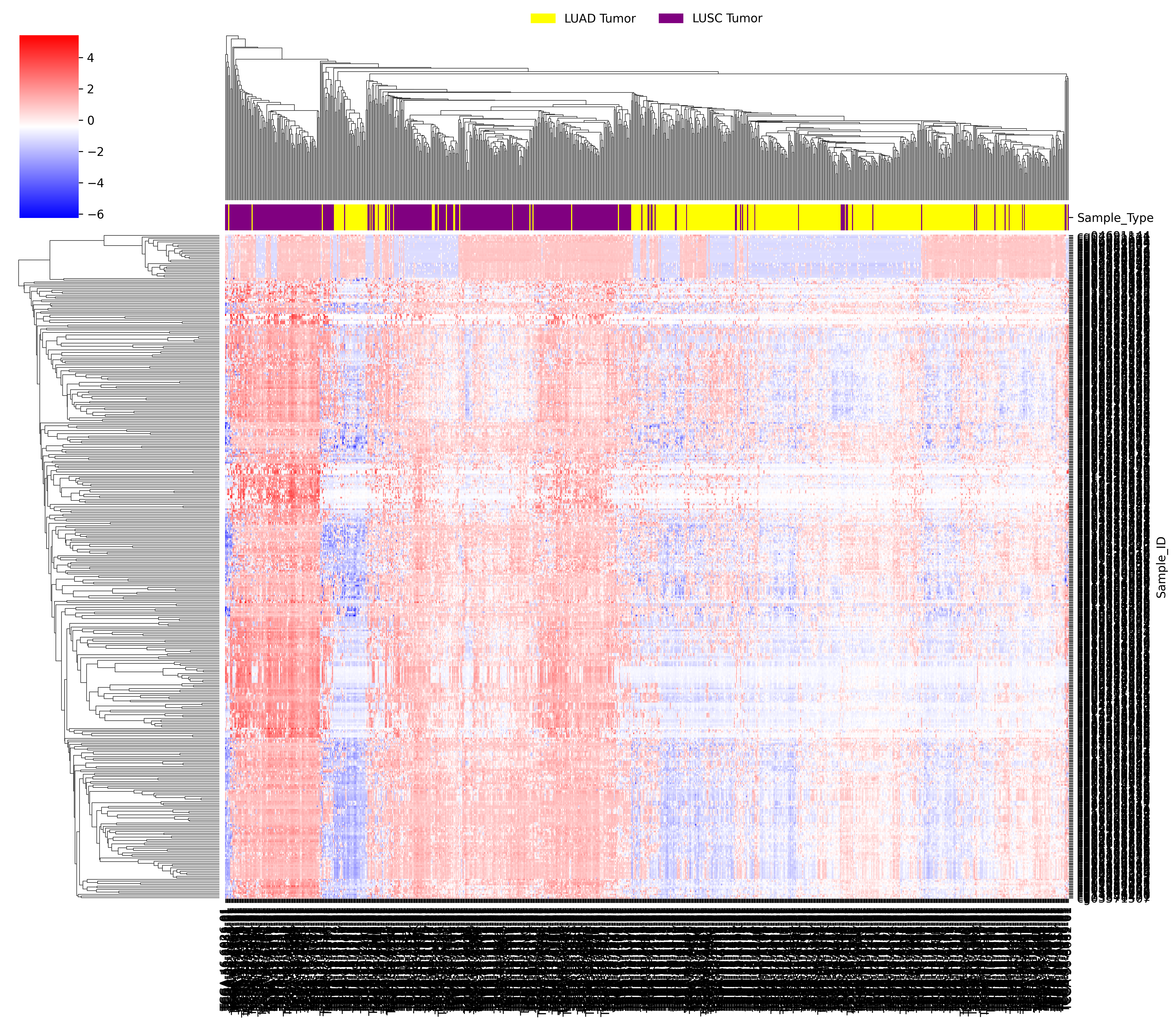}
	\includegraphics[scale=0.15]{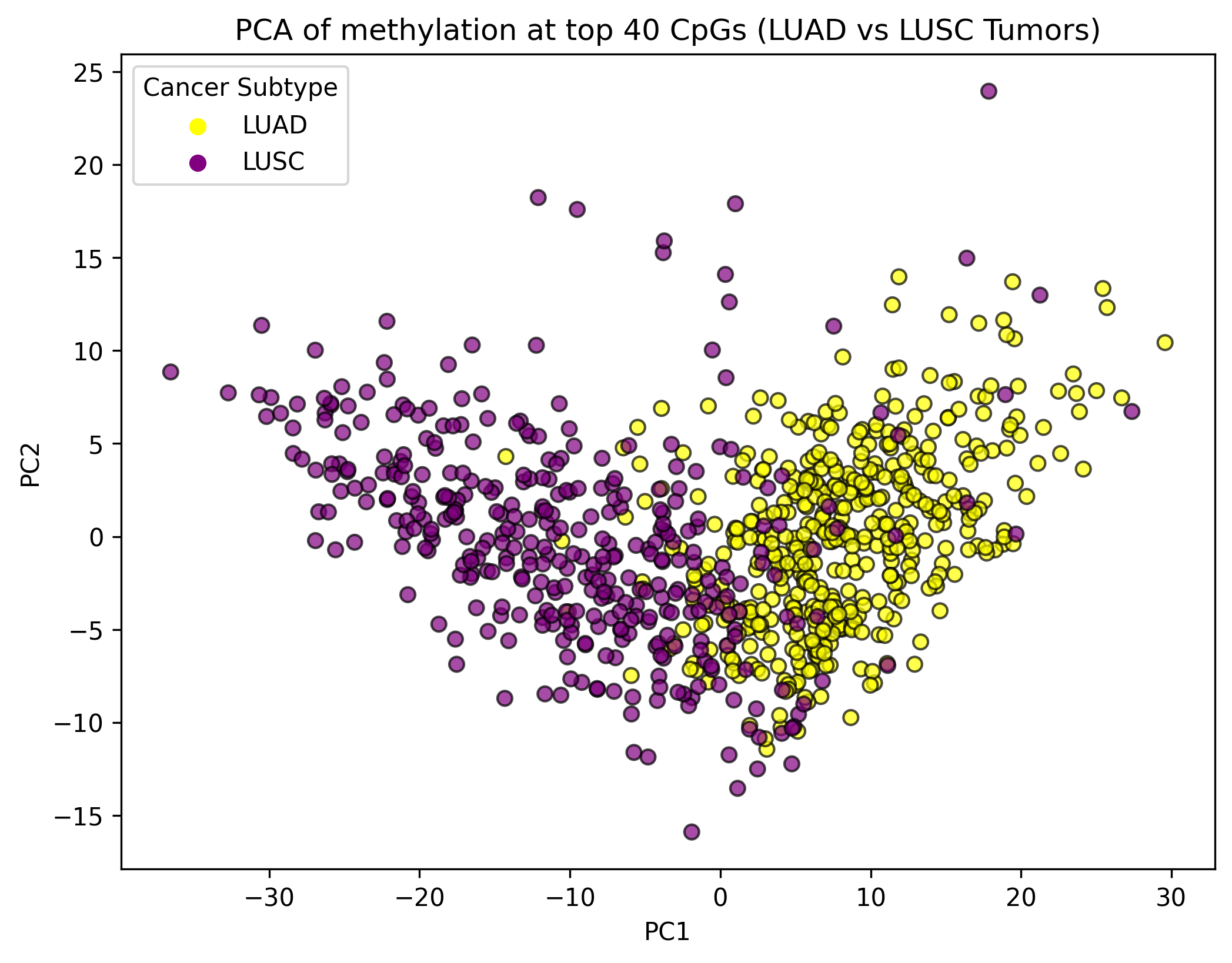}

     \includegraphics[scale=0.30]{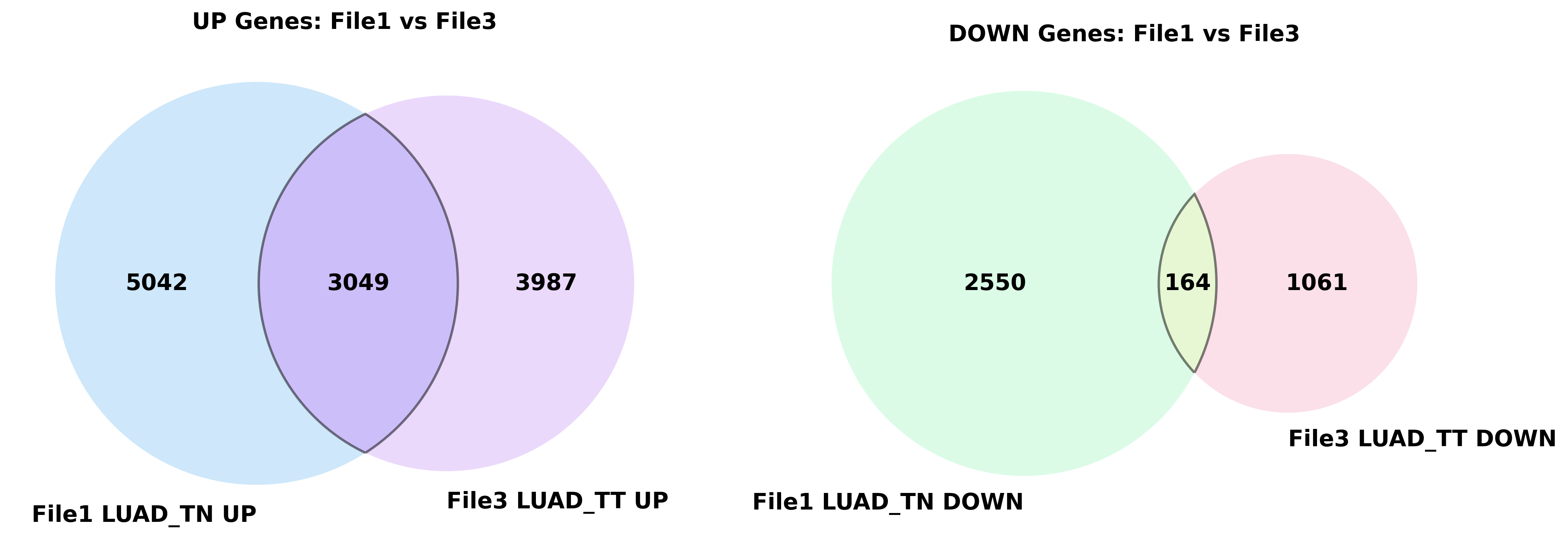}
	\includegraphics[scale=0.15]{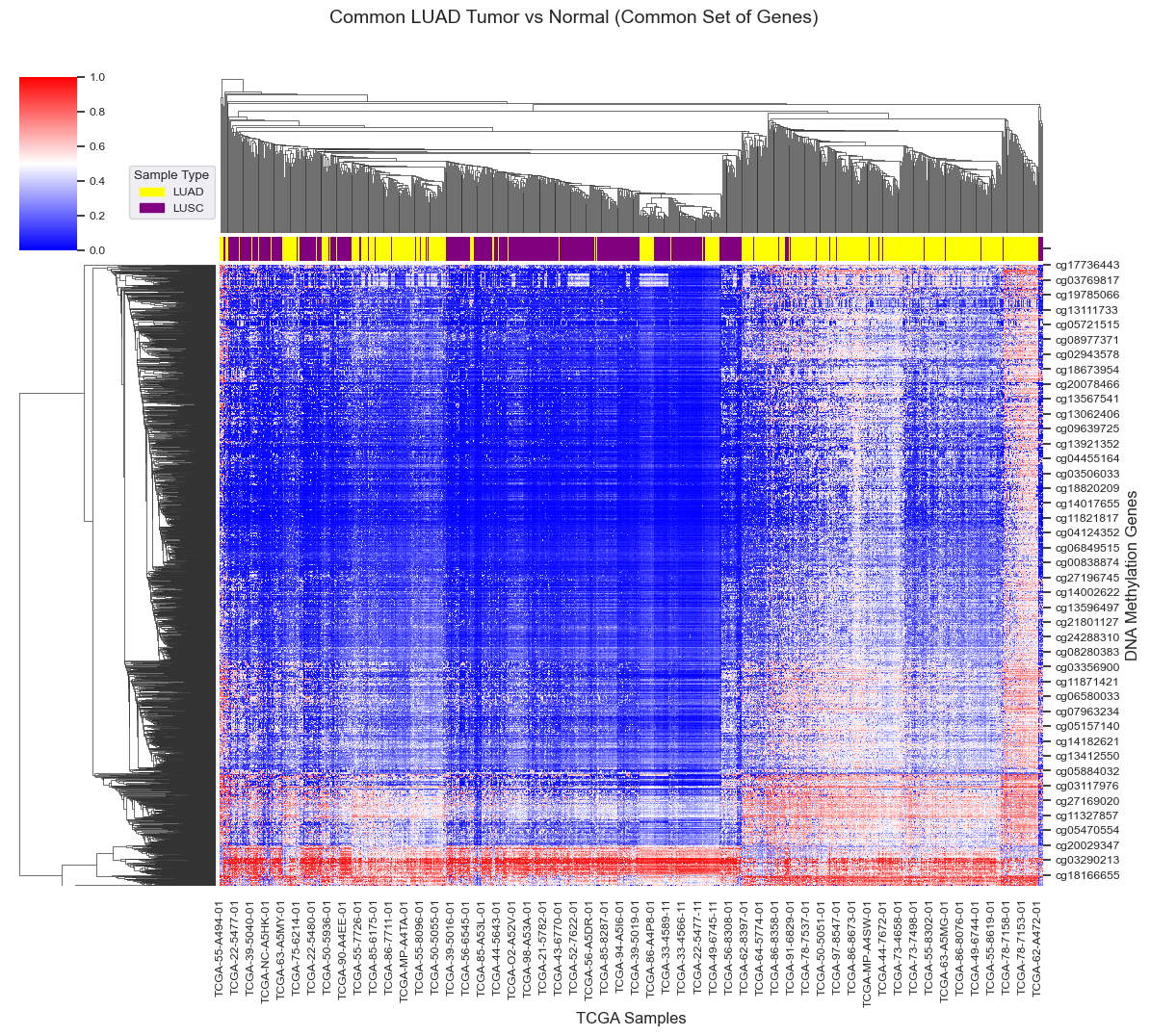}

    \includegraphics[scale=0.17]{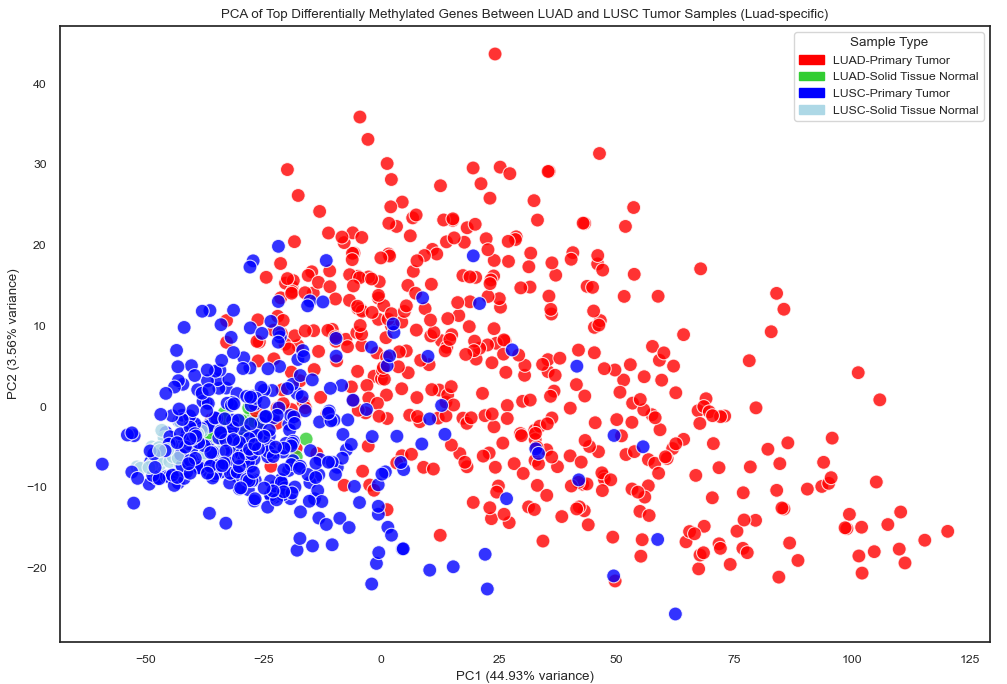}
    \includegraphics[scale=0.17]{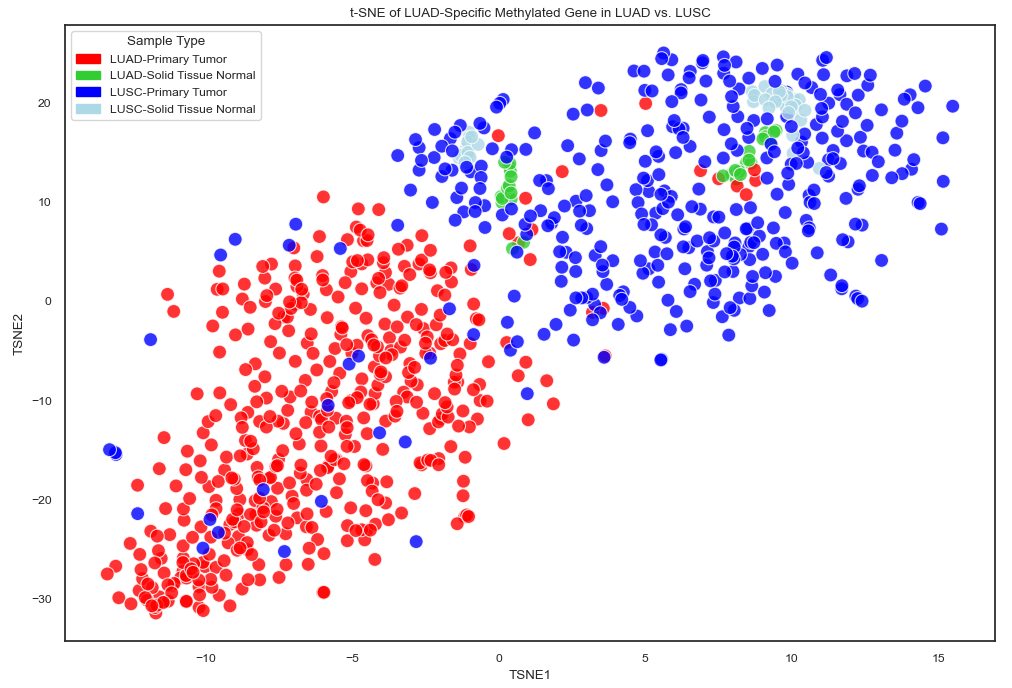}
    \includegraphics[scale=0.13]{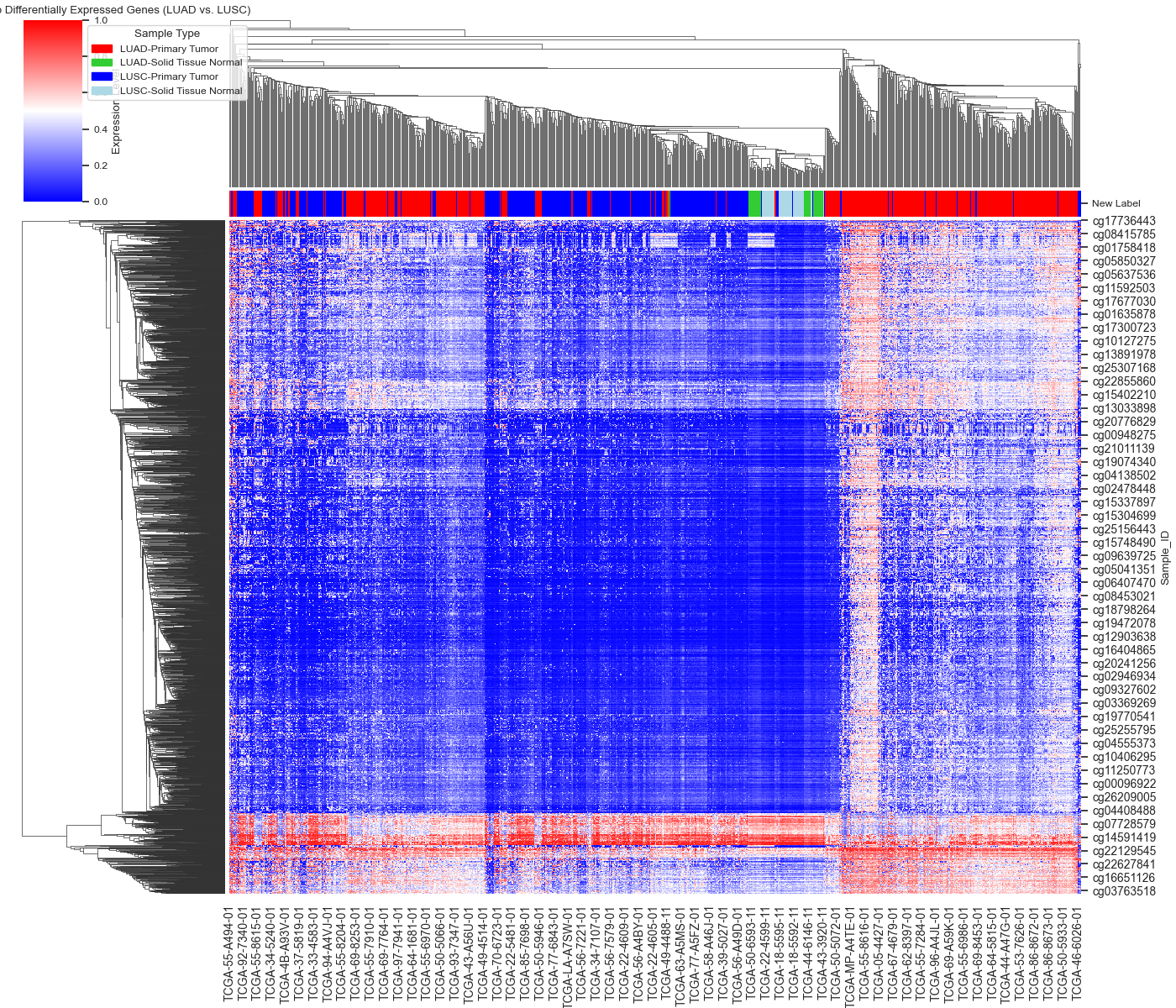}
    \caption{\textbf{Performance of differentially methylated genes for: CASE-1} (a) LUAD tumor vs. normal,  (b) LUSC tumor vs. normal, and (c) LUSC tumor vs. LUAD tumor. \textbf{LUAD-specific} Case 1 criteria were applied: p-value $<$ 0.05 and Delta $>$ 0.1 for hypermethylated genes and $<$ -01 for hypomethylated genes. The results are visualized using a volcano plot, PCA plot, and hierarchical clustering heatmap.}

	\label{Figure11}
\end{figure*}

\noindent \text{Case 1:} LUAD-Specific Genes (p-value $<$ 0.05, DB $>$ 0.1 for Hypermethylation, DB $<$ -0.1 for Hypomethylation) shown as shown as Fig~.\ref{Figure11} \noindent \text{Case 1:} LUSC-Specific Genes (p-value $<$ 0.05, DB $>$ 0.1 for Hypermethylation, DB $<$ -0.1 for Hypomethylation)  as Fig~.\ref{Figure11}

The RNA dataset comprised three distinct sample groups, including F1\_Luad\_TN (20,197 $\times$ 5), F2\_Lusc\_TN (20,243 $\times$ 5), and F3\_Luad\_Lusc\_TT (20,258 $\times$ 5), reflecting transcriptomic profiles across LUAD and LUSC subtypes (Table~\ref{tab:Table4}). Feature selection and differential analysis identified 617 LUAD-specific genes and 1,086 LUAD-dominant shared genes, highlighting subtype-specific transcriptional signatures. Further stratification of LUAD-biased genes revealed 258 shared, 506 opposite, and 1,690 neutral genes, indicating varying degrees of expression bias within LUAD samples.

\begin{figure*}[!ht]
	\centering
	\begin{tabular}{c}
	  
    \includegraphics[scale=0.18]{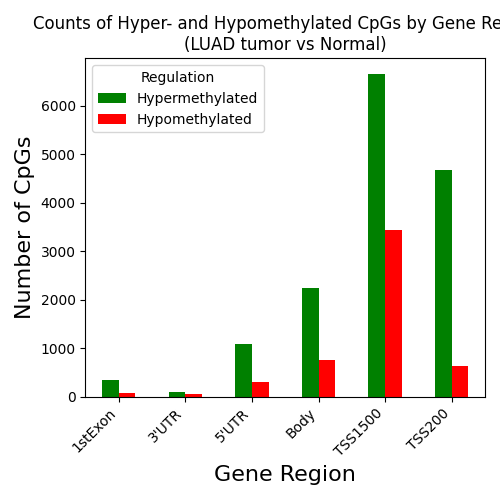}  
    \includegraphics[scale=0.18]{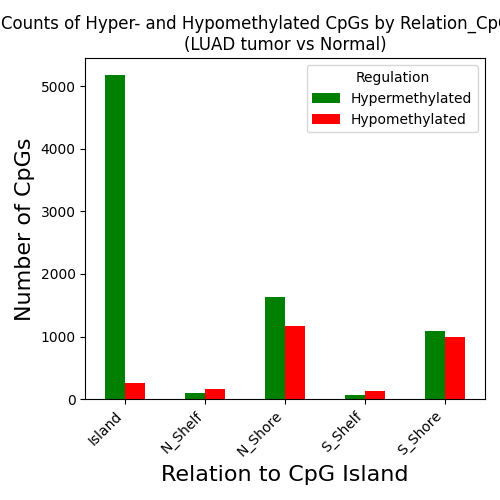}  
    \includegraphics[scale=0.18]{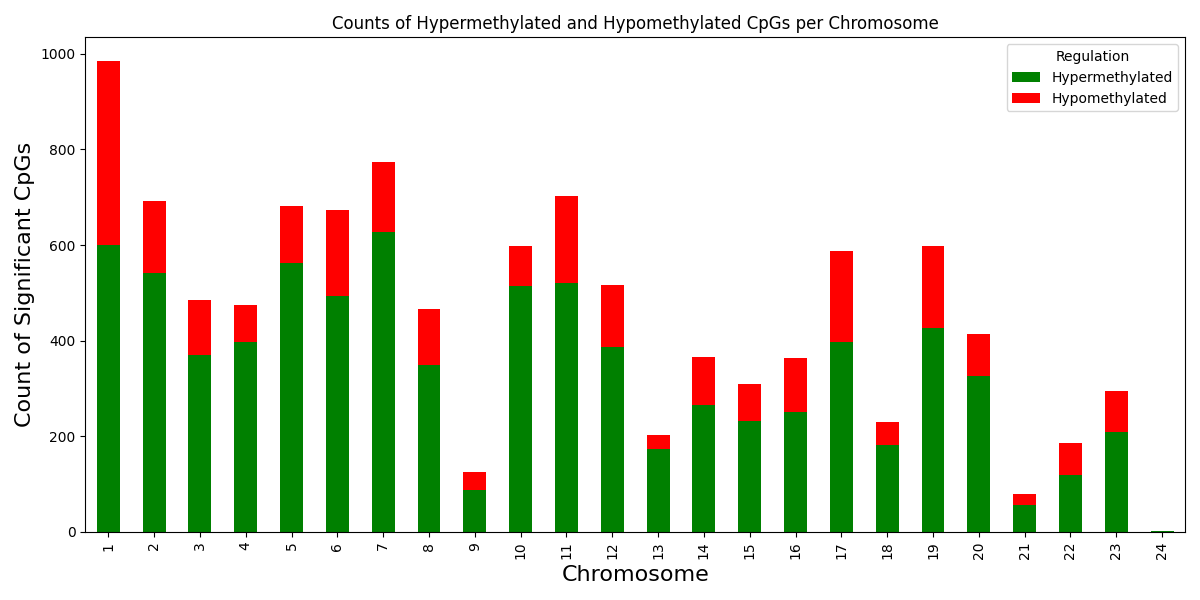}\\
    \includegraphics[scale=0.18]{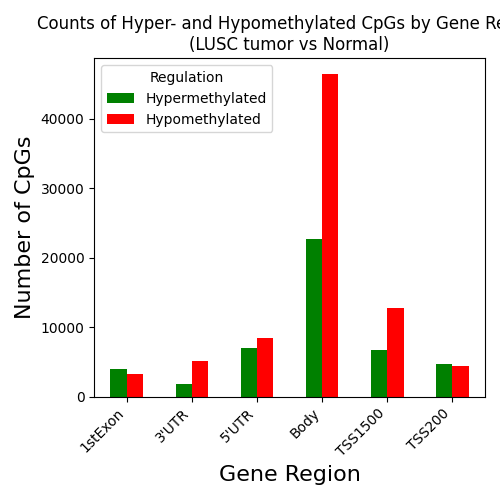}  
    \includegraphics[scale=0.18]{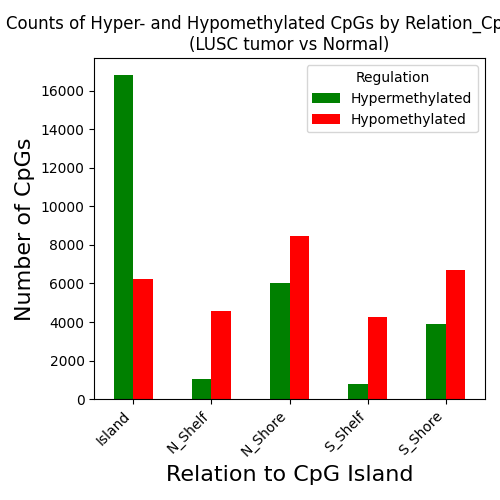}  
    \includegraphics[scale=0.18]{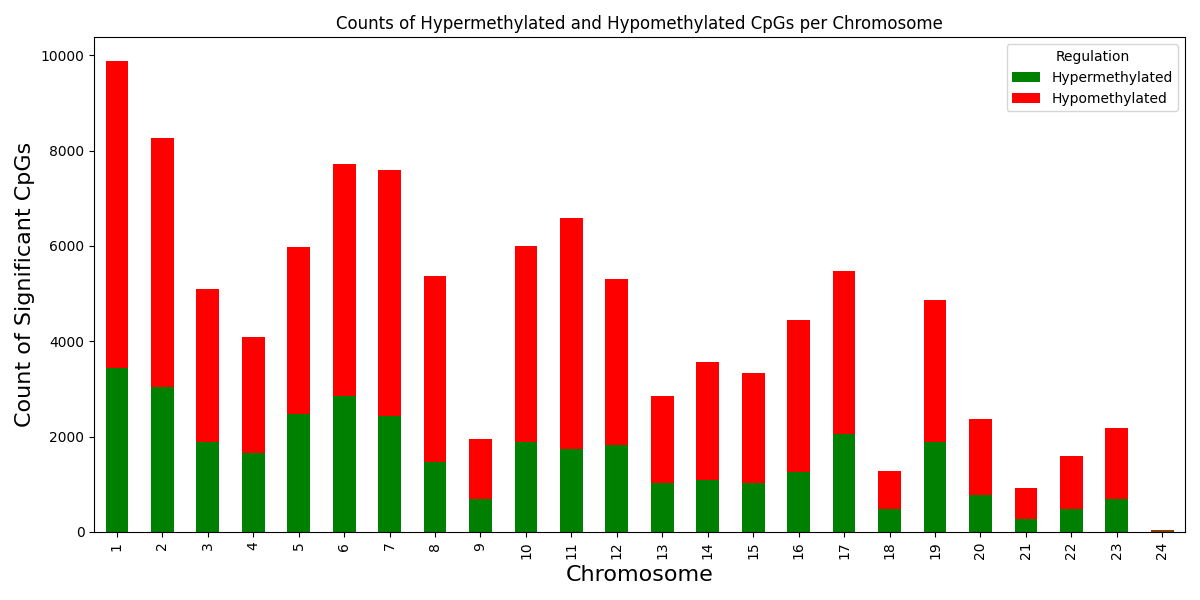} \\
    \includegraphics[scale=0.18]{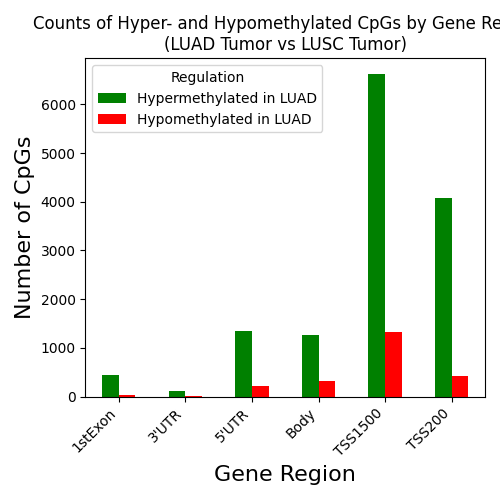}  \includegraphics[scale=0.18]{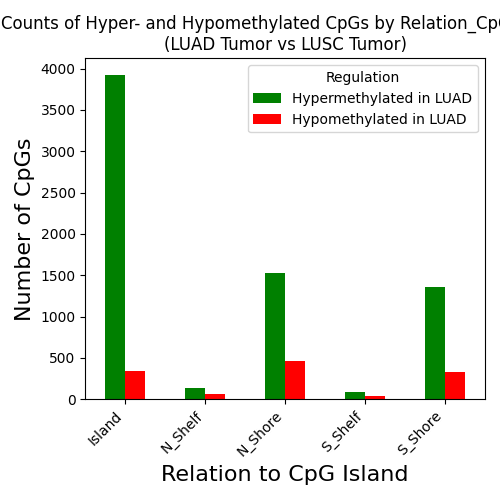}
    \includegraphics[scale=0.18]{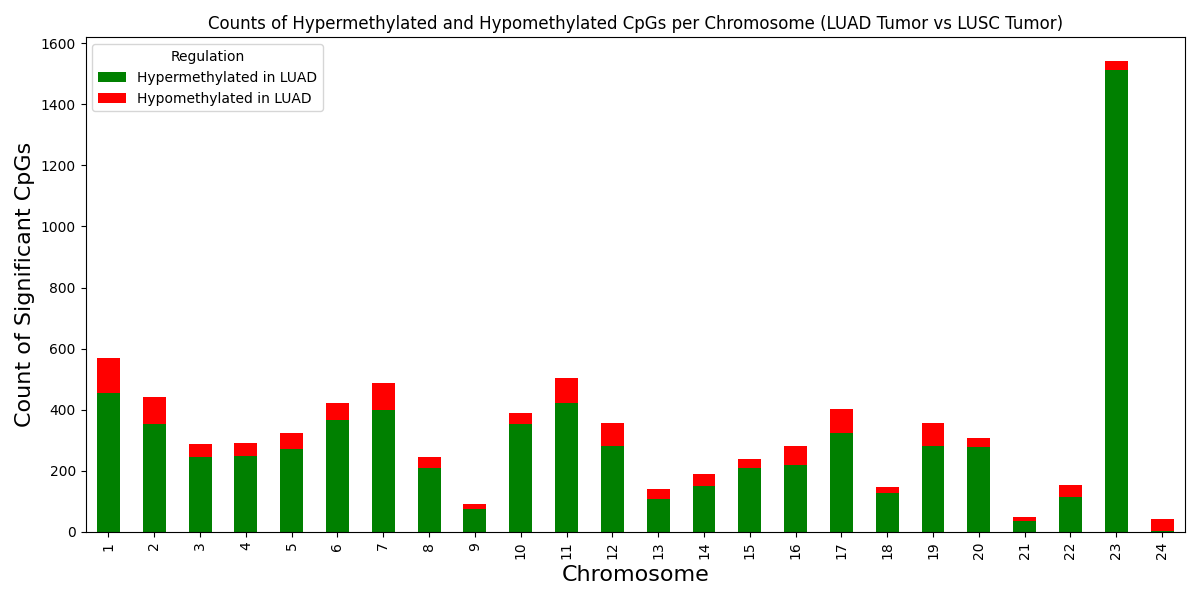} 
\end{tabular}
\caption{Distribution of hyper- and hypomethylated CpGs in lung adenocarcinoma (LUAD) and lung squamous cell carcinoma (LUSC). Differentially methylated CpGs (DMCs) are summarized across gene regions (TSS1500, TSS200, 5'UTR, 1st exon, gene body, and 3'UTR) and CpG island-related regions (islands, shores, shelves, and open sea) for LUAD and LUSC tumors versus normal tissues, along with their chromosomal distributions. Comparative analysis between LUAD and LUSC highlights distinct methylation patterns across genomic contexts and chromosomes.}
    \label{Figure12}
\end{figure*}

The DNA methylation dataset encompassed larger feature dimensions, with F1\_Luad\_TN (98,847 $\times$ 19), F2\_Lusc\_TN (396,065 $\times$ 19), and F3\_Luad\_Lusc\_TT (98,847 $\times$ 18), which were further expanded to include additional CpG sites (F1\_exp: 155,856 $\times$ 19; F2\_exp: 516,791 $\times$ 19; F3\_exp: 155,856 $\times$ 18) as summarized in Table~\ref{tab:Table4}. Column filtering and integration resulted in a consolidated DNA dataset (F1+F2+F3: 155,856 $\times$ 17) and a multi-omic summary of 22,784 features across 28 dimensions. As illustrated in Fig~.\ref{Figure12}, differentially methylated CpGs exhibit distinct patterns across gene regions, CpG island contexts, and chromosomes in LUAD and LUSC.

\subsection{Performance of Sample$_{1}$ Hypermethylated and Downregulated}

\noindent \textbf{Performance with 64 Features ($DNA_{32}$ + $RNA_{32}$) for Sample$_{1}$:} The first set of experiments combined $DNA_{32}$ and $RNA_{32}$ features, resulting in a total of 64 features. Using 6 qubits, 6 epochs, and a batch size of 32 with a learning rate of 0.03, the model achieved a \textbf{training accuracy of 95.45\%} and a \textbf{test accuracy of 94.55\%}, with a corresponding \textbf{AUC of 0.9964}, indicating highly accurate discrimination between classes. The confusion matrix showed that most misclassifications were concentrated in the minority class, with only 1 false negative. Moreover, the learning rate was reduced to 0.003, the model maintained robust performance (\textbf{test ACC: 93.94\%, AUC: 0.9846}), though with a slightly higher loss, suggesting that lower learning rates may lead to more stable but slightly slower convergence. Increasing the batch size to 64 and training with 50 epochs further improved performance, reaching a \textbf{test accuracy of 95.76\%} and an \textbf{AUC of 0.9881}, demonstrating that longer training and larger batch sizes can enhance model generalization.

\noindent \textbf{Performance with 128 Features (DNA$_{64}$ + RNA$_{64}$) Sample$_{1}$:} In addition, we evaluated the model with DNA$_{128}$ and RNA$_{128}$ features, total 256 features. Using 7 qubits and 50 training epochs, the model with a batch size of 32 achieved a \textbf{test accuracy of 93.94\%} and an \textbf{AUC of 0.9874}, slightly lower than the smaller 64-feature model for this configuration. However, increasing the batch size to 64 significantly improved the test accuracy to \textbf{95.15\%} and the AUC to \textbf{0.9936}, highlighting the sensitivity of model performance to batch size in higher-dimensional feature spaces. Notably, training losses remained consistently low (0.12–0.17), indicating effective convergence across all runs.

\noindent \textbf{Performance with 256 Features (DNA$_{128}$ + RNA$_{128}$) Sample$_{1}$:} Finally, the model was tested on 256 combined features using 8 qubits and 50 epochs. With a batch size of 32, the test accuracy was \textbf{93.33\%} with an \textbf{AUC of 0.9954}, and with batch size 64, performance improved to \textbf{95.15\% accuracy} and \textbf{AUC of 0.9838}, showing that even with doubled feature dimensions, the model maintained robust generalization when sufficient batch sizes were used. Across all experiments, the F1-scores ranged from 0.94 to 0.96, confirming balanced performance between precision and recall.

\subsection{Performance of Sample$_{2}$ (Hypomethylated + Downregulated Genes)}		
\noindent \textbf{Performance with 64 Features ($DNA_{32}$ + $RNA_{32}$) for Sample$_{2}$:} 
The initial set of 64 features (32 DNA + 32 RNA) was trained using 6 qubits, 6 epochs, and a learning rate of 0.003 with a batch size of 32. The model achieved a training accuracy of 92.72\% and a test accuracy of 93.33\%, with an AUC of 0.9688. The confusion matrix showed that LUAD (1) samples were correctly classified in most cases (87 out of 91), while LUSC (0) samples were slightly misclassified (7 out of 74), indicating that the model slightly favored LUAD detection at this configuration. Increasing the batch size to 64 and training for 50 epochs resulted in a slight decrease in test accuracy (90.91\%) and AUC (0.9690), highlighting the sensitivity of smaller learning rates with larger batch sizes. Although, by increasing the learning rate to 0.03 and keeping batch size at 64, the model improved test accuracy to 93.94\% and AUC to 0.9715, demonstrating that appropriate learning rate tuning enhances convergence. Further optimization (6 qubits, 50 epochs, batch size 64) yielded a test accuracy of 94.55\% and AUC of 0.9835, confirming robust generalization.

\noindent \textbf{Performance with 128 Features ($DNA_{64}$ + $RNA_{64}$) for Sample$_{2}$:} 
Further, expanding the feature set to 128 DNA and RNA genes and training with 7 qubits and 50 epochs, the model achieved a test accuracy of 94.55\% and AUC of 0.9710. The confusion matrix indicated perfect subtype classification of LUAD samples (91/91) and minimal misclassification of LUSC (9/74), suggesting that increased features improved model sensitivity for both targets.

\noindent \textbf{Performance with 256 Features ($DNA_{128}$ + $RNA_{128}$) for Sample$_{2}$:} 
Finally, with 256 combined features with 8 qubits, the model demonstrated high performance across both batch sizes. For batch size 32, test accuracy was 94.55\% and AUC 0.9905, while increasing batch size to 64 further improved test accuracy to 95.76\% with AUC 0.9884. Confusion matrix analysis showed that LUAD samples were correctly classified in 89–91 cases, and LUSC misclassifications were minimal (2–5 samples), indicating a balanced classification across both cancer subtypes. F1-scores ranged from 0.949 to 0.963, confirming strong precision-recall balance.

\begin{table}[h!]
\centering
\caption{QNN Performance Metrics for Sample 3 (Hypomethylated+Upregulated and Hypermethylated+Downregulated Genes)}
\begin{tabular}{lcccc}
\hline
Metric & QNN-32 & QNN-64 & QNN-128 & QNN-256 \\
\hline
DNA Features & 8 & 16 & 32 & 64 \\
RNA Features & 8 & 16 & 32 & 64 \\
Total Features & 32 & 64 & 128 & 256 \\
Qubits & 5 & 6 & 7 & 8 \\
Layers & 6 & 6 & 6 & 6 \\
Epochs & 50 & 50 & 50 & 50 \\
LeRate & 0.03 & 0.03 & 0.03 & 0.03 \\
Train Acc & 0.9605 & 0.9605 & 0.9621 & 0.9484 \\
Test Acc & 0.9697 & 0.9697 & 0.9697 & 0.9636 \\
Train F1 & 0.9640 & 0.9648 & 0.9663 & 0.9548 \\
Test F1 & 0.9721 & 0.9727 & 0.9730 & 0.9677 \\
Train Prec & 0.9667 & 0.9468 & 0.9446 & 0.9205 \\
Test Prec & 0.9886 & 0.9674 & 0.9574 & 0.9474 \\
Train Rec & 0.9613 & 0.9834 & 0.9890 & 0.9917 \\
Test Rec & 0.9560 & 0.9780 & 0.9890 & 0.9890 \\
Train AUC & 0.9844 & 0.9890 & 0.9875 & 0.9918 \\
Test AUC & 0.9930 & 0.9947 & 0.9903 & 0.9960 \\
\hline
\end{tabular}
\label{tab:Table3}
\end{table}

\begin{table}[h!]
\footnotesize
\centering
\caption{Summary of RNA and DNA data processing and feature selection}
\label{tab:Table4} 
\begin{tabular}{|l|l|}
\hline
\multicolumn{2}{|c|}{\textbf{RNA Data}} \\
\hline
\textbf{Description} & \textbf{Dimensions} \\
\hline
F1\_Luad\_TN & (20197, 5) \\
F2\_Lusc\_TN & (20243, 5) \\
F3\_Luad\_Lusc\_TT & (20258, 5) \\
LUAD-specific genes identified & 617 \\
LUAD-dominant shared genes & 1086 \\
luad\_biased\_shared & (258, 14) \\
luad\_biased\_opposite & (506, 14) \\
luad\_biased\_neutral & (1690, 14) \\
LUAD-biased neutral genes & 1690 \\
\hline
\multicolumn{2}{|c|}{\textbf{DNA Data}} \\
\hline
\textbf{Description} & \textbf{Dimensions} \\
\hline
F1\_Luad\_TN & (98847, 19) \\
F2\_Lusc\_TN & (396065, 19) \\
F3\_Luad\_Lusc\_TT & (98847, 18) \\
Expanded CpGs (F1\_exp) & (155856, 19) \\
Expanded CpGs (F2\_exp) & (516791, 19) \\
Expanded CpGs (F3\_exp) & (155856, 18) \\
F1\_dna after column filtering & (155856, 7) \\
F3\_dna after column filtering & (155856, 9) \\
Final DNA data (F1 only) & (155856, 14) \\
F2\_subset & (516791, 5) \\
Final DNA data (F1+F2+F3) & (155856, 17) \\
Multi-omic summary & (22784, 28) \\
After top CpG selection: genes / CpGs & 2694 / 2626 \\
After arbitration: genes / CpGs & 2626 / 2626 \\
Duplicate CpGs & 0 \\
Likely Tumor suppressors (Hyper+Down) & 388 \\
Likely Activated Oncogenes (Hypo+UP) & 262 \\
Conflicting & 1977 \\
Other & 1 \\
\hline
\end{tabular}
\end{table}

\begin{table*}[htbp]
\caption{DNA methylation summary for LUAD and LUSC patients NC = No change, Hyper = Hypermethylated, Hypo = Hypomethylated.}
\centering
\scriptsize
\setlength{\tabcolsep}{2pt}
\renewcommand{\arraystretch}{1}
\begin{tabular}{|l|l|c|c|c|c|c|c|l|l|c|c|c|}
\hline
Gene & CpG & db$\_$F1 & AdjP$\_$F1 & is$\_$promoter & db$\_$F3 & AdjP$\_$F3 & Dir$\_$F3 & region & RelC$\_$Island & Dir$\_$F2 & db$\_$F2 & AdjP$\_$F2 \\
\hline
NTRK2 & cg01009697 & 0.0904 & 0.0021 & TRUE & 0.1195 & 4.99e-45 & Hyper & TSS1500;1stExon;5'UTR & Island & NC & 0.0108 & 0.0676 \\
ADAM23 & cg06123396 & 0.1455 & 2.22e-06 & TRUE & 0.0061 & 0.6526 & NC & TSS1500 & N$\_$Shore & Hyper & 0.1578 & 2.82e-06 \\
DMRT2 & cg00934355 & 0.1079 & 0.000157 & TRUE & -0.2026 & 4.80e-53 & Hypo & TSS1500;TSS1500 & N$\_$Shore & Hyper & 0.3523 & 2.04e-24 \\
SYN2 & cg02700894 & 0.2614 & 2.26e-13 & TRUE & 0.2083 & 3.39e-52 & Hyper & TSS1500;TSS1500 & N$\_$Shore & NC & 0.0920 & 0.0013 \\
SOST & cg01898628 & -0.1032 & 0.000127 & TRUE & 0.2022 & 3.57e-65 & Hyper & TSS200 & S$\_$Shelf & Hypo & -0.3336 & 1.29e-29 \\
PAK7 & cg02167438 & 0.1941 & 6.23e-08 & TRUE & 0.1703 & 5.67e-51 & Hyper & TSS200;TSS200 & Island & NC & 0.0430 & 0.0026 \\
PRIMA1 & cg00745231 & 0.0892 & 0.000519 & TRUE & 0.1170 & 1.57e-52 & Hyper & TSS1500 & Island & NC & 0.0140 & 0.0398 \\
ALOX12 & cg03762994 & 0.0247 & 0.4452 & TRUE & 0.1022 & 2.25e-23 & Hyper & TSS200 & Island & NC & -0.0649 & 0.0044 \\
GPC3 & cg14867863 & 0.0237 & 0.5322 & TRUE & 0.1215 & 2.92e-26 & Hyper & TSS1500;TSS1500 & Island & NC & -0.0147 & 0.5945 \\
EMILIN3 & cg07609788 & 0.1944 & 2.32e-13 & TRUE & 0.1723 & 9.38e-66 & Hyper & TSS200 & Island & NC & 0.0593 & 0.0026 \\
OXGR1 & cg03371199 & 0.1504 & 6.19e-07 & TRUE & 0.1476 & 1.12e-48 & Hyper & TSS200 & Island & NC & 0.0252 & 0.1325 \\
D4S234E & cg17183546 & 0.1783 & 8.17e-11 & TRUE & 0.1660 & 1.66e-70 & Hyper & TSS200;TSS200 & Island & NC & 0.0448 & 0.00132 \\
\hline
\end{tabular}
\label{tab:Table5}
\end{table*}

\begin{table*}[t]
\caption{GO Molecular Function (GO:MF) Term Enrichment Results}
\centering
\scriptsize
\begin{tabular}{l c c c l}
\toprule
\textbf{Term} & \textbf{P-value} & \textbf{AdjP-value} & \textbf{CScore} & \textbf{Genes} \\
\midrule

\parbox[t]{5.2cm}{ION CHANNEL REGULATOR ACTIVITY (GO:0099106)} &
1.29E-05 & 0.003262 & 69.5474 &
\parbox[t]{8.8cm}{FKBP1A; KCNE3; NRXN1; NPY; KCNB2; KCNIP4; DRD2; SGK1; FGF12; GPLD1} \\

\parbox[t]{5.2cm}{TRANSMITTER-GATED MONOATOMIC ION CHANNEL ACTIVITY INVOLVED IN REGULATION OF POSTSYNAPTIC MEMBRANE POTENTIAL (GO:1904315)} &
1.74E-05 & 0.003262 & 141.4516 &
\parbox[t]{8.8cm}{GABRA2; CHRNA3; GRIN2A; CHRNA4; GRIK3; GRIA3} \\

\parbox[t]{5.2cm}{POTASSIUM CHANNEL REGULATOR ACTIVITY (GO:0015459)} &
1.40E-04 & 0.017553 & 75.4627 &
\parbox[t]{8.8cm}{KCNE3; KCNB2; KCNIP4; SGK1; DRD2; LRRC55} \\

\parbox[t]{5.2cm}{NEUROTRANSMITTER RECEPTOR ACTIVITY INVOLVED IN REGULATION OF POSTSYNAPTIC MEMBRANE POTENTIAL (GO:0099529)} &
7.74E-04 & 0.072780 & 79.0932 &
\parbox[t]{8.8cm}{GABRA2; GRIN2A; GRIK3; GRIA3} \\

\parbox[t]{5.2cm}{LIGAND-GATED MONOATOMIC ION CHANNEL ACTIVITY (GO:0015276)} &
0.001148 & 0.076700 & 66.4258 &
\parbox[t]{8.8cm}{GABRA2; CHRNA3; CHRNA4; GRIA3} \\

\parbox[t]{5.2cm}{VOLTAGE-GATED POTASSIUM CHANNEL ACTIVITY (GO:0005249)} &
0.001224 & 0.076700 & 36.6675 &
\parbox[t]{8.8cm}{KCNH4; KCNE3; KCNA1; KCNB2; LRRC55; KCNJ2} \\

\bottomrule
\end{tabular}
\label{tab:Table6}
\end{table*}

\begin{table*}[t]
\caption{GO Biological Process enrichment analysis results}
\centering
\scriptsize
\begin{tabular}{l c c c l}
\toprule
\textbf{Term} & \textbf{P-value} & \textbf{AdjP-value} & \textbf{CScore} & \textbf{Genes} \\
\midrule

\parbox[t]{5.2cm}{CHEMICAL SYNAPTIC TRANSMISSION (GO:0007268)} &
1.65E-08 & 2.85E-05 & 97.4850 &
\parbox[t]{8.8cm}{GABRA2; CHRNA3; SYT3; CHRNA4; NRXN1; GRIK3; SYT9; SYN2; SHISA6; GRIN2A; NPY; PENK; PCDHB6; SLC17A7; SLC12A6; DRD2; GRIA3; DRD5; SNCA} \\

\parbox[t]{5.2cm}{SYNAPSE ASSEMBLY (GO:0007416)} &
6.45E-06 & 0.0055798 & 92.6622 &
\parbox[t]{8.8cm}{GABRA2; NLGN4Y; NLGN1; NRXN1; LRRC4; PCDHB6; NRCAM; DRD2; SDK2} \\

\parbox[t]{5.2cm}{REGULATION OF SYNAPTIC TRANSMISSION, GLUTAMATERGIC (GO:0051966)} &
1.22E-05 & 0.0070445 & 117.6408 &
\parbox[t]{8.8cm}{NLGN1; GRIN2A; TSHZ3; NRXN1; GRIK3; DRD2; SLC38A2} \\

\parbox[t]{5.2cm}{ANTEROGRADE TRANS-SYNAPTIC SIGNALING (GO:0098916)} &
3.08E-05 & 0.0133321 & 47.5698 &
\parbox[t]{8.8cm}{GRIN2A; SYT3; CHRNA4; NRXN1; NPY; PENK; PCDHB6; SLC12A6; SYT9; SYN2; DRD5; SNCA} \\

\parbox[t]{5.2cm}{SYNAPTIC VESICLE CLUSTERING (GO:0097091)} &
6.53E-05 & 0.0164971 & 636.9464 &
\parbox[t]{8.8cm}{NLGN1; NRXN1; SYN2} \\

\parbox[t]{5.2cm}{SYNAPTIC VESICLE LOCALIZATION (GO:0097479)} &
6.53E-05 & 0.0164971 & 636.9464 &
\parbox[t]{8.8cm}{NLGN1; NRXN1; SYN2} \\

\parbox[t]{5.2cm}{SYNAPSE ORGANIZATION (GO:0050808)} &
6.68E-05 & 0.0164971 & 48.2403 &
\parbox[t]{8.8cm}{NLGN1; NRXN1; NEDD4; PCDHB6; LRRC4; NRCAM; DRD2; SYN2; SDK2; SNCA} \\

\parbox[t]{5.2cm}{AXON GUIDANCE (GO:0007411)} &
1.05E-04 & 0.0226957 & 43.3623 &
\parbox[t]{8.8cm}{NELL2; EFNB1; ROBO3; EPHA7; NRXN1; BOC; CNTN1; NRCAM; NEO1; NTN1} \\

\parbox[t]{5.2cm}{NEURON PROJECTION MORPHOGENESIS (GO:0048812)} &
1.37E-04 & 0.0257632 & 40.6531 &
\parbox[t]{8.8cm}{NLGN1; NRN1; NRXN1; SLITRK4; NTF3; NRCAM; DRD2; SGK1; FGFR2; SHANK1} \\

\parbox[t]{5.2cm}{CELL SURFACE RECEPTOR PROTEIN TYROSINE KINASE SIGNALING PATHWAY (GO:0007169)} &
1.70E-04 & 0.0257632 & 29.1602 &
\parbox[t]{8.8cm}{FOXC2; NTRK2; EPHA7; CHRNA3; IRS4; NRG1; GHR; TIAM1; EFNB1; FLRT2; NTF3; FGF12; GPLD1; FGFR2} \\

\parbox[t]{5.2cm}{RESPONSE TO MAGNESIUM ION (GO:0032026)} &
1.79E-04 & 0.0257632 & 342.1910 &
\parbox[t]{8.8cm}{KCNA1; RYR3; SNCA} \\

\parbox[t]{5.2cm}{POSITIVE REGULATION OF VESICLE FUSION (GO:0031340)} &
1.79E-04 & 0.0257632 & 342.1910 &
\parbox[t]{8.8cm}{SYT3; SYT9; SNCA} \\

\parbox[t]{5.2cm}{AXONOGENESIS (GO:0007409)} &
2.34E-04 & 0.0303830 & 32.7293 &
\parbox[t]{8.8cm}{ROBO3; EFNB1; EPHA7; NRXN1; SLITRK4; BOC; CNTN1; NRCAM; DRD2; NEO1; FGFR2} \\

\parbox[t]{5.2cm}{EXCITATORY POSTSYNAPTIC POTENTIAL (GO:0060079)} &
2.46E-04 & 0.0303830 & 129.5869 &
\parbox[t]{8.8cm}{CHRNA3; GRIN2A; DRD2; BEGAIN} \\

\parbox[t]{5.2cm}{POSITIVE REGULATION OF SYNAPTIC TRANSMISSION, GLUTAMATERGIC (GO:0051968)} &
2.97E-04 & 0.0342507 & 119.6009 &
\parbox[t]{8.8cm}{NLGN1; GRIN2A; TSHZ3; NRXN1} \\

\parbox[t]{5.2cm}{POSITIVE REGULATION OF EXCITATORY POSTSYNAPTIC POTENTIAL (GO:2000463)} &
3.55E-04 & 0.0384121 & 110.8005 &
\parbox[t]{8.8cm}{NLGN1; GRIN2A; NRXN1; SHANK1} \\

\parbox[t]{5.2cm}{MODULATION OF CHEMICAL SYNAPTIC TRANSMISSION (GO:0050804)} &
3.96E-04 & 0.0394601 & 38.4433 &
\parbox[t]{8.8cm}{NLGN1; CLSTN2; NTF3; GRIK3; LRRC4; DRD2; SNCAIP; GRIA3} \\

\parbox[t]{5.2cm}{POSITIVE REGULATION OF MAPK CASCADE (GO:0043410)} &
4.16E-04 & 0.0394601 & 22.7732 &
\parbox[t]{8.8cm}{MAP4K2; NTRK2; NDRG4; NRXN1; NRG1; GDF6; ADRA2C; ROBO1; PYCARD; GHR; SLCO3A1; HAND2; NTF3; FGFR2; DRD5} \\

\parbox[t]{5.2cm}{POSITIVE REGULATION OF CELL JUNCTION ASSEMBLY (GO:1901890)} &
4.33E-04 & 0.0394601 & 52.4298 &
\parbox[t]{8.8cm}{SLC48A1; SFRP1; NLGN1; CLSTN2; NRXN1; SLITRK4} \\

\bottomrule
\end{tabular}
\label{tab:Table7}
\end{table*}

\begin{table*}[t]
\caption{Significant GO Cellular Component terms}
\centering
\scriptsize
\begin{tabular}{l c c c l}
\toprule
\textbf{Term} & \textbf{P-value} & \textbf{AdjP-value} & \textbf{CScore} & \textbf{Genes} \\
\midrule

\parbox[t]{5.2cm}{NEURON PROJECTION (GO:0043005)} &
1.15E-07 & 2.20E-05 & 56.61 &
\parbox[t]{8.8cm}{ROBO3; CHRNA3; NLGN1; KCNE3; CHRNA4; KCNA1; GRIK3; ADCY2; UNC80; PENK; NTF3; SCN9A; BOC; NEFM; NRCAM; DRD2; RGS6; SNCA; NTRK2; EPHA7; DST; KCNB2; SSTR2; PRSS12; CNTN1; RGS12; SHANK1} \\

\parbox[t]{5.2cm}{GLUTAMATERGIC SYNAPSE (GO:0098978)} &
3.95E-07 & 3.77E-05 & 120.50 &
\parbox[t]{8.8cm}{NLGN4Y; EFNB1; NLGN1; CTTNBP2; NRXN1; SLITRK4; LRRC4; NRG1; DRD2; PRSS12; SHISA6} \\

\parbox[t]{5.2cm}{EXCITATORY SYNAPSE (GO:0060076)} &
1.18E-06 & 6.51E-05 & 303.47 &
\parbox[t]{8.8cm}{NLGN4Y; NLGN1; SLC17A7; FGFR2; SHISA6; SHANK1} \\

\parbox[t]{5.2cm}{DENDRITE (GO:0030425)} &
1.36E-06 & 6.51E-05 & 59.52 &
\parbox[t]{8.8cm}{GABRA2; NTRK2; EPHA7; CHRNA3; NLGN1; KCNE3; CHRNA4; KCNA1; KCNB2; GRIK3; ADCY2; PRSS12; PENK; NTF3; RGS12; DRD2; SHANK1} \\

\parbox[t]{5.2cm}{AXON (GO:0030424)} &
5.64E-06 & 2.15E-04 & 56.62 &
\parbox[t]{8.8cm}{ROBO3; NTRK2; DST; GRIK3; PRSS12; UNC80; NTF3; SCN9A; BOC; CNTN1; NEFM; NRCAM; DRD2; SNCA} \\

\parbox[t]{5.2cm}{POTASSIUM CHANNEL COMPLEX (GO:0034705)} &
6.38E-05 & 0.00203 & 63.09 &
\parbox[t]{8.8cm}{KCNH4; KCNE3; KCNA1; KCNB2; KCNIP4; GRIK3; LRRC55; KCNJ2} \\

\parbox[t]{5.2cm}{IONOTROPIC GLUTAMATE RECEPTOR COMPLEX (GO:0008328)} &
1.45E-04 & 0.00396 & 101.25 &
\parbox[t]{8.8cm}{GRIN2A; GRIK3; VWC2; GRIA3; SHISA6} \\

\parbox[t]{5.2cm}{POSTSYNAPTIC DENSITY (GO:0014069)} &
1.96E-04 & 0.00447 & 37.23 &
\parbox[t]{8.8cm}{NTRK2; CHRNA3; GRIN2A; NLGN1; GRIK3; LRRC4; GRIA3; SHISA6; SHANK1; ADD2} \\

\parbox[t]{5.2cm}{POSTSYNAPTIC SPECIALIZATION MEMBRANE (GO:0099634)} &
2.11E-04 & 0.00447 & 66.34 &
\parbox[t]{8.8cm}{NLGN4Y; NLGN1; GRIN2A; GRIK3; LRRC4; GRIA3} \\

\parbox[t]{5.2cm}{VOLTAGE-GATED POTASSIUM CHANNEL COMPLEX (GO:0008076)} &
2.55E-04 & 0.00487 & 50.87 &
\parbox[t]{8.8cm}{KCNH4; KCNE3; KCNA1; KCNB2; KCNIP4; LRRC55; KCNJ2} \\

\parbox[t]{5.2cm}{ASYMMETRIC SYNAPSE (GO:0032279)} &
8.39E-04 & 0.01456 & 30.78 &
\parbox[t]{8.8cm}{NLGN4Y; NTRK2; CHRNA3; GRIN2A; NLGN1; SHISA6; SHANK1; ADD2} \\

\parbox[t]{5.2cm}{COLLAGEN-CONTAINING EXTRACELLULAR MATRIX (GO:0062023)} &
0.00211 & 0.03219 & 15.75 &
\parbox[t]{8.8cm}{SLC48A1; ANXA8L1; DST; COL12A1; COL23A1; NID1; LTBP1; ADAM19; SFRP1; LOX; EMILIN3; SOST; MATN4; FGFR2} \\

\parbox[t]{5.2cm}{ASYMMETRIC, GLUTAMATERGIC, EXCITATORY SYNAPSE (GO:0098985)} &
0.00219 & 0.03219 & 268.91 &
\parbox[t]{8.8cm}{NLGN4Y; SHISA6} \\

\parbox[t]{5.2cm}{SYNAPTIC MEMBRANE (GO:0097060)} &
0.00329 & 0.04483 & 40.92 &
\parbox[t]{8.8cm}{NLGN4Y; NLGN1; GRIN2A; SHISA6} \\

\bottomrule
\end{tabular}
\label{tab:Table8}
\end{table*}

\begin{table*}[t]
\caption{KEGG pathway enrichment analysis results}
\centering
\scriptsize
\begin{tabular}{l c c c l}
\toprule
\textbf{Term} & \textbf{P-value} & \textbf{AdjP-value} & \textbf{CScore} & \textbf{Genes} \\
\midrule

\parbox[t]{5.2cm}{NEUROACTIVE LIGAND SIGNALING} &
2.82E-07 & 5.98E-05 & 85.7676 &
\parbox[t]{8.8cm}{GABRA2; CHRNA3; CHRNA4; GRIK3; ADCY2; ADRA2C; GNAI1; POMC; GRIN2A; GNAL; PENK; SLC17A7; DRD2; GRIA3; DRD5} \\

\parbox[t]{5.2cm}{CAMP SIGNALING PATHWAY} &
1.45E-04 & 0.015335 & 34.0689 &
\parbox[t]{8.8cm}{VAV3; POMC; TIAM1; GRIN2A; NPY; ADCY2; DRD2; SSTR2; GLI3; GNAI1; GRIA3; DRD5} \\

\parbox[t]{5.2cm}{GLUTAMATERGIC SYNAPSE} &
2.95E-04 & 0.016042 & 41.8246 &
\parbox[t]{8.8cm}{GRIN2A; GRIK3; ADCY2; SLC17A7; SLC38A2; GRIA3; GNAI1; SHANK1} \\

\parbox[t]{5.2cm}{NICOTINE ADDICTION} &
3.19E-04 & 0.016042 & 76.3962 &
\parbox[t]{8.8cm}{GABRA2; GRIN2A; CHRNA4; SLC17A7; GRIA3} \\

\parbox[t]{5.2cm}{AXON GUIDANCE} &
3.78E-04 & 0.016042 & 31.4799 &
\parbox[t]{8.8cm}{EFNB1; ROBO3; EPHA7; LIMK2; BOC; LRRC4; NEO1; NTN1; GNAI1; ROBO1} \\

\parbox[t]{5.2cm}{HORMONE SIGNALING} &
4.88E-04 & 0.017245 & 27.2347 &
\parbox[t]{8.8cm}{GHR; GNA13; POMC; PENK; IRS4; ADCY2; DRD2; SSTR2; ADRA2C; GNAI1; DRD5} \\

\parbox[t]{5.2cm}{CELL ADHESION MOLECULE (CAM) INTERACTION} &
5.81E-04 & 0.017587 & 30.7575 &
\parbox[t]{8.8cm}{NLGN4Y; CDH4; NLGN1; NRXN1; SLITRK4; CNTN1; LRRC4; NRCAM; NEO1} \\

\parbox[t]{5.2cm}{NEUROACTIVE LIGAND-RECEPTOR INTERACTION} &
0.001382708 & 0.036642 & 17.6518 &
\parbox[t]{8.8cm}{GABRA2; CHRNA3; CHRNA4; GRIK3; ADRA2C; SSTR2; GHR; POMC; GRIN2A; NPY; PENK; DRD2; GRIA3; DRD5} \\

\parbox[t]{5.2cm}{GLYCOSAMINOGLYCAN BIOSYNTHESIS - CHONDROITIN SULFATE / DERMATAN SULFATE} &
0.003671633 & 0.078931 & 61.7191 &
\parbox[t]{8.8cm}{CHST7; DSE; CHST3} \\

\parbox[t]{5.2cm}{DOPAMINERGIC SYNAPSE} &
0.003723162 & 0.078931 & 21.0228 &
\parbox[t]{8.8cm}{GRIN2A; GNAL; PPP2R2B; DRD2; GRIA3; GNAI1; DRD5} \\

\parbox[t]{5.2cm}{IGSF CAM SIGNALING} &
0.005161797 & 0.099482 & 13.7053 &
\parbox[t]{8.8cm}{VAV3; NLGN4Y; ROBO3; TUBB6; NLGN1; KCNA1; SLITRK4; CNTN1; NRCAM; NTN1; ROBO1} \\

\bottomrule
\end{tabular}
\label{tab:Table9}
\end{table*}

\subsection{Performance of Sample 3 Hypomethylated+UPregulated and Hypermethylated+Downregulated}		

The performance of the QNN was evaluated on Sample-3, which combines both hypomethylated–upregulated and hypermethylated–downregulated genes, in order to assess the effect of integrating complementary epigenetic and transcriptomic signals. Experiments were conducted by progressively increasing the number of DNA and RNA features from 32 to 256, while monitoring classification accuracy, F1-score, precision, recall, and AUC.

Using the smallest subset of 32 combined features, the QNN achieved a training accuracy of 96.05\% and a test accuracy of 96.97\%, with a test AUC of 0.9930. The high F1-score (0.9721) indicates balanced precision and recall. These results demonstrate that even a compact feature set containing biologically relevant genes can effectively discriminate between LUAD and LUSC. The strong performance at this scale highlights the capacity of the QNN to capture discriminative patterns with limited dimensionality.

\noindent \textbf{Performance with 64 Features (16 DNA + 16 RNA)}
When the feature size was increased to 64, the model maintained similarly high performance, with test accuracy of 96.97\% and AUC of 0.9947. Compared to the 32-feature setting, recall increased, indicating improved sensitivity for LUAD detection, while precision remained high. This suggests that incorporating additional genes improves class separability without introducing significant noise. The stability of metrics across these two subsets indicates that the QNN generalizes well at moderate feature dimensionality.

\noindent \textbf{Performance with 128 Features (32 DNA + 32 RNA)}
Further, 128 combined features, the QNN achieved a test accuracy of 96.97\% and a test AUC of 0.9903, maintaining strong discriminative power. The recall increased to 0.9890, reflecting improved detection of LUAD samples, while precision slightly decreased, suggesting a small increase in false positives for LUSC. This trade-off indicates that expanding the feature set enhances sensitivity but may slightly affect specificity. Overall, the balanced F1-score (0.9730) confirms robust classification behavior.

\noindent \textbf{Performance with 256 Features (64 DNA + 64 RNA)}
The largest feature subset (256 features) yielded a test accuracy of 96.36\% and the highest AUC of 0.9960. Recall reached 0.9890, showing excellent identification of LUAD samples, while precision remained high (0.9474). Although the training accuracy slightly decreased compared to smaller subsets, the improvement in AUC indicates stronger ranking capability and better overall class separation. These results demonstrate that combining larger numbers of DNA and RNA features enhances the model’s ability to capture complex multi-omic interactions.

Across all subsets, increasing the number of features required a corresponding increase in the number of qubits (from 5 to 8), which allowed the quantum circuit to represent higher-dimensional data. The number of layers and epochs was kept constant to ensure fair comparison, indicating that performance gains are primarily driven by feature integration rather than deeper architectures. The results show that the QNN benefits from richer feature representations, particularly when both methylation and expression information are combined.

Across all configurations, our QNN consistently achieved high classification performance, with \textbf{test accuracies exceeding 93\% and AUCs approaching 0.996\%}, even as the feature dimensionality increased from 64 to 256. The results indicate that:  
\begin{itemize}
    \item \textbf{Feature Scaling:} Increasing DNA+RNA features improved representation; optimal batch size and learning rate were critical.
    \item \textbf{Stability \& Convergence:} Lower learning rates stabilized training; larger batches enhanced generalization in high-dimensional spaces.
    \item \textbf{Quantum Efficiency:} 6–8 qubits effectively captured complex multi-omic interactions, demonstrating QNN suitability for biomedical data.
    \item \textbf{Balanced Performance:} High F1 and AUC values indicate strong, unbiased performance across classes.
\end{itemize}

The workflow consisted of pairing equal numbers of DNA and RNA features (e.g., 32+32, 64+64, 128+128, and 256+256), concatenating them into a single feature vector, encoding them into qubits (6–9 qubits depending on feature size), and training the QNN under controlled learning rates, batch sizes, and epochs. Our experiments demonstrate that \textbf{QUBID framework can robustly classify multi-omic DNA and RNA data} across varying feature dimensions, achieving high accuracy and reliability. The combination of quantum layers with carefully tuned hyperparameters allows for effective learning from both low- and high-dimensional feature sets, making it a promising approach for integrative genomics analysis. By repeating this procedure across increasing feature sizes, the workflow enabled direct comparison between compact and expanded gene sets. 
\begin{figure*}[!ht]
	\centering
	\begin{tabular}{c}   
    \includegraphics[scale=0.3]{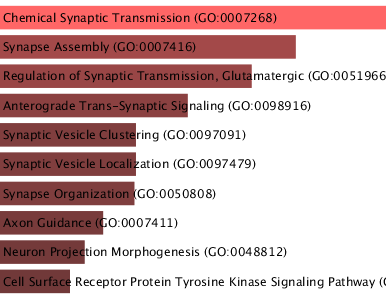}  
    \includegraphics[scale=0.3]{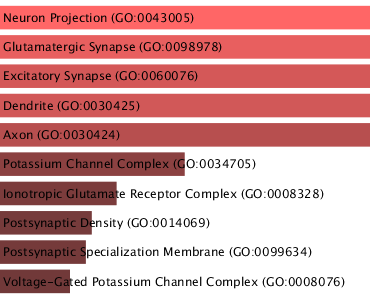}  
    \includegraphics[scale=0.3]{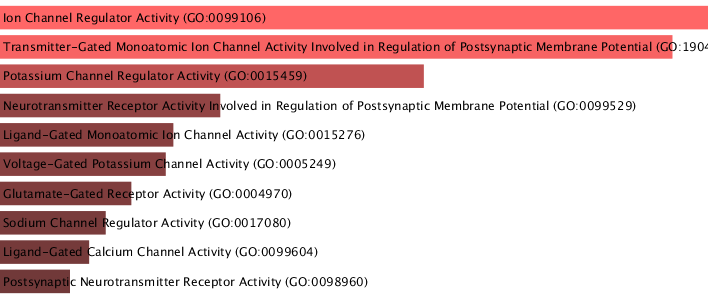}  
    \includegraphics[scale=0.3]{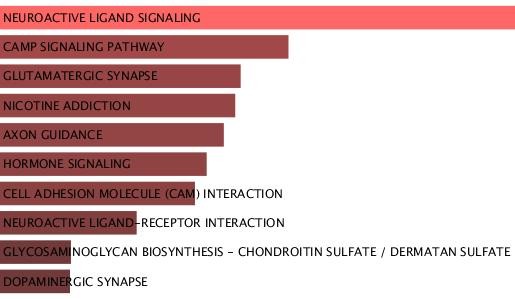}  \\
	\end{tabular}
	\caption{\textbf{Gene Ontology (GO) and KEGG Pathway Enrichment Analysis of classical phase.}  Bar plots representing the top enriched terms in (A) Biological Process, (B) Cellular Component, (C) Molecular Function categories, and (D) KEGG pathways for the analyzed multi-omics dataset. The x-axis shows the enrichment score or significance, and the y-axis lists the top terms or pathways. These visualizations highlight the most significantly associated biological functions, cellular locations, molecular activities, and pathways in the dataset.}
    \label{Figure13}
\end{figure*}

\begin{figure*}[!ht]
	\centering
	\begin{tabular}{c}      
    \includegraphics[scale=0.2]{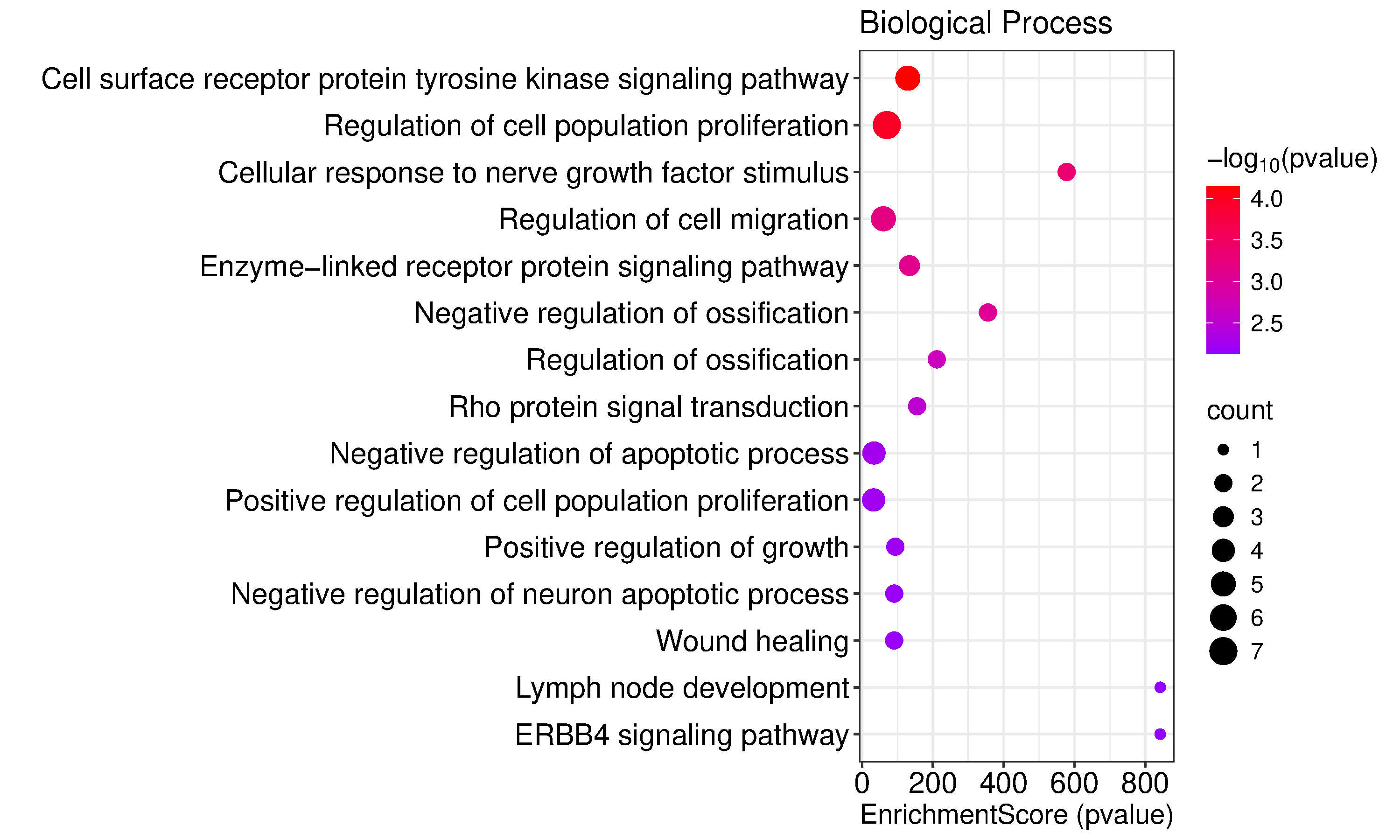}  
    \includegraphics[scale=0.2]{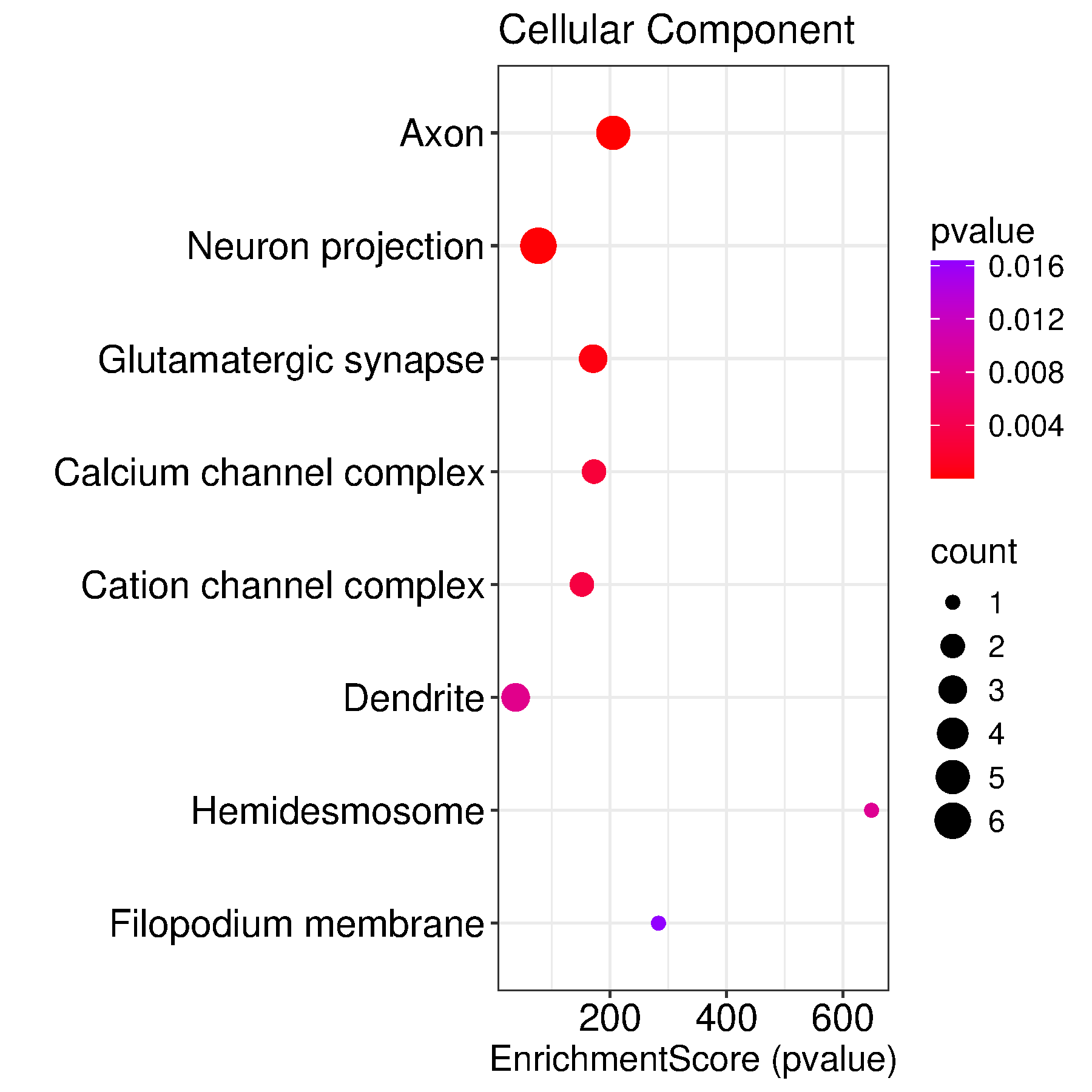}  
    \includegraphics[scale=0.2]{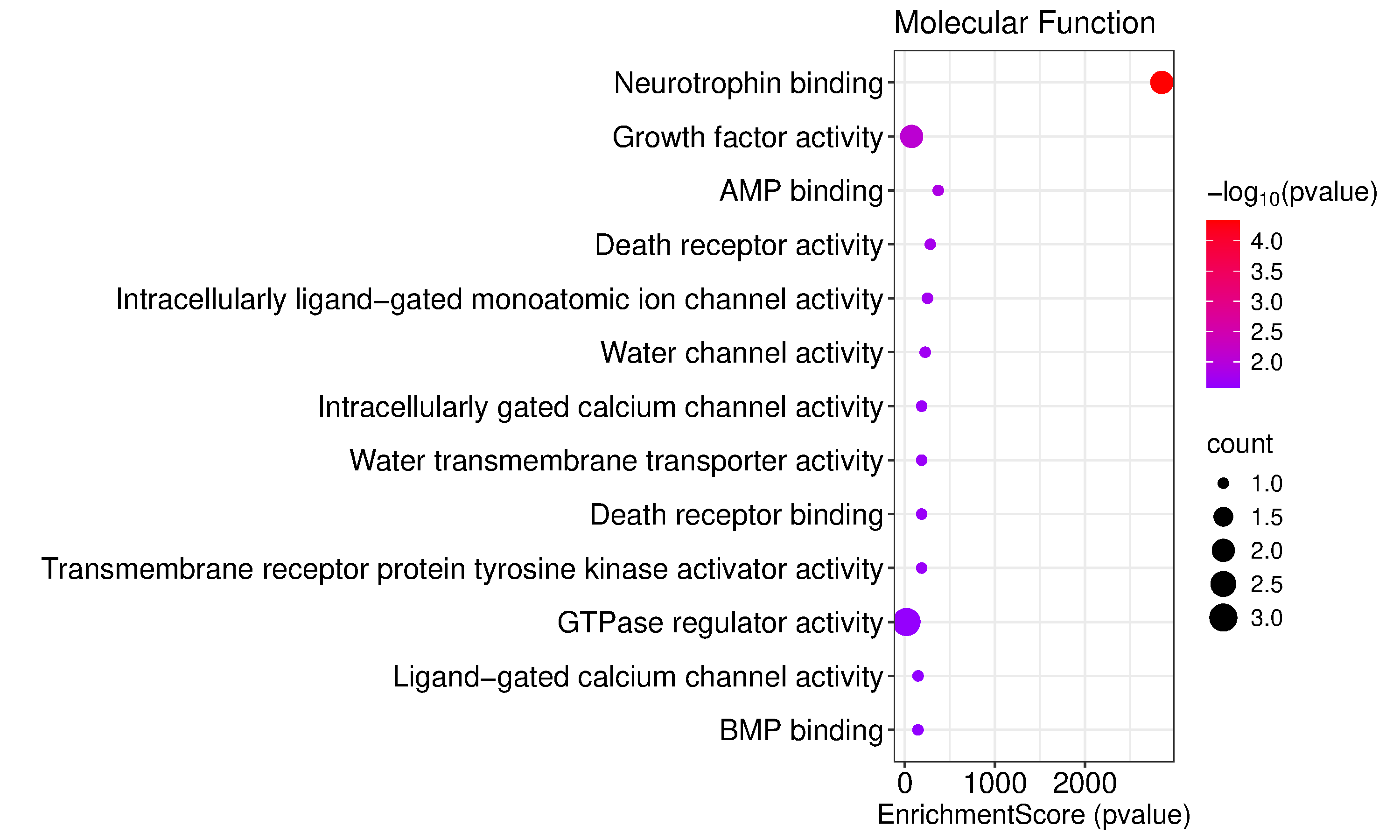}  
    \includegraphics[scale=0.2]{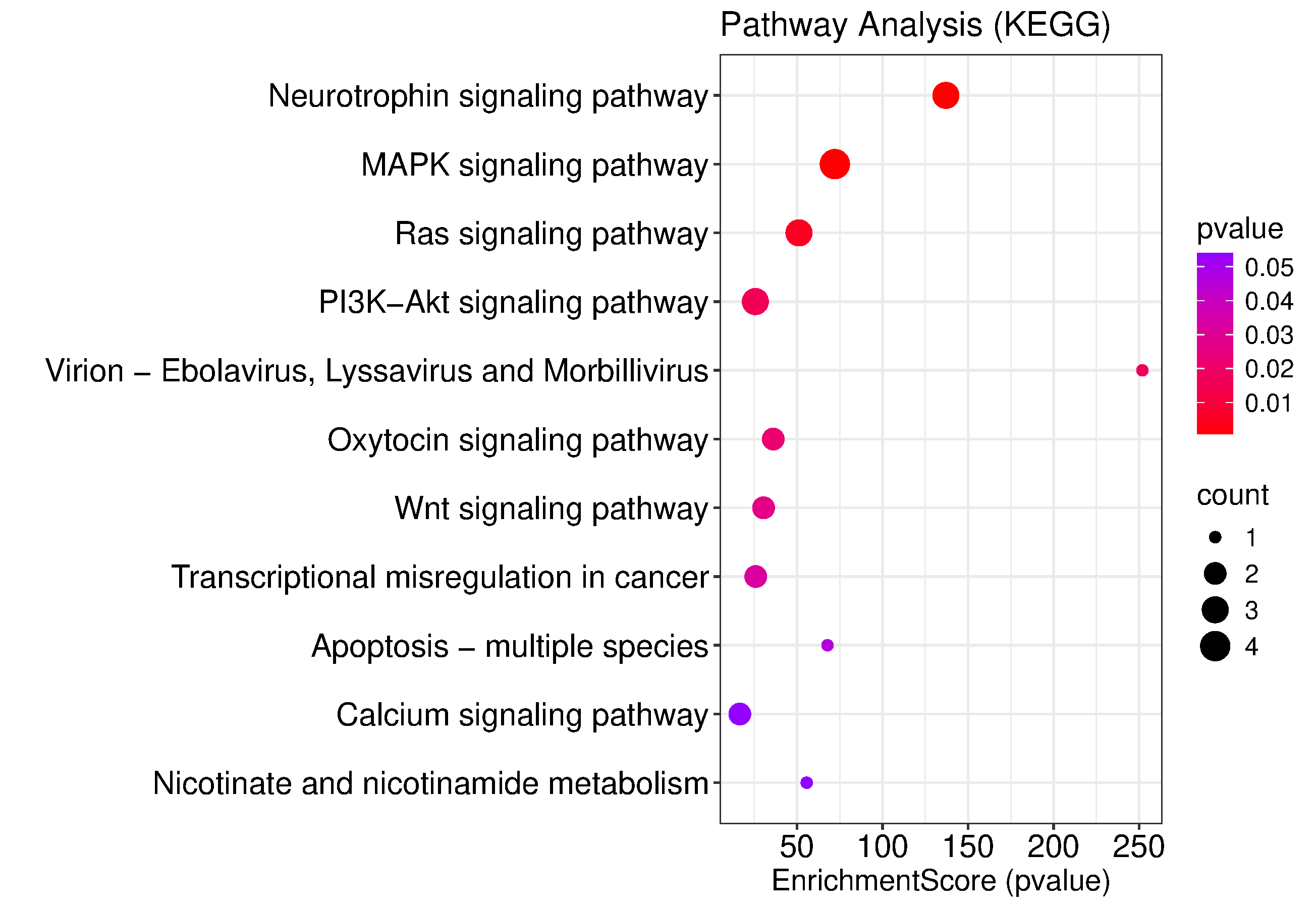} \\ 
	\end{tabular}
	\caption{\textbf{Gene Ontology and KEGG pathway enrichment analysis were conducted on the top-ranked features identified by the QNN model.}  Bar plots representing the top enriched terms in (A) Biological Process, (B) Cellular Component, (C) Molecular Function categories, and (D) KEGG pathways for the analyzed multi-omics dataset. The x-axis shows the enrichment score or significance, and the y-axis lists the top terms or pathways. These visualizations highlight the most significantly associated biological functions, cellular locations, molecular activities, and pathways in the dataset.}
    \label{Figure14}
\end{figure*}

\begin{figure}[!ht]
	\centering
	\begin{tabular}{c}      
    \includegraphics[scale=0.28]{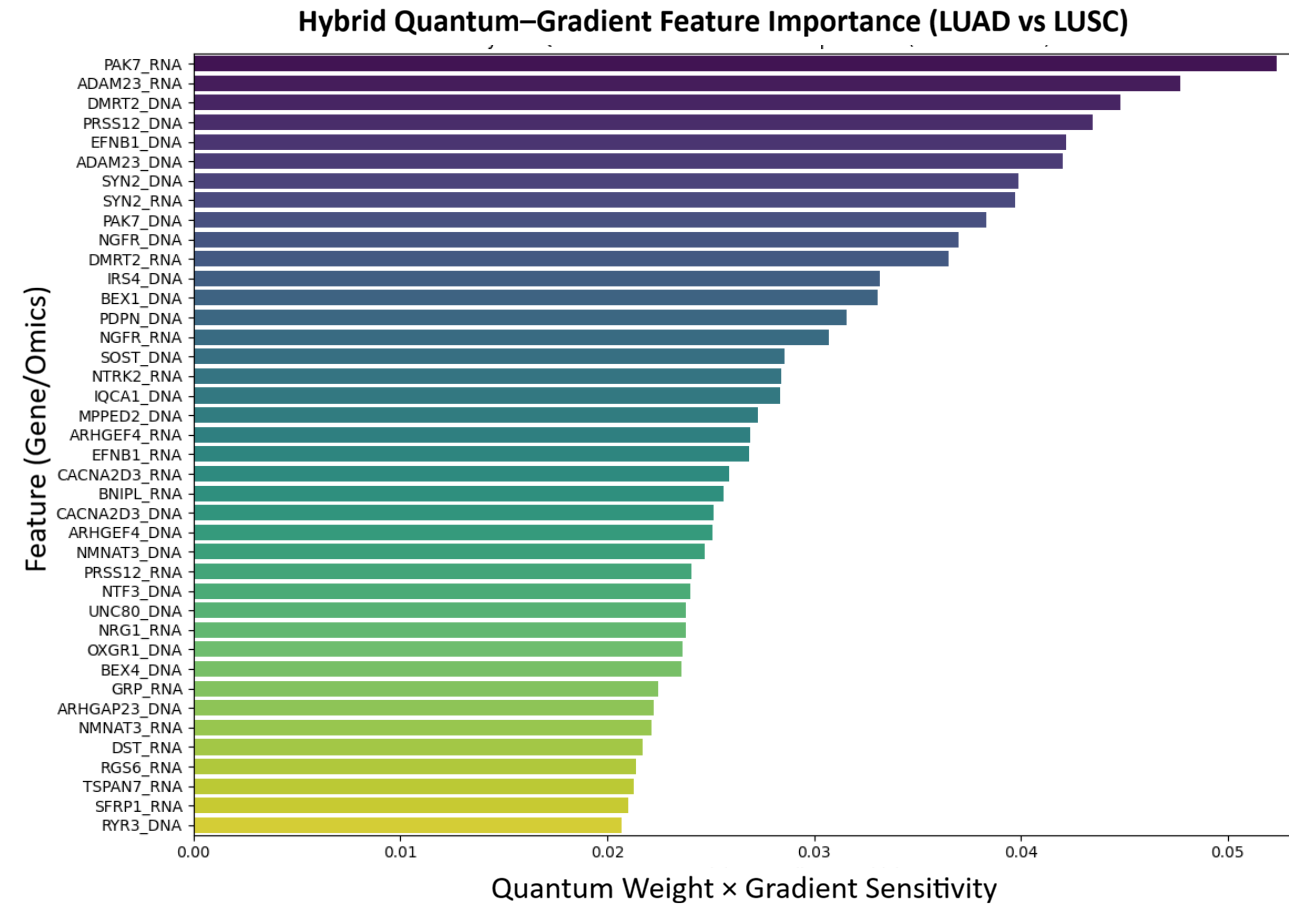}  
	\end{tabular}
	\caption{\textbf{Top-ranked features identified by the QNN model}.Bar plot showing the top 40 features ranked by hybrid quantum–gradient importance. Each bar represents a feature, with height indicating the combined contribution of quantum encoding weight and gradient-based sensitivity to LUAD vs. LUSC classification}
    \label{Figure15}
\end{figure}

\begin{table*}[ht]
\centering
\tiny
\caption{KEGG pathway enrichment analysis of top-ranked QNN genes.}
\begin{tabular}{|l|l|l|l|l|l|}
\hline
ID & Description & \textbf{P-value} & \textbf{AdjP-val}  & Genes & Enrich.Score \\
\hline
hsa04722 & Neurotrophin signaling pathway & 0.0008 & 0.0247 & NGFR/ NTRK2/ NTF3 & 3.0828 \\
hsa04010 & MAPK signaling pathway & 0.0010 & 0.024 & NGFR/ NTRK2/ CACNA2D3/ NTF3 & 2.9780 \\
hsa04014 & Ras signaling pathway & 0.005 & 0.0913 & NGFR/ NTRK2/ NTF3 & 2.2345 \\
hsa04151 & PI3K-Akt signaling pathway & 0.0182 & 0.1786 & NGFR/ NTRK2/ NTF3 & 1.7381 \\
hsa03265 & Virion - Ebolavirus, Lyssavirus and Morbillivirus & 0.0190 & 0.1786 & NGFR & 1.7212 \\
hsa04921 & Oxytocin signaling pathway & 0.0246 & 0.193 & CACNA2D3/ RYR3 & 1.6080 \\
\hline

GO:0007169 & Cell surface receptor protein tyrosine kinase signaling pathway & 7.22E-05 & 0.0205 & EFNB1, NTRK2, NTF3, IRS4, NRG1 & 128.97 \\
GO:0042127 & Regulation of cell population proliferation & 1.13E-04 & 0.0205 & BEX4, NTRK2, SFRP1, PDPN, NTF3, NRG1, BNIPL & 69.69 \\
GO:1990090 & Cellular response to nerve growth factor stimulus & 4.49E-04 & 0.0540 & NTRK2, NTF3 & 578.21 \\
GO:0030334 & Regulation of cell migration & 6.70E-04 & 0.0555 & BEX4, NGFR, SFRP1, PDPN, NTF3 & 60.11 \\ \hline

GO:0030424 & Axon & 1.64E-05 & 7.05E-04 & UNC80, NTRK2, DST, NTF3, PRSS12 & 205.37 \\
GO:0043005 & Neuron projection & 1.57E-04 & 0.00338 & UNC80, NTRK2, DST, NTF3, PRSS12, RGS6 & 76.74 \\
GO:0098978 & Glutamatergic synapse & 4.73E-04 & 0.00678 & EFNB1, NRG1, PRSS12 & 170.75 \\
GO:0034704 & Calcium channel complex & 2.65E-03 & 0.0274 & CACNA2D3, RYR3 & 172.30 \\ \hline

GO:0043121 & Neurotrophin binding & 4.55E-05 & 0.00327 & NGFR, NTRK2 & 2851.75 \\
GO:0008083 & Growth factor activity & 8.37E-03 & 0.1350 & NTF3, NRG1 & 75.46 \\
GO:0016208 & AMP binding & 1.34E-02 & 0.1350 & MPPED2 & 370.92 \\
GO:0005035 & Death receptor activity & 1.64E-02 & 0.1350 & NGFR & 282.99 \\
\hline
\multicolumn{6}{l}{\footnotesize \textbf{Notes:} p-value: raw p-value; AdjP-val: Adjusted P-value; Enrich.Score: Combined enrichment score.} \\
\hline
\end{tabular}
\label{tab:Table10}
\end{table*}

\noindent \textbf{Experimental Hardware and Software}
\noindent We implemented the experiments on the NCSU Hazel HPC cluster at North Carolina State University (NC State) with one GPU node, 32 cores per node, an A100 GPU, and 40 GB memory per node. All experiments were implemented in \texttt{Python 3.11.13} with key packages: \texttt{NumPy-1.23.5} and \texttt{Pandas-2.3.3} for data handling, \texttt{torch-2.9.1} for model construction and training,\texttt{TensorFlow-2.12.0}, \texttt{TensorCircuit-0.11.0} and PyTorch version: 2.9.1, GPU 1: NVIDIA A100, for quantum computation and amplitude encoding, \texttt{scikit-learn} for train-test splitting and evaluation metrics \texttt{Matplotlib} and \texttt{Seaborn} for visualization.

The DNA methylation profiles of LUAD and LUSC patients revealed diverse epigenetic alterations across key genes (Table~\ref{tab:Table5}). Several promoter-associated CpG sites showed significant differential methylation, with examples including NTRK2 (hypermethylated in F3), DMRT2 (hypomethylated in F3), and SYN2 (hypermethylated in F3). While some CpGs exhibited consistent changes across multiple datasets, others displayed subtype-specific or conflicting patterns, reflecting the complex regulatory landscape. Overall, these data highlight both hyper- and hypo-methylation events in promoter regions, particularly within CpG islands and shores, providing candidate loci for further functional validation and potential biomarkers in LUAD and LUSC.

The RNA expression analysis across LUAD and LUSC patients revealed distinct transcriptional patterns with both subtype-specific and shared alterations (Table~\ref{tab:rna-omic}). Several genes, including NTRK2, ADAM23, and DMRT2, displayed LUAD-biased opposite expression patterns, reflecting downregulation in one dataset and upregulation in another. Other genes, such as SYN2 and GPC3, were LUAD-biased shared, showing consistent dysregulation across multiple datasets. Notably, many of these genes are annotated as tumor suppressors (TSS), highlighting potential functional relevance in LUAD pathogenesis. Overall, these results identify key differentially expressed genes that may serve as candidate biomarkers and provide a transcriptomic layer for multi-omics integration.

In this Fig~.\ref{Figure13} Gene Ontology (GO) and KEGG pathway enrichment analysis was performed for the top 100 genes using classical phase. Bar plots show the most significantly enriched (A) Biological Processes, (B) Cellular Components, (C) Molecular Functions, and (D) KEGG pathways.
The GO enrichment analysis revealed that the identified genes were significantly associated with synaptic signaling and ion channel–related functions (Tables \ref{tab:Table6}–\ref{tab:Table8}). In the Molecular Function category (Table \ref{tab:Table6}), the top enriched terms included ion channel regulator activity, transmitter-gated monoatomic ion channel activity, potassium channel regulator activity, and voltage-gated potassium channel activity, indicating a strong involvement in modulation of membrane potential and neurotransmitter-mediated signaling. In the Biological Process category (Table \ref{tab:Table7}), the most significant enrichment was observed for chemical synaptic transmission, synapse assembly, regulation of glutamatergic synaptic transmission, axon guidance, and neuron projection morphogenesis. These results highlight a prominent role of the selected genes in synaptic organization, vesicle dynamics, excitatory postsynaptic potential, and neuronal development. Consistently, Cellular Component analysis (Table \ref{tab:GO_CO_terms}) demonstrated significant enrichment in neuron projection, glutamatergic synapse, excitatory synapse, dendrite, axon, postsynaptic density, and potassium channel complexes. Collectively, these findings suggest that the gene set is predominantly enriched in synapse-related structures and ion channel complexes, supporting their functional relevance in neuronal communication and excitatory signaling pathways.

Furthermore, Table~\ref{tab:Table9} presents the (Kyoto Encyclopedia of Genes and Genomes, KEGG) pathway enrichment analysis of the identified genes. The results demonstrate most significant enrichment in neurobiologically relevant pathways, particularly Neuroactive ligand–receptor interaction pathway (P = 2.82E–07, adj. P = 5.98E–05), cAMP signaling pathway, Glutamatergic synapse, Dopaminergic synapse, and Nicotine addiction, all showing strong statistical significance (adjusted P < 0.05). The enrichment of neuroactive ligand–receptor interaction and synaptic signaling pathways has been previously reported to contribute to lung cancer susceptibility and progression, particularly in the context of chromosome 15q25.1-associated risk loci \cite{28}. These findings collectively suggest that dysregulation of neuroactive signaling and receptor-mediated communication may play a critical role in disease pathogenesis. Additionally, pathways related to Axon guidance, Hormone signaling, and Cell Adhesion Molecule (CAM) interaction were enriched, suggesting involvement in synaptic transmission, neuronal connectivity, and receptor-mediated signaling mechanisms. 

\paragraph{Hybrid Quantum–Gradient Feature Importance.}
In this study, we quantified \emph{feature importance} using a hybrid quantum--gradient approach that integrates contributions from both the QNN weights and classical gradient sensitivity. Each qubit in the QNN has trainable rotation parameters $w_{l,j}$, where $l$ indexes layers and $j$ indexes qubits. The \emph{quantum contribution} of each qubit is computed as the sum of absolute weights across layers:$QI_j = \sum_{l} |w_{l,j}|$
Simultaneously, \emph{gradient-based sensitivity} measures how the model loss $\mathcal{L}$ changes with respect to each input feature $x_k$, computed as the mean absolute gradient:
$
GI_k = \frac{1}{N}\sum_{i=1}^{N} \left| \frac{\partial \mathcal{L}}{\partial x_{ik}} \right|.
$
Qubits are mapped to features, and a \emph{hybrid importance score} is obtained multiplicatively:
$
HI_k = \tilde{QI}_k \times GI_k.
$
This score captures both the \emph{learned significance of the quantum encoding} and the \emph{classical sensitivity of the model to each input feature}. The resulting bar plot ranks features by hybrid importance, highlighting the top contributors for distinguishing LUAD versus LUSC.

Using the hybrid quantum-gradient feature importance framework, we identified the top 40 multiomic features contributing to the classification of LUAD and LUSC in our QNN model represents Sample~1, where 256 multiomic features (128 DNA methylation and 128 RNA expression) were used under the tumor suppressor scenario (hypermethylation and downregulation) as shown in Fig~.\ref{Figure15}. The QNN model was implemented with 8 qubits, a batch size of 64, and a learning rate of 0.03. Each feature importance score was computed as the product of the quantum weight contribution and the gradient-based sensitivity, defined as 
$HI_k = \tilde{QI}_k \times GI_k$, where $\tilde{QI}_k$ represents the mapped qubit importance for feature $k$ and $GI_k$ denotes the mean absolute gradient with respect to the loss function.
The resulting ranked features include multiple DNA methylation and RNA expression markers such as \textit{PAK7}, \textit{ADAM23}, \textit{DMRT2}, \textit{PRSS12}, \textit{EFNB1}, \textit{SYN2}, \textit{NGFR}, \textit{IRS4}, \textit{BEX1}, and \textit{PDPN}. 
Notably, several genes appear across multiple omics layers (e.g., DNA methylation and RNA expression), indicating consistent importance across molecular modalities. 
For instance, \textit{PAK7} and \textit{ADAM23} showed the highest hybrid importance scores, suggesting strong contributions from both quantum encoding weights and gradient sensitivity. 
These top-ranked biomarkers highlight biologically relevant multiomic signals that the QNN model leverages to discriminate between LUAD and LUSC, demonstrating the capability of the hybrid quantum–classical framework to identify robust cancer-associated molecular features.
To assess the biological relevance of the top-ranked genes identified from Sample~1, KEGG pathway enrichment analysis was performed as shown Table~\ref{tab:Table10} and Fig~.\ref{Figure14}. The analysis revealed significant enrichment in several cancer-related signaling pathways, including the \textit{Neurotrophin signaling pathway} (adjusted $p=0.0247$), \textit{MAPK signaling pathway} (adjusted $p=0.024$), \textit{Ras signaling pathway}, and \textit{PI3K--Akt signaling pathway}. Key genes such as \textit{NGFR}, \textit{NTRK2}, and \textit{NTF3} were consistently involved across multiple pathways, indicating their central role in neurotrophin-mediated signaling and downstream oncogenic cascades. Additional enrichment was observed in pathways associated with calcium signaling and cellular communication, involving genes such as \textit{CACNA2D3} and \textit{RYR3}. These pathways collectively regulate essential processes including cell proliferation, survival, differentiation, and migration, suggesting that the top QNN-identified genes capture biologically meaningful mechanisms relevant to LUAD and LUSC subtype differentiation. Neurotrophin signaling plays a critical role in lung adenocarcinoma. Neurotrophin-3 (NTF3) promotes cancer stemness, migration, and tumorigenicity through activation of its receptor TrkC (NTRK3), with high NTF3 expression correlating with poor survival in lung cancer patients \cite{29}. Activation of the TrkB receptor (NTRK2) enhances metastasis via Akt signaling, and its knockdown significantly reduces metastatic potential \cite{30}. In addition, the low-affinity receptor p75/NGFR contributes to tumor progression by modulating survival and apoptosis signals. Overall, dysregulated neurotrophin signaling supports tumor growth, invasion, and angiogenesis, and pharmacological targeting of these receptors is being explored as a therapeutic strategy in NSCLC \cite{31}.
Moreover, NTRK2 expression was found to be significantly decreased in LUAD tissues compared to normal lung, and both its expression level and DNA methylation status were significantly correlated with clinical features and prognosis of LUAD patients \cite{32}. Functional analysis further showed that NTKR2 expression increases after AKT pathway inhibition, suggesting mechanistic ties to PI3K–Akt signaling. These results support the relevance of NTRK2 both as a potential diagnostic/prognostic biomarker and as a gene with functional impact on cancer signaling pathways consistent with the QNN and KEGG enrichment findings in this study.
CACNA2D3, a voltage-dependent calcium channel subunit on chromosome 3p21.1, acts as a tumor suppressor in multiple cancers, including osteosarcoma, neuroblastoma, lung, renal, esophageal, gastric, and breast cancers. Its expression is frequently downregulated in highly metastatic cell lines (e.g., MG63-A1) and clinical tumors, often due to promoter hypermethylation, and this downregulation correlates with poor prognosis \cite{33}.
Additionaly, ADAM23 is frequently downregulated in NSCLC, with its reduced expression strongly associated with promoter hypermethylation. This epigenetic silencing suggests that ADAM23 functions as a tumor suppressor, potentially influencing tumor progression and metastasis.

\begin{table}[ht]
\tiny
\centering
\caption{\textbf{Classical Machine Learning Model Comparison across Different Feature Sizes for Sample1, Sample2 and Sample3}}
\begin{tabular}{cccccccccc}
\hline
\textbf{Genes} & \textbf{Model} & \textbf{Input} & \textbf{Params} & \textbf{Train Acc} & \textbf{Test Acc} & \textbf{Test F1} & \textbf{Test AUC} & \textbf{Time (s)} \\
\hline

8   & NN  & 16  & 289   & 0.9029 & 0.9273 & 0.9375 & 0.9782 & 0.17 \\
8   & SVM & 16  & 9264  & 0.8998 & 0.9273 & 0.9375 & 0.9832 & 0.22 \\
8   & RF  & 16  & 48    & 0.9014 & 0.9394 & 0.9474 & 0.9904 & 0.05 \\

16  & NN  & 32  & 1089  & 0.9348 & 0.9394 & 0.9451 & 0.9844 & 0.24 \\
16  & SVM & 32  & 19040 & 0.9014 & 0.9212 & 0.9326 & 0.9895 & 0.24 \\
16  & RF  & 32  & 96    & 0.9211 & 0.9212 & 0.9319 & 0.9868 & 0.10 \\

32  & NN  & 64  & 4225  & 0.9439 & 0.9394 & 0.9457 & 0.9902 & 0.39 \\
32  & SVM & 64  & 38144 & 0.8816 & 0.9091 & 0.9231 & 0.9903 & 0.30 \\
32  & RF  & 64  & 192   & 0.9363 & 0.9515 & 0.9579 & 0.9939 & 0.19 \\

64  & NN  & 128 & 16641 & 0.9605 & 0.9636 & 0.9677 & 0.9917 & 0.91 \\
64  & SVM & 128 & 76160 & 0.8498 & 0.8606 & 0.8878 & 0.9887 & 0.34 \\
64  & RF  & 128 & 384   & 0.9545 & 0.9394 & 0.9474 & 0.9963 & 0.38 \\

128 & NN  & 256 & 66049 & 0.9879 & 0.9636 & 0.9677 & 0.9945 & 1.61 \\
128 & SVM & 256 & 152320& 0.7648 & 0.7636 & 0.8235 & 0.9886 & 0.49 \\
128 & RF  & 256 & 768   & 0.9484 & 0.9273 & 0.9375 & 0.9967 & 0.76 \\

\hline

8   & NN  & 16  & 289    & 0.8498 & 0.8788 & 0.8947 & 0.9534 & 0.10 \\
8   & SVM & 16  & 6976   & 0.9135 & 0.9212 & 0.9297 & 0.9786 & 0.12 \\
8   & RF  & 16  & 48     & 0.9241 & 0.9333 & 0.9405 & 0.9763 & 0.04 \\

16  & NN  & 32  & 1089   & 0.9044 & 0.9212 & 0.9305 & 0.9690 & 0.13 \\
16  & SVM & 32  & 15104  & 0.9150 & 0.9333 & 0.9412 & 0.9797 & 0.12 \\
16  & RF  & 32  & 96     & 0.9332 & 0.9455 & 0.9514 & 0.9911 & 0.08 \\

32  & NN  & 64  & 4225   & 0.9287 & 0.9455 & 0.9514 & 0.9788 & 0.21 \\
32  & SVM & 64  & 29824  & 0.9272 & 0.9394 & 0.9468 & 0.9847 & 0.13 \\
32  & RF  & 64  & 192    & 0.9256 & 0.9212 & 0.9305 & 0.9886 & 0.14 \\

64  & NN  & 128 & 16641  & 0.9605 & 0.9576 & 0.9622 & 0.9840 & 0.53 \\
64  & SVM & 128 & 64256  & 0.9378 & 0.9455 & 0.9519 & 0.9893 & 0.18 \\
64  & RF  & 128 & 384    & 0.9363 & 0.9576 & 0.9622 & 0.9938 & 0.34 \\

128 & NN  & 256 & 66049  & 0.9757 & 0.9758 & 0.9780 & 0.9903 & 1.58 \\
128 & SVM & 256 & 147456 & 0.9014 & 0.9333 & 0.9430 & 0.9930 & 0.30 \\
128 & RF  & 256 & 768    & 0.9332 & 0.9455 & 0.9519 & 0.9952 & 0.63 \\

\hline
8   & NN & 32  & 1089   & 0.6753 & 0.6545 & 0.7615 & 0.9032 & 0.16 \\
8   & SVM           & 32  & 19008  & 0.5493 & 0.5515 & 0.7109 & 0.9866 & 0.10 \\
8   & RF  & 32  & 96     & 0.9408 & 0.9394 & 0.9462 & 0.9941 & 0.07 \\
16  & NN & 64  & 4225   & 0.9090 & 0.9091 & 0.9198 & 0.9617 & 0.23 \\
16  & SVM           & 64  & 38080  & 0.5493 & 0.5515 & 0.7109 & 0.9823 & 0.09 \\
16  & RF & 64  & 192    & 0.9545 & 0.9697 & 0.9733 & 0.9984 & 0.13 \\
32  & NN & 128 & 16641  & 0.9378 & 0.9515 & 0.9570 & 0.9912 & 0.57 \\
32  & SVM           & 128 & 76288  & 0.6267 & 0.6121 & 0.7398 & 0.9909 & 0.12 \\
32  & RF  & 128 & 384    & 0.9697 & 0.9697 & 0.9733 & 0.9997 & 0.32 \\
64  & NN & 256 & 66049  & 0.9545 & 0.9758 & 0.9783 & 0.9911 & 1.63 \\
64  & SVM           & 256 & 152320 & 0.8968 & 0.8727 & 0.8955 & 0.9911 & 0.19 \\
64  & RF  & 256 & 768    & 0.9575 & 0.9576 & 0.9630 & 0.9990 & 0.51 \\
128 & NN & 512 & 263169 & 0.9666 & 0.9758 & 0.9783 & 0.9947 & 6.08 \\
128 & SVM           & 512 & 299008 & 0.9484 & 0.9455 & 0.9524 & 0.9902 & 0.43 \\
128 & RF  & 512 & 1536   & 0.9590 & 0.9273 & 0.9375 & 0.9960 & 0.97 \\
\hline

\end{tabular}
\label{tab:Table11}
\end{table}

Across all samples, performance improves with increasing feature size, with optimal results at 64–128 genes, where Test Accuracy reaches ~0.96–0.98, F1-score ~0.96–0.98, and AUC consistently ~0.99, indicating strong classification capability as shown in (Table \ref{tab:Table11}). Random Forest (RF) shows the most efficient performance, achieving high accuracy (up to 0.9697) and very high AUC (~0.99) with minimal parameters (48–1536) and lowest computation time (~0.04–0.97 s). Neural Networks (NN) achieve the highest predictive performance (Test Acc/F1 up to 0.9758/0.9783) but require significantly larger model complexity (up to 263,169 parameters) and higher computation time (~6.08 s). In contrast, SVM exhibits instability at higher feature dimensions, with accuracy dropping to ~0.76–0.86 in Sample 1, despite maintaining high AUC (~0.98–0.99), suggesting sensitivity to feature scaling and dimensionality. Overall, AUC remains the most stable metric (~0.99 across models), while accuracy and F1 better reflect model discrimination and robustness, highlighting RF as the best trade-off and NN as the best high-capacity model.

For Sample 3 (32–256 features), the QNN outperformed classical models in test accuracy and AUC while using far fewer parameters. QNN-32 (32 features, 5 qubits, 6 layers) achieved a test AUC of 0.9930, slightly higher than Random Forest (0.990) and NN (0.9788), and QNN-256 (256 features, 8 qubits, 6 layers) reached 0.9960, surpassing RF (0.9900) and NN (0.9900). Classical models required substantially more parameters for the same features (e.g., NN: 66,049; RF: 1,536), whereas QNN parameters are calculated as (number of qubits × 2) × number of layers, resulting in only 60 for QNN-32 and 96 for QNN-256. These results highlight that QNN efficiently scales to high-dimensional multiomic data, providing superior discriminative power with minimal trainable parameters while maintaining competitive F1, precision, and recall metrics.

\subsection{\textbf{Model Compilation Parameters}}

In this paper, we have created the function \texttt{MetricsCallback} to evaluate the performance of a machine learning model during training. The evaluation metrics include accuracy, precision, recall, and F1-score. These metrics are widely used in classification tasks to assess the model's ability to classify instances belonging to different classes correctly. All methods use the Adaptive Moment Estimation (Adam) optimizer; the learning rate is set to 0.03, and the batch size is 32 and 64. The details of the classification metrics are given:

\textbf{Accuracy}: It is used to evaluate how often the predictions match the actual labels and defined as the ratio of correctly predicted instances to the total instances:

\begin{equation}
	\text{Accuracy} = \frac{1}{N} \sum_{i=1}^{N} \mathbf{1}(\hat{y}_i = y_i) 
\end{equation}

\noindent where \(\mathbf{1}(\cdot)\) is the indicator function that returns 1 if the condition inside is true and 0 otherwise.

\textbf{Precision Score}: It is defined as the ratio of true positive predictions to the total number of positive predictions made by the model as
\begin{equation}
	\text{Precision} = \frac{{\sum_{i=1}^{N} TP_i}}{{\sum_{i=1}^{N} (TP_i + FP_i)}} 
\end{equation}
where \( TP_i \) represents the number of true positive predictions for class \( i \) and \( FP_i \) represents the number of false positive predictions for class \( i \).

\textbf{Recall Score}: It is also known as sensitivity, and measures the ability of the model to correctly identify positive instances out of all actual positive instances. Mathematically, it is represented as:
\begin{equation} \text{Recall} = \frac{{\sum_{i=1}^{N} TP_i}}{{\sum_{i=1}^{N} (TP_i + FN_i)}}
\end{equation}
where \( FN_i \) represents the number of false negative predictions for class \( i \).

\textbf{F1-score}: It is the harmonic mean of precision and recall, providing a balance between the two metrics. It is calculated using the formula:
\begin{equation} F_1 = 2 \times \frac{{\text{Precision} \times \text{Recall}}}{{\text{Precision} + \text{Recall}}}  \end{equation}

\section{Conclusion}
\noindent 
Recent advances in quantum computing have highlighted its potential to transform computational biology, particularly in the areas of cancer diagnostics, biomarker discovery, and precision medicine. Quantum Machine Learning, which exploits uniquely quantum mechanical resources such as superposition, entanglement, and interference, offers a promising framework for analyzing the high-dimensional and complex datasets that characterize modern biomedical research. These capabilities may enable a more efficient exploration of large feature spaces and the identification of subtle molecular patterns that are difficult to capture using classical machine learning approaches. In this work, we investigated the application of hybrid classical–quantum neural networks  for the classification of lung cancer subtypes using integrated multi-omics data. Our analysis demonstrates that the combination of Hypermethylated + Downregulated and Hypomethylated + Upregulated genes (Sample3) provides a more informative and reliable feature set for QNN-based classification of lung cancer subtypes. While the individual gene sets (Sample1 and Sample2) achieved strong predictive performance, their integration in Sample3 consistently produced superior results across all evaluation metrics, including accuracy, F1-score, precision, recall, and area under the receiver operating characteristic curve (AUC). These results indicate that incorporating complementary epigenetic and transcriptional signals enables the model to capture a richer and more discriminative molecular representation of tumor biology.

The improved performance observed with the integrated dataset suggests that combining DNA methylation and gene expression features provides a more comprehensive characterization of the molecular differences between LUAD and LUSC. By leveraging multi-omics data, the QNN models were able to learn patterns that reflect both regulatory and transcriptional alterations associated with tumor subtype differentiation. This integrative approach enhances model robustness and generalizability, highlighting the value of multi-modal biological data in quantum machine learning frameworks. Importantly, our findings also demonstrate the feasibility of hybrid quantum–classical learning models for real-world biomedical applications using currently available noisy intermediate-scale quantum (NISQ) devices and simulators. Although quantum hardware is still in its early stages, the results suggest that QML approaches can already provide meaningful insights in complex biological classification tasks when combined with appropriate feature engineering and classical preprocessing pipelines.

To the best of our knowledge, this study represents the first investigation of LUAD–LUSC subtype differentiation using The Cancer Genome Atlas (TCGA) dataset within a hybrid classical–quantum lung cancer classification framework. Beyond demonstrating proof-of-concept performance, this work provides a foundation for future studies exploring the integration of larger multi-omics datasets, more expressive quantum circuit architectures, and advanced quantum feature maps. As quantum hardware continues to mature, such approaches may play an increasingly important role in next-generation biomedical analytics, enabling improved biomarker discovery, disease stratification, and ultimately more personalized cancer diagnostics and treatment strategies.


\subsection*{Data Availability}

\noindent All data generated and/or analyzed during this study are included in this article. The datasets were obtained from The Cancer Genome Atlas (TCGA) Research Network (\url{https://www.cancer.gov/tcga}) through the UCSC Xena platform (\url{https://xenabrowser.net/datapages/}). \noindent Multi-omics data were collected from the TCGA-LUAD cohort, including: (1) DNA methylation, (2) RNA sequencing (RNA-seq), and (3) clinical and phenotype data.
\noindent (1) \textbf{DNA Methylation:} Illumina HumanMethylation450 platform (TCGA Hub), comprising 485,578 CpG probes across 492 samples. DNA methylation levels are represented as beta values ranging from 0 to 1, where higher values indicate hypermethylation and lower values indicate hypomethylation.
\noindent (2) \textbf{RNA-seq:} Gene expression data generated using the Illumina HiSeq 2000 platform (University of North Carolina TCGA). The dataset includes 20,531 genes across 576 samples, quantified as log$_2$(x+1) transformed RSEM normalized counts.
\noindent (3) \textbf{Clinical Data:} Phenotype (TCGA-LUAD phenotype) and survival information were downloaded from the TCGA cohort via UCSC Xena.

\section*{Conflict of Interest}
\noindent The authors declare that they have no conflicts of interest.

\section*{Acknowledgements}
\noindent The authors would like to acknowledge the partial funding from the U.S. Department of Energy (DOE) Office of Basic Energy Sciences, Award No. DE-SC0026309.

\section*{Author Contributions}
S.K. and H.G. conceived and supervised the project.  M.K.S and A.S.B. designed the framework, and performed the simulations.  M.K.S. wrote the initial manuscript and contributed to the development and visualization. The Data feature engineering  and feature selection phases was conducted by M.K.S. Resources and Funding acquisition by S.K. All authors contributed to analyzing the results and  finalizing the manuscript.

\section*{Data and software Availability}
The authors confirm that the data supporting the findings of this study are available from the corresponding author, upon reasonable request.

\bibliography{arxiv}

@article{1,
	author = {P. Board},
	title = {Non-small cell lung cancer treatment (PDQ)},
	journal = {PDQ Cancer Information Summaries [Internet]},
	year = {2024}
}

@article{35,
  title={Implementing entangled states on a quantum computer},
  author={Bhatia, Amandeep Singh and Saggi, Mandeep Kaur},
  journal={arXiv preprint arXiv:1811.09833},
  year={2018}
}

@article{36,
	author = {J. Biamonte and P. Wittek and N. Pancotti and P. Rebentrost and N. Wiebe and S. Lloyd},
	title = {Quantum machine learning},
	journal = {Nature},
	volume = {549},
	number = {7671},
	pages = {195--202},
	year = {2017}
}

@article{2,
	author = {D. Ristè and M. P. Da Silva and C. A. Ryan and A. W. Cross and A. D. C{\'o}rcoles and J. A. Smolin and J. M. Gambetta and J. M. Chow and B. R. Johnson},
	title = {Demonstration of quantum advantage in machine learning},
	journal = {npj Quantum Information},
	volume = {3},
	number = {1},
	pages = {16},
	year = {2017}
}

@article{3,
	author = {A. S. Bhatia and M. K. Saggi and S. Kais},
	title = {Quantum machine learning predicting ADME-Tox properties in drug discovery},
	journal = {Journal of Chemical Information and Modeling},
	volume = {63},
	number = {21},
	pages = {6476--6486},
	year = {2023}
}

@article{4,
	author = {A. S. Bhatia and S. Kais and M. A. Alam},
	title = {Federated quanvolutional neural network: a new paradigm for collaborative quantum learning},
	journal = {Quantum Science and Technology},
	volume = {8},
	number = {4},
	pages = {045032},
	year = {2023}
}

@inproceedings{5,
	author = {A. S. Bhatia and S. Kais and M. Alam},
	title = {Handling privacy-sensitive clinical data with federated quantum machine learning},
	booktitle = {APS March Meeting Abstracts},
	volume = {2023},
	pages = {T70--007},
	year = {2023}
}

@article{6,
	author = {K. Prousalis and N. Konofaos},
	title = {A quantum pattern recognition method for improving pairwise sequence alignment},
	journal = {Scientific reports},
	volume = {9},
	number = {1},
	pages = {7226},
	year = {2019}
}

@article{7,
	author = {A. S. Boev and A. S. Rakitko and S. R. Usmanov and A. N. Kobzeva and I. V. Popov and V. V. Ilinsky and E. O. Kiktenko and A. K. Fedorov},
	title = {Genome assembly using quantum and quantum-inspired annealing},
	journal = {Scientific Reports},
	volume = {11},
	number = {1},
	pages = {13183},
	year = {2021}
}

@article{8,
	author = {E. H. Houssein and Z. Abohashima and M. Elhoseny and W. M. Mohamed},
	title = {Hybrid quantum-classical convolutional neural network model for COVID-19 prediction using chest X-ray images},
	journal = {Journal of Computational Design and Engineering},
	volume = {9},
	number = {2},
	pages = {343--363},
	year = {2022}
}

@article{9,
	author = {A. S. Bhatia and M. K. Saggi and A. Kumar and S. Jain},
	title = {Matrix product state--based quantum classifier},
	journal = {Neural computation},
	volume = {31},
	number = {7},
	pages = {1499--1517},
	year = {2019}
}

@incollection{10,
  title={Federated quantum machine learning for drug discovery and healthcare},
  author={Saggi, Mandeep Kaur and Bhatia, Amandeep Singh and Kais, Sabre},
  booktitle={Annual Reports in Computational Chemistry},
  volume={20},
  pages={269--322},
  year={2024},
  publisher={Elsevier}
}

@article{11,
	author = {E. Akpinar and M. Oduncuoglu},
	title = {Beyond Limits: Charting New Horizons in Glioma Tumor Classification through Hybrid Quantum Computing with The Cancer Genome Atlas (TCGA) Data},
	journal = {in press},
	year = {2024}
}

@article{12,
	author = {J. W. Chen and J. Dhahbi},
	title = {Lung adenocarcinoma and lung squamous cell carcinoma cancer classification, biomarker identification, and gene expression analysis using overlapping feature selection methods},
	journal = {Scientific reports},
	volume = {11},
	number = {1},
	pages = {13323},
	year = {2021}
}

@article{13,
	author = {Z. Huang and L. Chen and C. Wang},
	title = {Classifying lung adenocarcinoma and squamous cell carcinoma using RNA-Seq data},
	journal = {Cancer Stud Mol Med Open J},
	volume = {3},
	number = {2},
	pages = {27--31},
	year = {2017}
}

@article{14,
  title={An integrative analysis of DNA methylation and gene expression to predict lung adenocarcinoma prognosis},
  author={Xu, Liexi and Huang, Zhengrong and Zeng, Zihang and Li, Jiali and Xie, Hongxin and Xie, Conghua},
  journal={Frontiers in Genetics},
  volume={13},
  pages={970507},
  year={2022},
  publisher={Frontiers Media SA}
}

@article{15,
	author = {Goldman, M. and Craft, B. and Hastie, M. and Repečka, K. and McDade, F. and Kamath, A. and Banerjee, A. and Luo, Y. and Rogers, D. and Brooks, A. and Others},
	title = {Visualizing and interpreting cancer genomics data via the Xena platform},
	journal = {Nature Biotechnology},
	volume = {38},
	pages = {675--678},
	year = {2020}
}

@article{16,
	author = {Esteller, M.},
	title = {Cancer epigenomics: DNA methylomes and histone-modification maps},
	journal = {Nature Reviews Genetics},
	volume = {8},
	pages = {286--298},
	year = {2007}
}

@article{17,
	author = {Network, C. and Others},
	title = {Comprehensive genomic characterization of squamous cell lung cancers},
	journal = {Nature},
	volume = {489},
	pages = {519},
	year = {2012}
}

@article{18,
  title={Machine learning approaches for biomarker discovery using gene expression data},
  author={Zhang, Xiaokang and Jonassen, Inge and Goks{\o}yr, Anders},
  journal={Exon Publications},
  pages={53--64},
  year={2021}
}

@article{19,
  author    = {Chaudhary, K. and Poirion, O. B. and Lu, L. and Garmire, L. X.},
  title     = {Deep learning-based multi-omics integration robustly predicts survival in liver cancer},
  journal   = {Clinical Cancer Research},
  year      = {2018},
  volume    = {24},
  number    = {6},
  pages     = {1248--1259}
}

@article{20,
  author    = {Wang, Y. and Zhang, L. and Liu, Y. and Li, J.},
  title     = {Machine learning approaches for biomarker discovery in lung adenocarcinoma using TCGA data},
  journal   = {Frontiers in Genetics},
  year      = {2023},
  volume    = {14},
  pages     = {116789}
}

@article{21,
  author    = {Li, Q. and Sun, Y. and Zhang, Y. and others},
  title     = {Support vector machine-based gene selection for lung cancer subtype classification using TCGA RNA-seq data},
  journal   = {BMC Bioinformatics},
  year      = {2023},
  volume    = {24},
  pages     = {112}
}

@article{22,
  author    = {Chen, R. and Li, J. and Sun, Y.},
  title     = {Autoencoder-based feature extraction for biomarker discovery in non-small cell lung cancer},
  journal   = {IEEE/ACM Transactions on Computational Biology and Bioinformatics},
  year      = {2022},
  volume    = {19},
  number    = {4},
  pages     = {1954--1965}
}

@article{23,
  author    = {Zhang, L. and Chen, X. and Li, Y. and others},
  title     = {Hybrid deep learning framework for biomarker identification in lung adenocarcinoma},
  journal   = {BMC Bioinformatics},
  year      = {2019},
  volume    = {20},
  pages     = {621}
}

@article{24,
  author    = {Huang, Z. and Yang, J. and Zhang, Q.},
  title     = {Multi-omics integration and machine learning for lung cancer subtype classification and biomarker discovery},
  journal   = {Scientific Reports},
  year      = {2023},
  volume    = {13},
  pages     = {3214}
}

@article{25,
  title={Multi-Omic and quantum machine learning integration for lung subtypes classification},
  author={Saggi, Mandeep Kaur and Bhatia, Amandeep Singh and Isaiah, Mensah and Gowher, Humaira and Kais, Sabre},
  journal={arXiv preprint arXiv:2410.02085},
  year={2024}
}

@inproceedings{26,
  title={MQML: Multi-Omic Quantum Machine Learning Based Cancer Classification, Biomarker Identification in Human Lung Adenocarcinoma},
  author={Saggi, Mandeep Kaur and Kais, Sabre},
  booktitle={2024 IEEE International Conference on Quantum Computing and Engineering (QCE)},
  volume={1},
  pages={1713--1720},
  year={2024},
  organization={IEEE}
}

@article{27,
	author = {Relli, V. and Trerotola, M. and Guerra, E. and Alberti, S.},
	title = {Abandoning the notion of non-small cell lung cancer},
	journal = {Trends In Molecular Medicine},
	volume = {25},
	pages = {585--594},
	year = {2019}
}

@article{28,
  title={Identification of susceptibility pathways for the role of chromosome 15q25. 1 in modifying lung cancer risk},
  author={Ji, Xuemei and Boss{\'e}, Yohan and Landi, Maria Teresa and Gui, Jiang and Xiao, Xiangjun and Qian, David and Joubert, Philippe and Lamontagne, Maxime and Li, Yafang and Gorlov, Ivan and others},
  journal={Nature Communications},
  volume={9},
  number={1},
  pages={3221},
  year={2018},
  publisher={Nature Publishing Group UK London}
}

@article{29,
  title={Neurotrophin-3 facilitates stemness properties and associates with poor survival in lung cancer},
  author={Peng, Ta-Jung and Chang Wang, Chien-Chih and Tang, Shye-Jye and Sun, Guang-Huan and Sun, Kuang-Hui},
  journal={Neuroendocrinology},
  volume={114},
  number={10},
  pages={921--933},
  year={2024},
  publisher={S. Karger AG}
}

@article{30,
  title={Neurotrophin receptor TrkB promotes lung adenocarcinoma metastasis},
  author={Sinkevicius, Kerstin W and Kriegel, Christina and Bellaria, Kelly J and Lee, Jaewon and Lau, Allison N and Leeman, Kristen T and Zhou, Pengcheng and Beede, Alexander M and Fillmore, Christine M and Caswell, Deborah and others},
  journal={Proceedings of the National Academy of Sciences},
  volume={111},
  number={28},
  pages={10299--10304},
  year={2014},
  publisher={National Academy of Sciences}
}

@article{31,
  title={Adenocarcinomas of the lung and neurotrophin system: a review},
  author={Ricci, Alberto and Salvucci, Claudia and Castelli, Silvia and Carraturo, Antonella and De Vitis, Claudia and D’Ascanio, Michela},
  journal={Biomedicines},
  volume={10},
  number={10},
  pages={2531},
  year={2022},
  publisher={MDPI}
}

@article{32,
  title={A tropomyosin receptor kinase family protein, NTRK2 is a potential predictive biomarker for lung adenocarcinoma},
  author={Wang, Xiang and Xu, Zhijie and Chen, Xi and Ren, Xinxin and Wei, Jie and Zhou, Shuyi and Yang, Xue and Zeng, Shuangshuang and Qian, Long and Wu, Geting and others},
  journal={PeerJ},
  volume={7},
  pages={e7125},
  year={2019},
  publisher={PeerJ Inc.}
}

@article{33,
  title={Methylation of the calcium channel regulatory subunit $\alpha$2$\delta$-3 (CACNA2D3) predicts site-specific relapse in oestrogen receptor-positive primary breast carcinomas},
  author={Palmieri, C and Rudraraju, B and Monteverde, M and Lattanzio, L and Gojis, O and Brizio, R and Garrone, O and Merlano, M and Syed, N and Lo Nigro, C and others},
  journal={British journal of cancer},
  volume={107},
  number={2},
  pages={375--381},
  year={2012},
  publisher={Nature Publishing Group}
}

@article{37,
  title={Quantum machine learning for electronic structure calculations},
  author={Xia, Rongxin and Kais, Sabre},
  journal={Nature communications},
  volume={9},
  number={1},
  pages={4195},
  year={2018},
  publisher={Nature Publishing Group UK London}
}

@article{38,
  title={Quantum machine-learning for eigenstate filtration in two-dimensional materials},
  author={Sajjan, Manas and Sureshbabu, Shree Hari and Kais, Sabre},
  journal={Journal of the American Chemical Society},
  volume={143},
  number={44},
  pages={18426--18445},
  year={2021},
  publisher={ACS Publications}
}

@article{39,
  title={Quantum machine learning for chemistry and physics},
  author={Sajjan, Manas and Li, Junxu and Selvarajan, Raja and Sureshbabu, Shree Hari and Kale, Sumit Suresh and Gupta, Rishabh and Singh, Vinit and Kais, Sabre},
  journal={Chemical Society Reviews},
  volume={51},
  number={15},
  pages={6475--6573},
  year={2022},
  publisher={Royal Society of Chemistry}
}

@article{40,
  title={Quantum Machine Learning for Complex Systems},
  author={Singh, Vinit and Bhatia, Amandeep Singh and Saggi, Mandeep Kaur and Sajjan, Manas and Kais, Sabre},
  journal={arXiv preprint arXiv:2602.20352},
  year={2026}
}

	
\end{document}